\documentclass[a4paper,11pt]{article}
\usepackage{jheppub} 
\usepackage{lineno}

\usepackage{amssymb}
\usepackage{amsmath}
\usepackage{mathtools}
\usepackage{epsfig}
\usepackage{hyperref}
\usepackage{breakurl}
\usepackage{bm}
\usepackage{color}
\usepackage{tikz}
\usepackage{todonotes}
\usepackage{cancel}
\usepackage[labelformat=simple]{subcaption}
\renewcommand\thesubfigure{(\arabic{subfigure})}
\usepackage[utf8]{inputenc}
\usepackage[section]{placeins}
\usetikzlibrary{snakes}
\usetikzlibrary{decorations}
\usetikzlibrary{trees}
\usetikzlibrary{decorations.pathmorphing}
\usetikzlibrary{decorations.markings}
\usetikzlibrary{external}
\usetikzlibrary{intersections}
\usetikzlibrary{shapes,arrows}
\usetikzlibrary{arrows.meta}
\usetikzlibrary{calc}
\usetikzlibrary{shapes.misc}
\usetikzlibrary{decorations.text}
\usetikzlibrary{backgrounds}
\usetikzlibrary{fadings}
\usetikzlibrary{tikzmark,calc,arrows,shapes,decorations.pathreplacing}
\usetikzlibrary{patterns.meta}
\makeatletter
\newcommand{\includegraphicslog}[2][]{%
  \typeout{FIGURE: #2}%
  \includegraphics[#1]{#2}%
}

\def\simgt{\mathrel{\lower2.5pt\vbox{\lineskip=0pt\baselineskip=0pt
           \hbox{$>$}\hbox{$\sim$}}}}
\def\simlt{\mathrel{\lower2.5pt\vbox{\lineskip=0pt\baselineskip=0pt
           \hbox{$<$}\hbox{$\sim$}}}}

\def\fig#1{Fig.~\ref{#1}}
\def\figs#1#2{Figs.~\ref{#1} and~\ref{#2}}
\def\eqn#1{Eq.~\eqref{#1}}

\def\sect#1{Sec.~\ref{#1}}

\def\tree{{\rm tree}}
\def\f{\tilde f}
\def\n{\tilde n}
\def\pol{\varepsilon}

\def\t{\tau}

\makeatother

\def\eqn#1{Eq.~\eqref{#1}}
\def\spa#1.#2{\left\langle#1\,#2\right\rangle}
\def\spb#1.#2{\left[#1\,#2\right]}
\def\sand#1.#2.#3{%
\left\langle#1{\vphantom1}\right|{#2}\left|#3\right]}%
\def\sandmp#1.#2.#3{%
\left\langle#1{\vphantom1}\right|{#2}\left|#3\right]}%
\def\sandpm#1.#2.#3{%
\left[#1{\vphantom1}\right|{#2}\left|#3\right\rangle}%
\def\sandmm#1.#2.#3{%
\left\langle#1{\vphantom1}\right|{#2}\left|#3\right\rangle}%
\def\sandpp#1.#2.#3{%
\left[#1{\vphantom1}\right|{#2}\left|#3\right]}%

\def\nn{\nonumber}

\newcommand{\be}{\begin{equation}}
\newcommand{\ee}{\end{equation}}

\newcommand{\Fig}[1]{Fig.~\ref{#1}}
\newcommand{\Eq}[1]{Eq.~\eqref{#1}}
\newcommand{\Eqs}[2]{Eqs.~\eqref{#1} and \eqref{#2}}

\def\NeqFour{{\mathcal N} = 4}

\def\NeqEight{{\mathcal N} = 8}

\renewcommand{\imath}{\mathrm{i}}

\newcommand{\cutindex}{\widehat}

\def\topbotatom#1{\hbox{\hbox to 0pt{$#1\bot$\hss}$#1\top$}}

\usetikzlibrary{snakes}
\usetikzlibrary{decorations}
\usetikzlibrary{trees}
\usetikzlibrary{decorations.pathmorphing}
\usetikzlibrary{decorations.markings}
\usetikzlibrary{external}
\usetikzlibrary{intersections}
\usetikzlibrary{shapes,arrows}
\usetikzlibrary{arrows.meta}
\usetikzlibrary{calc}
\usetikzlibrary{shapes.misc}
\usetikzlibrary{decorations.text}
\usetikzlibrary{backgrounds}
\usetikzlibrary{fadings}
\usetikzlibrary{tikzmark,calc,arrows,shapes,decorations.pathreplacing}

\tikzset{
	graviton/.style={decorate,line width=0.1mm, decoration={snake,amplitude=.3mm, segment length=0.8mm}},
	photon/.style={decorate, decoration={snake}, draw=red},
	scalar/.style={postaction={decorate},
	},
	massive/.style={postaction={decorate},
		line width=.75mm,
	},
	massless/.style={postaction={decorate},
        line width=.4mm,
	},
	masslessWithDot/.style={postaction={decorate},
        line width=.4mm,
		decoration={
			markings,
			mark=at position 0.5 with {\fill circle (2pt);}}
	},
	massiveWithDot/.style={postaction={decorate},
		line width=0.5mm,
		decoration={
			markings,
			mark=at position 0.5 with {\fill circle (2pt);}}
	},
    masslessWithArrow/.style={postaction={decorate},
		line width=0.4mm,
		decoration={
			markings,
			mark=at position 0.5 with {\arrow{latex}}}
	},
	massiveWithArrow/.style={postaction={decorate},
		line width=0.75mm,
		decoration={
			markings,
			mark=at position 0.5 with {\arrow{latex}}}
	},
	massiveLin/.style={postaction={decorate},
		double,
		thick,
		fill=white
	},
	massivePhi/.style={postaction={decorate},
		line width=0.75mm,
		dashed
	},
	masslessPhi/.style={postaction={decorate},
		dashed
	},
	unitaryCut/.style={postaction={draw,densely dashed,blue,thin},
		line width = 0.2cm,white
	},
	gluon/.style={decorate, draw=magenta,
		decoration={coil,amplitude=4pt, segment length=5pt}},
	partial ellipse/.style args={#1:#2:#3}{
		insert path={+ (#1:#3) arc (#1:#2:#3)}
	},
	cross/.style={cross out, draw=black, minimum size=2*(#1-\pgflinewidth), inner sep=0pt, outer sep=0pt},
	branchCut/.style={postaction={decorate},
		snake=zigzag,
		decoration = {snake=zigzag,segment length = 2mm, amplitude = 2mm}	
	}
	cross/.default={1pt}
}

\colorlet{mred}{black!30!red}
\colorlet{mgreen}{black!30!green}
\colorlet{mblue}{black!30!blue}
\colorlet{morange}{blue!70!red}

\newcommand{\hexagon}[6]{
\begin{tikzpicture}[scale=1]
		\coordinate (e1) at (-0.5, 0.87);
  		\coordinate (e2) at (-1, 0);	
		\coordinate (e3) at (-0.5, -0.87);
		\coordinate (e4) at (0.5, -0.87);  
        \coordinate (e5) at (1,0);  
        \coordinate (e6) at (0.5, 0.87);
		
		\coordinate (v1) at (-0.38,0.65);
  		\coordinate (v2) at (-0.75,0.);
    	\coordinate (v3) at (-0.38,-0.65);
		\coordinate (v4) at (0.38,-0.65);
        \coordinate (v5) at (0.75,0.);
        \coordinate (v6) at (0.38,0.65);

		\draw[massless] (e1) -- (v1) ;
		\draw[massless] (e2) -- (v2) ;
		\draw[massless] (e3) -- (v3) ;
		\draw[massless] (e4) -- (v4) ;
        \draw[massless] (e5) -- (v5) ;
        \draw[massless] (e6) -- (v6) ;

        \draw[massless] (v1) -- (v2) -- (v3) -- (v4) -- (v5) -- (v6) -- (v1);

        \node[above left]  at (e1) {\large $#1$};
        \node[left]  at (e2) {\large $#2$};
        \node[below left]  at (e3) {\large $#3$};
        \node[below right] at (e4) {\large $#4$};
        \node[right] at (e5) {\large $#5$};
        \node[above right] at (e6) {\large $#6$};

\end{tikzpicture}
}

\newcommand{\scbox}[4]{
\begin{tikzpicture}[scale=.9]
		\coordinate (e1) at (-0.5,0.);
  		\coordinate (e2) at (-0.5,-1.);	
		\coordinate (e3) at (1.5,-1.);
		\coordinate (e4) at (1.5,0.);

		\coordinate (v1) at (0,0);
  		\coordinate (v2) at (0,-1);
    	\coordinate (v3) at (1,-1);
		\coordinate (v4) at (1,0);

		\draw[massless] (e1) -- (v1) ;
		\draw[massless] (e2) -- (v2) ;
		\draw[massless] (e3) -- (v3) ;
		\draw[massless] (e4) -- (v4) ;

        \draw[massless] (v1) -- (v2) -- (v3) -- (v4) -- (v1);

        \node[left]  at (e1) {\large $#1$};
        \node[left]  at (e2) {\large $#2$};
        \node[right] at (e3) {\large $#3$};
        \node[right] at (e4) {\large $#4$};

\end{tikzpicture}
}

\newcommand{\scboxlabeled}[4]{
\begin{tikzpicture}[scale=1]
		\coordinate (e1) at (-0.5,0.);
  		\coordinate (e2) at (-0.5,-1.);	
		\coordinate (e3) at (1.5,-1.);
		\coordinate (e4) at (1.5,0.);

		\coordinate (v1) at (0,0);
  		\coordinate (v2) at (0,-1);
    	\coordinate (v3) at (1,-1);
		\coordinate (v4) at (1,0);

		\draw[massless] (e1) -- (v1) ;
		\draw[massless] (e2) -- (v2) ;
		\draw[massless] (e3) -- (v3) ;
		\draw[massless] (e4) -- (v4) ;

        \draw[masslessWithArrow] (v4) -- (v1) node [midway,fill=white] {$a_1$};
        \draw[masslessWithArrow] (v1) -- (v2) node [midway,fill=white] {$a_2$};
        \draw[masslessWithArrow] (v2) -- (v3) node [midway,fill=white] {$a_3$};
        \draw[masslessWithArrow] (v3) -- (v4) node [midway,fill=white] {$a_4$};

        \node[left]  at (e1) {\large $#1$};
        \node[left]  at (e2) {\large $#2$};
        \node[right] at (e3) {\large $#3$};
        \node[right] at (e4) {\large $#4$};

\end{tikzpicture}
}

\newcommand{\boxegA}[4]{
\begin{tikzpicture}[scale=1]
		\coordinate (e1) at (-0.5,0.);
  		\coordinate (e2) at (-0.5,-1.);	
		\coordinate (e3) at (1.5,-1.);
		\coordinate (e4) at (1.5,0.);

		\coordinate (v1) at (0,0);
  		\coordinate (v2) at (0,-1);
    	\coordinate (v3) at (1,-1);
		\coordinate (v4) at (1,0);

		\coordinate (l1) at (.5,.1);
  		\coordinate (l2) at (-.1,-.5);
    	\coordinate (l3) at (.5,-1.1);
		\coordinate (l4) at (1.1,-.5);
  
		\draw[massless] (e1) -- (v1) ;
		\draw[massless] (e2) -- (v2) ;
		\draw[massless] (e3) -- (v3) ;
		\draw[massless] (e4) -- (v4) ;

        \draw[massless] (v4) -- (v1);
        \draw[massless] (v1) -- (v2);
        \draw[massless] (v2) -- (v3) ;
        \draw[masslessWithDot] (v3) -- (v4) ;

        \node[left]  at (e1) {\large $p_{#1}$};
        \node[left]  at (e2) {\large $p_{#2}$};
        \node[right] at (e3) {\large $p_{#3}$};
        \node[right] at (e4) {\large $p_{#4}$};

        \node[above]  at (l1) {\large $\ell$};
        \node[left]  at (l2) {\large $\ell{-}p_{#1}$};
        \node[below] at (l3) {\large $\ell{-}p_{#1 #2}$};
        \node[right] at (l4) {\large $\ell{-}p_{#1 #2 #3}$};

\end{tikzpicture}
}

\newcommand{\boxegB}[4]{
\begin{tikzpicture}[scale=1]
		\coordinate (e1) at (-0.5,0.);
  		\coordinate (e2) at (-0.5,-1.);	
		\coordinate (e3) at (1.5,-1.);
		\coordinate (e4) at (1.5,0.);

		\coordinate (v1) at (0,0);
  		\coordinate (v2) at (0,-1);
    	\coordinate (v3) at (1,-1);
		\coordinate (v4) at (1,0);

		\coordinate (l1) at (.5,.1);
  		\coordinate (l2) at (-.1,-.5);
    	\coordinate (l3) at (.5,-1.1);
		\coordinate (l4) at (1.1,-.5);
  
		\draw[massless] (e1) -- (v1) ;
		\draw[massless] (e2) -- (v2) ;
		\draw[massless] (e3) -- (v3) ;
		\draw[massless] (e4) -- (v4) ;

        \draw[massless] (v4) -- (v1);
        \draw[masslessWithDot] (v1) -- (v2);
        \draw[massless] (v2) -- (v3) ;
        \draw[massless] (v3) -- (v4) ;

        \node[left]  at (e1) {\large $p_{#1}$};
        \node[left]  at (e2) {\large $p_{#2}$};
        \node[right] at (e3) {\large $p_{#3}$};
        \node[right] at (e4) {\large $p_{#4}$};

        \node[above]  at (l1) {\large $\ell$};
        \node[left]  at (l2) {\large $\ell{-}p_{#1}$};
        \node[below] at (l3) {\large $\ell{-}p_{#1 #2}$};
        \node[right] at (l4) {\large $\ell{-}p_{#1 #2 #3}$};

\end{tikzpicture}
}

\newcommand{\sctri}[4]{
\begin{tikzpicture}[scale=.9]
		\coordinate (e1) at (-0.5,0.);
  		\coordinate (e2) at (-0.5,-1.);	
		\coordinate (e3) at (1.5,-1.);
		\coordinate (e4) at (1.5,0.);

		\coordinate (v1) at (0.5,0);
    	\coordinate (v2) at (0,-1);
		\coordinate (v3) at (1,-1);

		\draw[massless] (e1) -- (v1) ;
		\draw[massless] (e2) -- (v2) ;
		\draw[massless] (e3) -- (v3) ;
		\draw[massless] (e4) -- (v1) ;

        \draw[massless] (v1) -- (v2) ;
        \draw[massless] (v2) -- (v3) ;
        \draw[massless] (v3) -- (v1) ;

        \node[left]  at (e1) {\large $#1$};
        \node[left]  at (e2) {\large $#2$};
        \node[right] at (e3) {\large $#3$};
        \node[right] at (e4) {\large $#4$};

\end{tikzpicture}
}

\newcommand{\sctrilabeled}[4]{
\begin{tikzpicture}[scale=1]
		\coordinate (e1) at (-0.5,0.);
  		\coordinate (e2) at (-0.5,-1.);	
		\coordinate (e3) at (1.5,-1.);
		\coordinate (e4) at (1.5,0.);

		\coordinate (v1) at (0.5,0);
    	\coordinate (v2) at (0,-1);
		\coordinate (v3) at (1,-1);
  
		\draw[massless] (e1) -- (v1) ;
		\draw[massless] (e2) -- (v2) ;
		\draw[massless] (e3) -- (v3) ;
		\draw[massless] (e4) -- (v1) ;

        \draw[masslessWithArrow] (v1) -- (v2) node [midway,fill=white] {$a_2$};
        \draw[masslessWithArrow] (v2) -- (v3) node [midway,fill=white] {$a_3$};
        \draw[masslessWithArrow] (v3) -- (v1) node [midway,fill=white] {$a_4$};

        \node[left]  at (e1) {\large $#1$};
        \node[left]  at (e2) {\large $#2$};
        \node[right] at (e3) {\large $#3$};
        \node[right] at (e4) {\large $#4$};

\end{tikzpicture}
}

\newcommand{\scbub}[4]{
\begin{tikzpicture}[scale=.9]
		\coordinate (e1) at (-0.5,0.);
  		\coordinate (e2) at (-0.5,-1.);	
		\coordinate (e3) at (1.5,-1.);
		\coordinate (e4) at (1.5,0.);

		\coordinate (v1) at (0.5,0);
		\coordinate (v3) at (0.5,-1);

        \draw[massless] (v1) to[out=180,in=180] (v3);
        \draw[massless] (v3) to[out=0,in=0] (v1);

		\draw[massless] (e1) -- (v1) ;
		\draw[massless] (e2) -- (v3) ;
		\draw[massless] (e3) -- (v3) ;
		\draw[massless] (e4) -- (v1) ;
        
        \node[left]  at (e1) {\large $#1$};
        \node[left]  at (e2) {\large $#2$};
        \node[right] at (e3) {\large $#3$};
        \node[right] at (e4) {\large $#4$};

\end{tikzpicture}
}

\newcommand{\scbublabeled}[4]{
\begin{tikzpicture}[scale=1]
		\coordinate (e1) at (-0.5,0.);
  		\coordinate (e2) at (-0.5,-1.);	
		\coordinate (e3) at (1.5,-1.);
		\coordinate (e4) at (1.5,0.);

		\coordinate (v1) at (0.5,0);
		\coordinate (v3) at (0.5,-1);

        \draw[massless] (v1) to[out=180,in=180] (v3);
        \draw[massless] (v3) to[out=0,in=0] (v1);

        \node[fill=white] at (.75,-.5) {$a_4$};
        \node[fill=white] at (.25,-.5) {$a_2$};

		\draw[massless] (e1) -- (v1) ;
		\draw[massless] (e2) -- (v3) ;
		\draw[massless] (e3) -- (v3) ;
		\draw[massless] (e4) -- (v1) ;
        
        \node[left]  at (e1) {\large $#1$};
        \node[left]  at (e2) {\large $#2$};
        \node[right] at (e3) {\large $#3$};
        \node[right] at (e4) {\large $#4$};

\end{tikzpicture}
}

\newcommand{\sctridangling}[4]{
\begin{tikzpicture}[scale=1]
		\coordinate (e1) at (-0.5,0.);
  		\coordinate (e2) at (-0.5,-1.);	
		\coordinate (e3) at (1.5,-1.);
		\coordinate (e4) at (1.5,0.);

		\coordinate (v1) at (-.25,-.5);
        \coordinate (v2) at (.25,-.5);
    	\coordinate (v3) at (1,-1);
		\coordinate (v4) at (1,0);

		\draw[massless] (e1) -- (v1) ;
		\draw[massless] (e2) -- (v1) ;
		\draw[massless] (e3) -- (v3) ;
		\draw[massless] (e4) -- (v4) ;

        \draw[massless] (v1) -- (v2) -- (v3) -- (v4) -- (v2);

        \node[left]  at (e1) {\large $#1$};
        \node[left]  at (e2) {\large $#2$};
        \node[right] at (e3) {\large $#3$};
        \node[right] at (e4) {\large $#4$};

\end{tikzpicture}
}

\newcommand{\sctadA}{
\begin{tikzpicture}[scale=1]
		\coordinate (e1) at (-0.5,0.);
  		\coordinate (e2) at (-0.5,-1.);	
		\coordinate (e3) at (1.5,-1.);
		\coordinate (e4) at (1.5,0.);

		\coordinate (v1) at (0,-.5);
    	\coordinate (v2) at (1,-.5);
		\coordinate (v3) at (0.5,0);
        \coordinate (v3p) at (0.5,-.5);

		\draw[massless] (e1) -- (v1) ;
		\draw[massless] (e2) -- (v1) ;
		\draw[massless] (e3) -- (v2) ;
		\draw[massless] (e4) -- (v2) ;

        \draw[massless] (v1) -- (v2) ;
        \draw[massless] (v3p) -- (v3) ;
        \draw[fill=white,line width=.4mm] (v3) circle (2.5mm);

        \node[left]  at (e1) {};
        \node[left]  at (e2) {};
        \node[right] at (e3) {};
        \node[right] at (e4) {};

\end{tikzpicture}
}

\newcommand{\sctadPinchA}{
\begin{tikzpicture}[scale=1]
		\coordinate (e1) at (-0.5,0.);
  		\coordinate (e2) at (-0.5,-1.);	
		\coordinate (e3) at (1.5,-1.);
		\coordinate (e4) at (1.5,0.);

		\coordinate (v1) at (0,-.5);
    	\coordinate (v2) at (1,-.5);
		\coordinate (v3) at (0.5,-.25);

		\draw[massless] (e1) -- (v1) ;
		\draw[massless] (e2) -- (v1) ;
		\draw[massless] (e3) -- (v2) ;
		\draw[massless] (e4) -- (v2) ;

        \draw[massless] (v1) -- (v2) ;
        \draw[fill=white,line width=.4mm] (v3) circle (2.5mm);

        \node[left]  at (e1) {};
        \node[left]  at (e2) {};
        \node[right] at (e3) {};
        \node[right] at (e4) {};

\end{tikzpicture}
}

\newcommand{\sctadB}{
\begin{tikzpicture}[scale=1]
		\coordinate (e1) at (-0.5,0.);
  		\coordinate (e2) at (-0.5,-1.);	
		\coordinate (e3) at (1.5,-1.);
		\coordinate (e4) at (1.5,0.);

		\coordinate (v1) at (0,-.5);
    	\coordinate (v2) at (1,-.5);
		\coordinate (v3) at (0.125,0.125);
        \coordinate (v3p) at (-.25,-.25);

		\draw[massless] (e1) -- (v1) ;
		\draw[massless] (e2) -- (v1) ;
		\draw[massless] (e3) -- (v2) ;
		\draw[massless] (e4) -- (v2) ;

        \draw[massless] (v1) -- (v2) ;
        \draw[massless] (v3p) -- (v3) ;
        \draw[fill=white,line width=.4mm] (v3) circle (2.5mm);

        \node[left]  at (e1) {};
        \node[left]  at (e2) {};
        \node[right] at (e3) {};
        \node[right] at (e4) {};

\end{tikzpicture}
}

\newcommand{\scbubdrop}[4]{
\begin{tikzpicture}[scale=1]
		\coordinate (e1) at (-0.5,-.5);
		\coordinate (e2) at (1.5,-1.);
        \coordinate (e3) at (1.5,-.5);	
		\coordinate (e4) at (1.5,0.);  

		\coordinate (v1) at (0,-.5);
    	\coordinate (v3) at (1,-.5);
  
		\draw[massless] (e1) -- (v1) ;
		\draw[massless] (e2) -- (v3) ;
		\draw[massless] (e3) -- (v3) ;
		\draw[massless] (e4) -- (v3) ;

        \draw[massless] (v1) to[out=90,in=90] (v3) to[out=-90,in=-90] (v1);

        \node[left]  at (e1) {\large $#1$};
        \node[right] at (e2) {\large $#2$};
        \node[right] at (e3) {\large $#3$};
        \node[right] at (e4) {\large $#4$};

\end{tikzpicture}
}

\newcommand{\scbubdanglingdrop}[4]{
\begin{tikzpicture}[scale=1]
		\coordinate (e1) at (-0.5,-.5);
		\coordinate (e2) at (1.9,-1.);
        \coordinate (e3) at (1.9,-.5);	
		\coordinate (e4) at (1.9,0.);  

		\coordinate (v1) at (0,-.5);
    	\coordinate (v3) at (1,-.5);
        \coordinate (v4) at (1.4,-.5);
  
		\draw[massless] (e1) -- (v1) ;
		\draw[massless] (e2) -- (v4) ;
		\draw[massless] (e3) -- (v4) ;
		\draw[massless] (e4) -- (v4) ;

        \draw[massless] (v1) to[out=90,in=90] (v3) to[out=-90,in=-90] (v1);
        \draw[massless] (v3) -- (v4) ;

        \node[left]  at (e1) {\large $#1$};
        \node[right] at (e2) {\large $#2$};
        \node[right] at (e3) {\large $#3$};
        \node[right] at (e4) {\large $#4$};

\end{tikzpicture}
}

\begin{document}


\title{Global Bases for Nonplanar Loop Integrands, 
Generalized Unitarity, \newline
and the Double Copy to All Loop Orders}

\author[1]{Zvi Bern,}
\affiliation[1]{
Mani L. Bhaumik Institute for Theoretical Physics,
University of California at Los Angeles, \newline
Los Angeles, CA 90095, USA}
\emailAdd{bern@physics.ucla.edu}

\author[1]{Enrico Herrmann,}
\emailAdd{eh10@g.ucla.edu}

\author[2,3]{Radu Roiban,}
\affiliation[2]{Institute for Gravitation and the Cosmos,
Pennsylvania State University,\newline 
University Park, PA 16802, USA}
\affiliation[3]{Institute for Computational and Data Sciences,
Pennsylvania State University,\newline 
University Park, PA 16802, USA}
\emailAdd{radu@phys.psu.edu}

\author[1]{Michael~S.~Ruf,}
\emailAdd{mruf@physics.ucla.edu}

\author[4]{Mao Zeng}
\affiliation[4]{Higgs Centre for Theoretical Physics, University of Edinburgh, Edinburgh, EH9 3FD, UK}
\emailAdd{mzeng@ed.ac.uk}

\abstract{We introduce a constructive method for defining a global loop-integrand basis for scattering amplitudes, encompassing both planar and nonplanar contributions. Our approach utilizes a graph-based framework to establish a well-defined, non-redundant basis of integrands. This basis, constructed from a chosen set of non-redundant graphs together with a selection of irreducible scalar products, provides clear insights into various physical properties of scattering amplitudes and proves useful in multiple contexts, such as on-shell Ward identities and manifesting gauge-choice independence. A key advantage of our integrand basis is its ability to streamline the generalized unitarity method. Specifically, we can directly read off the coefficients of basis elements without resorting to ans\"atze or solving linear equations. This novel approach allows us to lift generalized unitarity cuts---expressed as products of tree amplitudes---to loop-level integrands, facilitating the use of the tree-level double copy to generate complete gravitational integrands at any loop order. This method circumvents the difficulties in identifying complete higher-loop-order gauge-theory integrands that adhere to the color-kinematics duality. Additionally, our cut-based organization is well-suited for expansion in hard or soft limits, aiding in the exploration of ultraviolet or classical limits of scattering amplitudes.}

\maketitle

%
\newpage
\section{Introduction}
%

Modern scattering amplitude methods have exposed a remarkably rich set of new structures and symmetries in both generic and particular field theories such as $\NeqFour$ super-Yang-Mills theory. Examples include the duality between color and kinematics and the associated double copy~\cite{Kawai:1985xq, Bern:2008qj, Bern:2010ue, Bern:2019prr}, geometric formulations~\cite{Arkani-Hamed:2010wgm, Arkani-Hamed:2013jha} such as the amplituhedron, positivity properties~\cite{Arkani-Hamed:2012zlh, Arkani-Hamed:2014dca}, string theory-like descriptions~\cite{Witten:2003nn, Roiban:2004yf, Cachazo:2013iea}, and dual conformal invariance~\cite{Drummond:2008vq}. There has also been impressive progress in calculations of scattering amplitudes and related quantities in gauge and gravity theories (see, e.g., Refs.~\cite{Berger:2008sj, Berger:2010zx, Arkani-Hamed:2010zjl, Bern:2005iz,   Bourjaily:2019iqr, Abreu:2020xvt,  Bern:2014sna,  Bern:2019nnu, Dixon:2022rse, Dixon:2023kop} for various examples).  A key concept is to focus on gauge-invariant quantities, particularly scattering amplitudes, to identify valuable novel features of the theories. Another important idea is to derive these invariants through iterative procedures such as generalized unitarity~\cite{Bern:1994zx, Bern:1994cg, Bern:1995db, Bern:1997sc, Britto:2004nc}, and on-shell recursion~\cite{Britto:2004ap, Britto:2005fq}, which recycle previously computed quantities and exploit uncovered structures. These approaches have led to many new computations in perturbative QCD, supersymmetric gauge theory, supergravity, etc. They have also led to new ways of classifying quantum field theories~\cite{Cheung:2016drk, Bern:2019prr}, and a renewed understanding of various properties, such as soft theorems (see, e.g.,~Refs.~\cite{Cachazo:2014fwa, Cheung:2014dqa, Cheung:2016drk, Kampf:2019mcd, Cheung:2021yog}). For a recent review of modern on-shell methods, see Ref.~\cite{Travaglini:2022uwo}. 

In this paper, we describe a reorganization of the generalized unitarity method for constructing loop integrands, which exposes certain uniqueness properties. 
The generalized unitarity method makes use of the observation that integrands of loop amplitudes are rational functions to all loop orders. Thus, they may be determined by knowing the location of the poles and their residues and suitably accounting for the theory's high-energy properties. 
The Feynman-diagrammatic representation of loop integrands indicates that its poles appear in kinematic configurations where internal propagators go on shell. The corresponding residues are, in general, products of lower-loop integrands summed over physical intermediate states. 
Such products are usually referred to as generalized cuts; applying this reasoning iteratively in $D$ dimensions using dimensional regularization, it follows that generalized cuts that are sums of products of \emph{tree} amplitudes can determine loop integrands. We will typically use the phrase {\em generalized cuts} only for such residues. 
A \emph{spanning set of cuts} (or, shorter, \emph{spanning cuts}) is a subset of all generalized cuts that can be used to reconstruct integrands in massless theories. For massive theories additional data is necessary; we will review this in Secs.~\ref{sec:global_basis} and \ref{sec:unitarity}. The unitarity method has been applied to a variety of problems, including QCD (see, e.g., Refs.~\cite{Bern:1997sc, Badger:2008cm, Berger:2008sj, Giele:2008ve, Ellis:2009zw, Berger:2010zx, Abreu:2020xvt, Abreu:2021asb}, super-Yang-Mills theory (see, e.g., Refs.~\cite{Bern:1994zx, Bern:1994cg, Drummond:2008bq, Bern:2012di, Carrasco:2021otn}, supergravity (see e.g. Refs.~\cite{Bern:1998ug, Bern:2009kd, Carrasco:2012ca, Bern:2014sna, Johansson:2017bfl}), and gravitational radiation~(see, e.g., Refs.~\cite{Bern:2019nnu, Bern:2019crd}) and has been useful for exposing various theoretical structures (see, e.g., Refs.~\cite{Anastasiou:2003kj, Bern:2005iz, Bern:2008qj, Bern:2010ue, Arkani-Hamed:2012zlh}).

A typical application of the generalized unitarity method begins by identifying a relevant spanning set of cuts and evaluating them by sewing together tree amplitudes.
One procedure for building an integrand starts with a diagrammatic representation of the integrand with ans\"atze for the diagrams' numerators. The ans\"atze are constrained, e.g., by imposing that a certain power counting is manifest in each diagram, or that the diagram numerators satisfy the duality between color and kinematics. The free coefficients are determined by comparing the generalized cuts of the integrand ansatz with the generalized cuts obtained by sewing together tree amplitudes, see, e.g.,~Refs.~\cite{Bern:2007ct, Bern:2012uc}. This entails solving a linear system that can grow substantially at higher loop orders but has the advantage that various desired properties can be imposed on the ansatz and, consequently, on the resulting amplitude.   

An alternative, constructive approach to finding an integrand starts by approximating it by one of the members of the spanning set of cuts and then systematically adding terms to reproduce the other spanning cuts, as in Ref.~\cite{Bern:2004cz}. That is, the information in the spanning cuts is (\emph{merged}) into an off-shell integrand. In this work, we significantly improve this traditional cut-merging procedure by enhancing it with a basis of planar and nonplanar loop integrands, where the overlap between different cuts is easily determined. Cut merging no longer requires solving large systems of linear equations. 

There are further variations on generalized unitarity, including the method of maximal cuts~\cite{Bern:2007ct, Bern:2008pv, Bern:2010tq}, single-cut constructions~\cite{Catani:2008xa, NigelGlover:2008ur, Bierenbaum:2010cy, Caron-Huot:2010fvq, Britto:2010um, Baadsgaard:2015twa}, and prescriptive unitarity~\cite{Bourjaily:2017wjl, Bourjaily:2019iqr}, which seek to diagonalize the generalized cuts.

Integrands are usually organized into Feynman-like diagrams, according to the propagators present. For each such diagram, one must choose the independent loop momenta---the `loop-momentum labels'. Generally, a natural choice of labels is unclear, let alone a uniform choice that can expose interrelations between different diagrams required, e.g., by the properties of the theory, such as gauge invariance.  
In special cases, such as the planar sector of multi-loop scattering amplitudes with a fixed ordering of external legs, a natural choice is dual momentum variables valid for all contributing diagrams~\cite{Drummond:2006rz, Drummond:2008vq, Arkani-Hamed:2010zjl}. This, for example, aids in formulating an on-shell recursion of the planar $\NeqFour$ super-Yang--Mills integrands to all loop orders~\cite{Arkani-Hamed:2010zjl}. Analogous choices beyond the planar sector are far less clear (see, however, Refs.~\cite{Bern:2015ple, Arkani-Hamed:2014via, Bern:2014kca, Bern:2018oao, Franco:2015rma}). Here, we partially address this issue.  While we do not find a canonical choice of loop momenta (or more generally loop integration variables), we demonstrate that \emph{any} choice of complete (and not overcomplete) integrand basis is sufficient to expose the global properties of integrands and to streamline their construction.
In particular, we define a basis for all terms in an integrand, including those associated with non-planar diagrams, by a four-step process that can be summarized as follows: 
\begin{enumerate}
\item 
\textbf{Initial list of top-level graphs:} 
Identify \emph{top-level graphs}, which have the maximum number of independent propagators carrying loop momentum. 

\item 
\textbf{Daughter-graph identification:} For all top-level graphs, iterate edge collapses to daughter graphs until one reaches \emph{bottom-level} graphs. Among the resulting graphs, choose a set of non-isomorphic (\emph{representative}) graphs with fixed external lines, not mapped into each other by graph isomorphisms. 

\item 
\textbf{Choice of basis of Lorentz invariants:}   For each representative graph, \emph{choose} a basis of loop-momentum dependent Lorentz-invariant scalar products in terms of inverse propagators and irreducible scalar products (ISPs).

\item 
\textbf{Global basis of loop integrands:} For each representative graph, list all possible products of propagators raised to arbitrary positive integer power multiplied by a generic monomial in the ISPs. Combining the resulting basis elements of all representative graphs yields the \emph{global basis} of non-planar integrands.

\end{enumerate}
Further details and qualifications of the construction are described in detail in \sect{sec:global_basis} below. We will show that the availability of a global integrand basis dramatically simplifies the construction of multi-loop integrands in general quantum field theories. Writing all generalized cuts in a global basis significantly simplifies their merging into a complete integrand by aligning terms in the cuts with terms in the integrand. This approach allows us to derive the full integrand by sequentially reading off the coefficients of the elements of the global basis from a spanning set of cuts, removing double counts, and ensuring consistency\footnote{That is, the coefficient of any basis element is identical when it appears in multiple generalized cuts, up to possible combinatorial factors.}. Moreover, all physically-equivalent loop integrands must be identical when written in the same global basis; thus, our construction leads to the manifest cancellation of gauge-fixing parameters and spurious singularities which are unrelated to those that can appear from applying integration-level identities~\cite{Passarino:1978jh, Tkachov:1981wb, Chetyrkin:1981qh, Laporta:2000dsw}. We demonstrate there benefit in subsequent sections.

The study of ultraviolet (UV) properties of gravitational theories has been an important motivation for the development of color-kinematics duality and the associated double-copy constructions~\cite{Bern:2008qj, Bern:2010yg, Bern:2019prr}. Indeed, if gauge-theory amplitudes in color-dual form are available, the double copy immediately constructs gravitational integrands and thus provides direct computational access to the coefficients of potential divergences. The duality and double copy have been proven to hold at tree level~\cite{Cachazo:2012uq, Chen:2011jxa, Feng:2010my, Stieberger:2009hq, Bjerrum-Bohr:2009ulz, Bern:2010yg}, and amplitudes manifesting the duality are available for any multiplicity. While many examples have been worked out at loop level (see, e.g., Refs.~\cite{Bern:2012uf, Boels:2012ew, Carrasco:2012ca, Bjerrum-Bohr:2013iza, Bern:2013yya, Chiodaroli:2013upa, Chiodaroli:2014xia, Badger:2015lda, Mafra:2015mja,  He:2015wgf, Mogull:2015adi, Chiodaroli:2015rdg, Yang:2016ear, Chiodaroli:2017ngp, Chiodaroli:2017ehv, He:2017spx, Johansson:2017bfl, Chiodaroli:2018dbu, Carrasco:2020ywq, Lin:2021kht, Li:2022tir, Edison:2022smn, Edison:2022jln}), finding such color-dual representations can be problematic~\cite{Bern:2017ucb}. Moreover, even when they have been constructed, they can obscure desirable properties of the theory, such as the power counting~\cite{Mogull:2015adi}. 

The basis-centric unitarity construction outlined here leverages the established existence of tree-level amplitudes that exhibit the duality and associated double copy~\cite{Bjerrum-Bohr:2010pnr, Mafra:2011kj, Cachazo:2013iea, Du:2017kpo, Edison:2020ehu}, providing a systematic way to overcome these challenges. It also presents a more efficient alternative to the generalized double-copy method~\cite{Bern:2017yxu, Bern:2017ucb}, which, while designed for a similar purpose, becomes increasingly difficult to apply beyond the next-to-next-to-maximal cut level.

Many individual aspects of our construction have already appeared in the literature. The organization of integrands into parent and daughter diagrams is a central part of the maximal-cut method~\cite{Bern:2007ct}, which determines contributions to the integrand according to the number of canceled propagators. 
The simplification of the cut-merging process in an integrand basis is part of the prescriptive unitarity method which aims to construct a basis so that the integrand terms can be determined one at a time~\cite{Bourjaily:2017wjl}. 
The organization of our integrand basis elements in terms of the inverse propagator basis is natural for subsequent integration-by-parts (IBP) reduction. For efficiency, it is also natural to remove as much redundancy in the set of graphs of a given problem, see, e.g., Refs.~\cite{Seidensticker:1999bb, Hoff:2015kub, Borowka:2017idc, Gerlach:2022qnc, Maheria:2022dsq, Shtabovenko:2023idz}. Cut-level color-kinematics duality and double copy as a means to bypass difficulties with finding gauge-theory loop integrands manifesting color-kinematics duality first appeared in Ref.~\cite{Bern:2015ooa}. Our approach combines these ideas into a streamlined process that applies to all quantum field theories and specifically enables the double-copy construction at any loop order whenever \emph{tree-level} color-kinematics duality is present.

The remainder of this work is structured as follows: In \sect{sec:global_basis}, we introduce the main result of our work: The systematic construction of a global basis for loop integrands. We abstractly state our algorithm in \sect{subsec:map_algo_scalar}, before discussing illustrative examples at one and two loops in \sect{subsec:mapping_explicit}. As concrete applications of global integrand bases, we discuss the gauge-choice independence of Feynman integrands (\sect{subsec:xiDrop}), on-shell Ward identities at integrand level (\sect{subsec:spinning_external_states}), and uniqueness properties of integrands (\sect{subsec:integrand_cleanup}). In \sect{sec:unitarity}, we utilize the global integrand basis to streamline cut merging in the context of generalized unitarity, resulting in a method to directly lift unitarity cuts to loop integrands without the need for linear algebra. We illustrate our general discussions on explicit examples in scalar QED (\sect{subsec:SQED_cut_construction}) and the nonlinear sigma model (NLSM) (\sect{subsec:NLSM}). For the NLSM, we furthermore show through two-loop order that the Adler zero can be exposed at the integrand level. As a final major application of our integrand basis construction, we harness the streamlined unitarity setup to export the proven tree-level double-copy property to all loop orders in \sect{sec:double_copy}. Our strategy circumvents the notoriously difficult task of constructing off-shell BCJ representations at higher loop order opening the door to the construction of high-loop (super-)gravity integrands with bespoke spectrum relevant both for classical physics and for the study of UV properties of supersymmetric gravity theories.  We give our conclusions and an outlook to future work in \sect{sec:conclusions}.

We provide a list of ancillary files with explicit integrands for examples discussed in this work. The files \texttt{SQED\textunderscore SSAA\textunderscore1L.m},  \texttt{SQED\textunderscore SSSS\textunderscore1L.m}, and \texttt{SQED\textunderscore SSSS\textunderscore2L.m} contain integrands for one-loop Compton scattering, and charged scalar scattering in QED at one- and two-loops, respectively. The files \texttt{StandardTopos\textunderscore SSSS\textunderscore1L.m} and \texttt{StandardTopos\textunderscore SSSS\textunderscore2L.m} contain all graph and labeling definitions relevant for one- and two-loop four-particle scattering of massless scalars. The relevant representative sectors for these examples are given in the files \texttt{RepresentativeSectors\textunderscore SSSS\textunderscore1L.m} and \texttt{RepresentativeSectors\textunderscore SSSS\textunderscore2L.m}. Finally, we provide the result for the two-loop four-scalar integrand in general relativity expanded in the classical limit and mapped to a global integrand basis in the file \texttt{GR\textunderscore SSSS\textunderscore 2L\textunderscore classical\textunderscore pot.m}.

%
\section{Global Bases of Planar and Nonplanar Loop Integrands}
\label{sec:global_basis}
%

Our understanding of perturbative scattering amplitudes in quantum field theory has significantly advanced in the last decades. The generalized unitarity method (see \sect{sec:unitarity}), in particular, provides a systematic means for building \emph{integrands} at any loop order~\cite{Bern:1994zx, Bern:1994cg, Britto:2004nc}. Early explorations at one-loop level~\cite{Bern:1992em, Bern:1996je, Bern:1996fj, Bern:1996ja, Roiban:2004ix, Bidder:2005ri} ultimately led to the discovery of tree-level recursion \cite{Britto:2004ap,Britto:2005fq} along many other insights, see, e.g.,~Refs.~\cite{Drummond:2006rz, Alday:2007hr, Drummond:2008vq, Beisert:2010jr, Arkani-Hamed:2012zlh, Arkani-Hamed:2013jha}. Although not originally described in this form, one key feature of generalized unitarity is that loop integrands---roughly `the sum of Feynman diagrams' prior to loop integration---are meaningful and interesting quantities. As we stressed in the introduction, the notion of global loop-momentum (or integration) variables is not well defined for amplitudes or Green's functions beyond the planar limit. The arbitrary choice of loop momenta for individual Feynman diagrams cannot be used (at the time of this writing) to globally align integration variables between different diagrams. In planar theories, there is a canonical choice due to the existence of dual variables, see, e.g., Refs.~\cite{Broadhurst:1993ib, Drummond:2006rz, Arkani-Hamed:2010zjl, Arkani-Hamed:2010pyv}. In a unitarity-based framework, integrands are determined by their generalized unitarity cuts, which in turn are products of tree amplitudes. For each such cut, the momentum variables are uniquely specified by the external and cut (on-shell) propagators. From this perspective, the difficulty with global loop-momentum variables for non-planar amplitudes is a consequence of the inability to correlate the unique momentum labels in different generalized cuts.

Given a sufficiently large basis of loop integrands $\mathfrak{B}$ (viewed as a vector space of rational functions), any loop amplitude integrand $\mathcal{A}$  can be represented as a linear combination of basis elements, i.e. schematically
\begin{align}
\mathcal{A}=\sum_{i} a_{i}\,\mathfrak{b}^i\,, \qquad \text{ where }\mathfrak{b}^i\in\mathfrak{B}\, ,
\label{schematic_unitarity_construction}
\end{align}
where we left implicit the form of the basis elements $\mathfrak{b}^i$.
In a unitarity-based framework, the coefficients $a_{i}$ are determined by matching all cuts, see, e.g.,~Refs.~\cite{Bern:1995db, Bern:2011qt}, and \sect{sec:unitarity} for more detailed discussions. A priori, one can even represent amplitudes in terms of an overcomplete set of integrands. 

As has been stressed in, e.g., Ref.~\cite{Bourjaily:2020qca}, a practically useful perspective is that there exists a finite-dimensional basis $\mathfrak{B}$ at any loop order in which all scattering amplitudes of a given theory can be represented.  Such bases are linked to Feynman-diagram-like graphs with loop-dependent numerator polynomials that are then fixed against unitarity cuts. Examples of such bases, suitable for representing all amplitudes in the Standard Model at one and two loops, have been discussed in various contexts~\cite{Ossola:2006us, Mastrolia:2011pr, Mastrolia:2012wf, Feng:2012bm, Mastrolia:2013kca, Kleiss:2012yv, Ita:2015tya, Bourjaily:2020qca} and implemented in practical codes, e.g.~\cite{Berger:2008sj, Abreu:2020xvt}. 

Most of the aforementioned approaches rely on a diagrammatic representation to define integrand bases (some of which are strictly four-dimensional and based on certain power-counting restrictions). Our approach to finding a global integrand basis, which we describe shortly, is also diagram based. As such, we do not solve the problem of finding canonical global variables analogous to the planar dual variables. We focus instead on defining a basis in which integrands are unique, i.e., a basis that is complete and {\em not} overcomplete, together with a practical implementation of expressing loop integrands in said basis.  
To obtain the global integrand basis, we apply an iterative mapping procedure to multi-loop integrands that systematically expresses all terms of an integrand in terms of a basis of representative graphs (allowing for doubled propagators as well as arbitrary but a priori fixed numerator polynomials in the irreducible scalar products). Our key observation is that, in such a basis, the coefficient of every term of the integrand basis is unique and independent of the method used to construct the integrand. 
Regardless of whether we begin with, for instance, light-cone gauge Feynman rules, covariant Feynman rules, or the unitarity method, mapping the integrand to a basis yields identical expressions.\footnote{It is worth noting that the yet unresolved question~\cite{Bassetto:1998uv} of whether the Mandelstam-Leibbrandt~\cite{Mandelstam:1982cb, Leibbrandt:1983pj, Leibbrandt:1987qv}, or the principle-part prescription is more suitable for regulating light-cone singularities~\cite{Heinrich:1997kv} can be avoided since \emph{all} such singularities are removed from the integrands by our mapping procedure.}
Our approach differs from the one recently proposed by Arkani-Hamed and collaborators \cite{Arkani-Hamed:2023lbd, Arkani-Hamed:2023mvg, Arkani-Hamed:2024yvu}; in particular, we work in an inherently nonplanar setup without having to resort to a $1/N_c$ genus expansion of colored theories which allows us to directly address gravitational theories. 
However, since our approach is inherently graph-based, we are not resolving tadpole, massless-bubble-on-external-lines, or related issues relevant to defining a putative loop-level recursion relation for non-supersymmetric theories. In massless theories in dimensional regularization, the aforementioned contributions are scaleless and integrate to zero. Upon taking into account their potential mixing of infrared and ultraviolet singularities, they can safely be ignored. 
For theories with massive particles such contributions are no longer scaleless and need to be determined by other means, e.g., by leveraging the knowledge of the ultraviolet and infrared structure~\cite{Bern:1995db, Badger:2017gta, Bern:2021ppb} of the theory. Nevertheless, an important property of our integrand basis is that term-by-term, various properties of $D$-dimensional amplitudes, such as gauge independence, on-shell Ward identities, or the Adler zeroes for the non-linear sigma model amplitudes, are manifest prior to integration. Moreover, the use of the global integrand basis in generalized cuts greatly simplifies cut merging into off-shell integrands and allows us to use the double copy to all loop orders. 

\subsection{Definition of Global Integrand Basis}
\label{subsec:map_algo_scalar}

Our construction applies to amplitudes and integrands in a diagrammatic representation, but we briefly comment on possible extensions to non-diagrammatic representations in \sect{sec:conclusions}. To set the stage for the discussions in later sections, we first state our general algorithm for the construction of a global integrand basis in a language that is relevant for the scattering of external scalar fields. We will explain in \sect{subsec:spinning_external_states} that the generalization to spinning external states does not pose conceptual challenges. At the core of our algorithm lie various incarnations of graph isomorphism, i.e., an algorithm (e.g.~implemented in the \texttt{igraph} package \cite{igraph, igraphm}) that determines whether two graphs are the same. Our general construction of an integrand basis for an $L$-loop $n$-point process proceeds as follows:

\begin{enumerate}
\item \label{algo:1_seeds} 
\textbf{Initial list of top-level graphs:} 
Determine the \emph{maximal cut} \cite{Bern:2007ct} residues of the integrand, i.e.\ the maximal set of independent Feynman propagators of diagrams that can go on-shell simultaneously in complexified loop-momentum space.  In gauge and gravity theories in $D$ dimensional space-time, these residues correspond to skeleton graphs with only cubic vertices and the maximum number of propagators. Depending on the process and the spectrum of the theory, the same skeleton graph might get dressed with various different mass and flavor information.  We refer to the resulting set of labeled graphs as \emph{top-level} graphs or top-level representative graphs.

\item \label{algo:2_subgraph}
\textbf{Daughter-graph identification:} For each top-level graph, collapse edges in all possible ways until one reaches \emph{bottom-level} graphs or the \emph{bottom sector}, i.e. $L$-loop graphs with only bubble-type subgraphs. This generates the set of \emph{daughter graphs} associated with each top-level graph. From this set, we explicitly remove daughter graphs with fewer loops than the top-level graph, daughter graphs containing tadpole subgraphs, and daughter graphs with massless bubbles on external lines.\footnote{\label{footnote:bub_tad}Dealing with such terms has been a long-standing issue, see, e.g.,~Refs.~\cite{NigelGlover:2008ur, Bierenbaum:2010cy, Elvang:2011ub, Caron-Huot:2010fvq,Elvang:2013cua}, and requires further regularization. This issue is also important in attempts to define loop-level recursion relations~\cite{Arkani-Hamed:2010zjl} for integrands in non-supersymmetric theories~\cite{Caron-Huot:2010fvq}. For massless theories such contributions are scaleless and integrate to zero in dimensional regularization. We note that in the surfacehedron formalism of Arkani-Hamed et al., see, e.g., Refs.~\cite{Arkani-Hamed:2023lbd, Arkani-Hamed:2023mvg, Arkani-Hamed:2024yvu}, there has been recent progress in making sense of such contributions. In our procedure, we simply set aside these issues by consistently dropping these topologies.} 
In particular applications, one can incorporate other criteria or restrictions that lead to an early termination of the collapsing procedure. We will see examples of such restriction when we discuss of classical physics in sections \ref{subsec:integrand_cleanup} and \ref{subsec:DoubleCopy}. Similarly, improved power counting of a theory can also lead to an early truncation of the required collapses, e.g.~in maximally supersymmetric Yang-Mills theory. While the top-level graphs are distinct, the collapse of their edges typically yields graphs that are isomorphic, i.e., they are identical upon relabeling internal lines.  From the union of all daughter graphs of all top-level graphs, choose a set of non-isomorphic daughter graphs which we take as \emph{representative} graphs.

\item \label{algo:3_ISP}
\textbf{Choice of basis of Lorentz invariants:}  Generically, for an $L$-loop, $n$-point scattering process, the space of loop-dependent Lorentz invariants has dimensionality $p=(n-1) L +  L(L+1)/2$.  For each representative graph, the space of loop-momentum dependent Lorentz dot products is spanned by all inverse propagators and a \emph{choice} of irreducible scalar products typically referred to as ISPs. 
ISPs are those Lorentz products between loop momenta and external momenta that cannot be expressed as a linear combination of inverse propagators and external kinematic invariants alone. 
To make an explicit choice of ISPs, it is necessary to choose independent loop momenta. Several different choices of ISP basis are possible for any choice of independent loop momenta, and different choices of momentum routing lead to further different choices of ISP basis. For amplitudes with external polarizations and spinors, the ISP basis also includes their Lorentz-invariant products with loop momenta, subject to momentum conservation and any other potential identities used to reduce them to a basis.\footnote{\label{footnote:dimension_dependence} We postpone to \sect{subsec:restricted_kinematics} a discussion of possible identities in a fixed space-time dimension, vis \`a vis the multiplicity of the integrand.}

\item \label{algo:4_global_basis}
\textbf{Global basis for the loop integrand:} For each representative graph, list all possible products of propagators raised to arbitrary positive integer power multiplied by a generic monomial in the ISPs. Remove from this list expressions related by relabeling the loop momenta, i.e., automorphism symmetry of the graphs. The union of the resulting basis for all representative graphs yields the \emph{global basis} of planar and non-planar integrands.
While the completeness of this basis follows almost trivially from the fact that the inverse propagators and ISPs are linearly independent and that we have removed redundancy between diagrams, the fact it is not overcomplete has powerful implications. See also footnote~\ref{footnote:dimension_dependence}. 

\end{enumerate}

\noindent
While we will not delve into it here, it is noteworthy that instead of employing this general construction, which uniformly applies to all field theories, we may also opt to build an integrand basis for $n$-point $L$-loop amplitudes in a specific theory, making use of its special properties. In particular, certain supersymmetric integrands can be reconstructed in integer dimensions instead of $D$ dimensions~\cite{Bern:1994zx, Bern:1994cg}, as usually required when using dimensional regularization. Additionally, the construction of daughter graphs can be truncated before reaching the true bottom sector if the theory's structure allows the collapse of a limited number of propagators.    

\subsection{Bookkeeping}

To establish a convenient naming convention for the integrand basis elements, we devise the following labeling scheme that tracks the relabeling of external kinematics (permutation sums) which manifests the symmetries of the relevant amplitude.  For each of the $n_{G}$ top-level representative graphs, modulo relabeling of external legs,  we assign a number $i \in \{1,\ldots,n_{G}\}$. For all of the $n_{M}$ relabelings of the external kinematics, we assign a number $k \in \{0,\ldots,n_{M}-1\}$. This way, every top-level graph has an assigned (not necessarily unique) index $i+ n_G \, k$. At first, we allow for possibly overcounting labels of graphs and remove the redundancy afterward. We will illustrate our bookkeeping conventions in explicit examples below. 

For each representative graph modulo external leg relabeling, choose \emph{one} top-level graph---or parent---which reproduces the given graph via edge collapse, and assign its index to the daughter graph. Furthermore,  we choose the basis of ISPs for the representative (daughter) graph as the ISPs of the top-level parent and the inverse propagators corresponding to the collapsed edges.
For all representatives obtained by external leg relabeling, we choose ISPs obtained by applying the corresponding relabeling. Once again, this construction manifestly accounts 
for all external-leg permutations that can appear in the amplitude.

The algorithm above does not yet cover the treatment of graphs with propagators that are not linearly independent or graphs that are one-particle reducible (or non-1PI), i.e., graphs that can be separated into two disconnected components by cutting a single line.  Both types of graphs may appear in generalized cuts and in loop integrands. A simple example of a graph with dependent propagators is 
\begin{align}
\label{eq:subleadingPole}
\vcenter{\hbox{
\begin{tikzpicture}[scale=1.9]		
        \coordinate (e1) at (-0.5,0.5);
  		\coordinate (e2) at (-0.5,-0.5);	
		\coordinate (e3) at (2,-.5);
		\coordinate (e4) at (2,0.5);  
        \coordinate (v1) at (-.2,0);
        \coordinate (v2) at (.5,0);
        \coordinate (v3) at (1,0);
		\coordinate (v4) at (1.7,0);
        \coordinate (v7) at (0.75,0);  
        \draw[massless] (e1) -- (v1) ;
		\draw[massless] (e2) -- (v1) ;
		\draw[massless] (e3) -- (v4) ;
		\draw[massless] (e4) -- (v4) ;
        \draw[massless] (v1) -- (v2) node [midway,fill=white] {$a_1$};
        \draw[massless] (v3) -- (v4) node [midway,fill=white] {$a_2$};      
        \draw[fill=white,line width=.4mm] (v7) circle (3mm); 
\end{tikzpicture}
}}
=
\vcenter{\hbox{
   \begin{tikzpicture}[scale=1.9]
        \coordinate (e1) at (-0.5,0.5);
  		\coordinate (e2) at (-0.5,-0.5);	
		\coordinate (e3) at (1.5,-.5);
		\coordinate (e4) at (1.5,0.5);  
		\coordinate (v1) at (-.2,0);
  		\coordinate (v2) at (-.2,-1);
    	\coordinate (v3) at (1.2,-1);
		\coordinate (v4) at (1.25,0);
        \coordinate (v5) at (0.65,0);
        \coordinate (v6) at (1.2,0);
        \coordinate (v7) at (0.95,0);
		\draw[massless] (e1) -- (v1) ;
		\draw[massless] (e2) -- (v1) ;
		\draw[massless] (e3) -- (v4) ;
		\draw[massless] (e4) -- (v4) ;
        \draw[massless] (v1) -- (v5) node [midway,fill=white] {$a_1{+}a_2$} ;
        \draw[fill=white,line width=.4mm] (v7) circle (3mm); 
\end{tikzpicture}
}}\,,
\end{align}
where momentum conservation ensures that the two propagators on the left-hand side of \Eq{eq:subleadingPole} are the same. Other examples with linearly-dependent propagators are encountered in asymptotic expansions. In such cases, the product of propagators for the graph obeys partial-fraction identities that allow a rewriting into a sum of terms, each of which is a product of independent factors in the denominator. Each partial fractioned term corresponds to an already-existing (relabeling of a) representative graph. 

At the top level, non-1PI contributions are given by cubic graphs with some number of internal momenta being determined by momentum conservation solely in terms of external kinematics (i.e., they do not give rise to maximal co-dimension residues in the space of loop kinematics). Separating the propagators that are independent of loop momenta yields a daughter graph of a 1PI top-level graph, which can be treated on par with other daughter graphs with the same number of collapsed propagators. Unlike those, however, the coefficients of daughter graphs with a non-1PI origin are non-local in the space of external kinematic invariants, e.g. 
\begin{align}
\label{eq:non1PI_eg}
  \vcenter{\hbox{\sctridangling{p_4}{p_3}{p_1}{p_2}}} 
  = 
  \frac{1}{s_{12}}\vcenter{\hbox{\sctri{p_3}{p_1}{p_2}{p_4}}}
  \longrightarrow 
  \frac{1}{s_{12}}\vcenter{\hbox{\scbox{p_3}{p_1}{p_2}{p_4}}}
  \,,
\end{align}
where the embedding of the triangle graph in the box graph amounts to multiplication and division by the propagator corresponding to the top horizontal edge of the box graph.
With the above specification of the kinematic data for the representative graphs and their natural embedding into integral families, it is convenient to label them as
\begin{align}
\label{integralbasis_firenotation}
I^{(r)}_{a_1\dots a_p} = \int\, \prod_{i=1}^L\frac{{\rm d}^D\ell_i}{(2\pi)^D} \frac{1}{\big[\rho^{(r)}_1\big]^{a_1} \times \cdots \times \big[\rho^{(r)}_p\big]^{a_p}} \,,
\end{align}
where $\rho^{(r)}_j$ are inverse propagators or ISPs of each top-level representative graph. As described earlier, we find it convenient to index integral families by the integer 
\begin{align}
\label{eq:general_index}
r=i+ n_G \, k \,, 
\end{align}
where $n_{G}$ is the number of top-level representative graphs modulo relabeling of external legs, $n_{M}$ is the number of relabelings of the external kinematics, and the parameter ranges are $i \in \{1,\ldots,n_{G}\}$, and $k \in \{0,\ldots,n_{M}-1\}$.  Furthermore, $p$ denotes the dimension of the loop-dependent space of Lorentz invariants. Any diagram that may appear in a Feynman diagram or unitarity cut-based calculation of $n$-point $L$-loop amplitudes in the chosen theory can be expressed in this global basis solely by using momentum conservation and isomorphism of graphs with labeled external lines. We point out, however, that the resulting integrand basis elements may not respect diagram symmetry manifestly or may not lead to an integrand that is manifestly crossing symmetric. 

For convenience, we introduce natural nomenclature for integrals of the form \eqref{integralbasis_firenotation}. We will call a \emph{sector} of integrals the class of integrals for which a given set of the $a_i$ are positive so that a sector can be represented by a graph. Naturally, a \emph{subsector} $s^\prime$ of a sector $s$, $s^\prime \prec s$, is a sector for which a subset of the indices defining $s$ is non-positive. Graphically, the graph representing subsector $s^\prime$ can be obtained from the graph representing sector $s$ by appropriate edge collapse.  Conversely, a \emph{supersector} $s^\prime$ of a sector $s$, $s^\prime \succ s$, is a sector such that $s$ is a subsector of $s^\prime$. 

The inverse propagator basis has been long used in the study of Feynman integrals~\cite{Baikov:1996rk}. Among its advantages is that it greatly simplifies the evaluation of cuts of integrals: a cut simply sets to zero the appropriate inverse propagators. We will use this basis below to simplify building loop integrands from their generalized cuts. For a particular cut of interest, we first find the representative graph corresponding to said cut and identify all representative supersectors. The cut is then obtained by setting to zero all integrals not contained in these supersectors. In this procedure, we have to account for symmetry factors related to integrals appearing multiple times on a cut. In \sect{sec:unitarity}, we give explicit examples of the cutting procedure.  

The idea of sector identification and identification of isomorphic integrals is, of course, not new and has been implemented in a number of private as well as publicly available codes for collider physics applications, e.g.~\texttt{TopoID}~\cite{Hoff:2015kub}, \texttt{pySecDec}~\cite{Borowka:2017idc},  \texttt{Exp}~\cite{Seidensticker:1999bb}, \texttt{Feynson}~\cite{Maheria:2022dsq}, \texttt{tapir}~\cite{Gerlach:2022qnc}, and \texttt{FeynCalc~10} (Sec.~4.3 in Ref.~\cite{Shtabovenko:2023idz}). 
Our primary addition to these ideas is the realization that the result of a mapping procedure that implements these identifications is a complete \emph{basis of integrands} which is not overcomplete. While the mapping procedure leads to simplifications when applied to families of Feynman diagrams, its full power is clear when it is applied to assembled amplitudes prior to integration. 
Further novel features, which are a consequence of the completeness of the integrand basis, are that cut merging is streamlined and, consequently, that double copy can be used for all loop orders whenever tree-level amplitudes obey the duality between color and kinematics.

\subsection{Explicit Examples}
\label{subsec:mapping_explicit}

The general algorithm outlined in the previous subsection is extremely general and applies to any perturbative quantum field theory computation. To familiarize the reader with these ideas and clarify some of the terminology, we discuss in detail here scalar scattering at one loop and some of the new aspects that appear starting at two loops.

\subsubsection{One-Loop Four-Point Example 
\label{subsubsec:oneLoopExample}}

To give a road map of the ideas presented here, we discuss in some detail perhaps one of the simplest examples, i.e. the one-loop four-scalar amplitude in a theory with only massless (but not necessarily only scalar) particles. Focusing on external scalars allows us to avoid unnecessary distractions due to the presence of polarization vectors or spinors. We will discuss the generalization to external particles with spin in \sect{subsec:spinning_external_states}. 
To encounter all the steps in the previous section, we assume that the power-counting and interaction vertices are such that we get all one-loop representative graphs shown in Eq.~(\ref{eq:1L_massless_representative_graphs}). 

For the four-scalar scattering amplitude, step \ref{algo:1_seeds} of the global integrand basis construction instructs us to list all the maximal cuts of the integrand. Having only four external states, the maximal cut topologies are given by the three box graphs depicted in \fig{fig:1-loop-boxes}. These are the top-level (representative) graphs. Modding out by external leg relabeling, we obtain $n_G=1$ graph.
%
\begin{figure}[tb!]
\centering
\begin{subfigure}[b]{0.3\textwidth}
\centering
    $\vcenter{\hbox{\scbox{p_1}{p_2}{p_3}{p_4}}}$
    \caption{\label{subfig:box_1234}}
\end{subfigure}
\quad
\begin{subfigure}[b]{0.3\textwidth}
\centering
    $\vcenter{\hbox{\scbox{p_1}{p_2}{p_4}{p_3}}}$
    \caption{\label{subfig:box_1243}}
\end{subfigure}
\quad
\begin{subfigure}[b]{0.3\textwidth}
\centering
    $\vcenter{\hbox{\scbox{p_1}{p_3}{p_2}{p_4}}}$
    \caption{\label{subfig:box_1324}}
\end{subfigure}
    \caption{
    \label{fig:1-loop-boxes}
    Inequivalent top-level graphs for massless one-loop four-point scattering with labeled external edges.} 
\end{figure}
%
All the other 21 permutations of the four external legs $\{p_1,\ldots,p_4\}$ lead to graphs that are isomorphic (with fixed external lines) to the ones shown in \fig{fig:1-loop-boxes} and are therefore not included in the list of top-level (representative) graphs. At this stage, we purposefully ignore the \emph{internal} leg labels, as they are not relevant for identifying the representative top-level graphs. 

\medskip

\begin{figure}[tbh!]
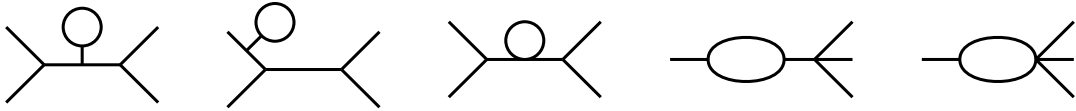

\centering
$\vcenter{\hbox{\sctadA}}$
$\vcenter{\hbox{\sctadB}}$
$\vcenter{\hbox{\sctadPinchA}}$
$\vcenter{\hbox{\scbubdanglingdrop{ }{ }{ }{ }}}$
$\vcenter{\hbox{\scbubdrop{ }{ }{ }{ }}}$
\caption{
\label{fig:1-loop-exclude-topos}
    Examples of discarded one-loop four-point tadpole and massless-bubble-on-external-leg graphs.} 
\end{figure}

In step \ref{algo:2_subgraph}, we start from the top-level representative graphs in \Fig{fig:1-loop-boxes} and perform edge collapse, e.g.,
\begin{align}
\label{eq:box_eg_collapses}
\hskip -.5cm
 \vcenter{\hbox{\scbox{p_1}{p_2}{p_3}{p_4}}}
 {\to}
 \begin{array}{c}
 \vcenter{\hbox{\sctri{p_1}{p_2}{p_3}{p_4}}} \vcenter{\hbox{\sctri{p_2}{p_3}{p_4}{p_1}}} \\[16pt]
 \vcenter{\hbox{\sctri{p_3}{p_4}{p_1}{p_2}}} \vcenter{\hbox{\sctri{p_4}{p_1}{p_2}{p_3}}} 
 \end{array}
 {\to}
 \begin{array}{c}
  \vcenter{\hbox{\scbub{p_1}{p_2}{p_3}{p_4}}}\\[16pt]
  \vcenter{\hbox{\scbub{p_1}{p_4}{p_3}{p_2}}}
 \end{array} 
 \hskip -.3cm
 .
\end{align}
Collapsing edges for the other two top-level graphs, and removing redundant graphs (keeping external lines fixed), leads to the following list of representative graphs:
\begin{align}
\label{eq:1L_massless_representative_graphs}
\hskip -.5cm
 \begin{array}{c}
 \vcenter{\hbox{\scbox{p_1}{p_2}{p_3}{p_4}}}\\[16pt]
 \vcenter{\hbox{\scbox{p_1}{p_2}{p_4}{p_3}}}\\[16pt]
 \vcenter{\hbox{\scbox{p_1}{p_3}{p_2}{p_4}}}
 \end{array}
 {\to}
 \begin{array}{c}
 \vcenter{\hbox{\sctri{p_1}{p_2}{p_3}{p_4}}} \vcenter{\hbox{\sctri{p_1}{p_2}{p_4}{p_3}}} \\[16pt]
 \vcenter{\hbox{\sctri{p_1}{p_3}{p_4}{p_2}}} \vcenter{\hbox{\sctri{p_2}{p_1}{p_3}{p_4}}} \\[16pt]
 \vcenter{\hbox{\sctri{p_2}{p_1}{p_4}{p_3}}} \vcenter{\hbox{\sctri{p_3}{p_1}{p_2}{p_4}}}
 \end{array}
 {\to} 
 \begin{array}{c}
  \vcenter{\hbox{\scbub{p_1}{p_2}{p_3}{p_4}}}\\[16pt]
  \vcenter{\hbox{\scbub{p_1}{p_2}{p_4}{p_3}}} \\[16pt]
  \vcenter{\hbox{\scbub{p_1}{p_3}{p_4}{p_2}}}
 \end{array}
 \hskip -.3cm
 .
\end{align}
The step of removing (or modding out by) identical (isomorphic with fixed external lines) graphs is crucial for our construction and allows us to eliminate the main source of redundancy in the construction of loop integrands. It is particularly important at higher loops, where bottom-level sectors tend to be shared by a large number of top-level graphs. 
A concrete example of such a redundancy at one-loop order is the representative triangle sector, which is a subsector of two boxes:
\begin{align}
\label{eq:triangle_from_box}
\vcenter{\hbox{\sctri{p_3}{p_1}{p_2}{p_4}}} & \prec \hskip .15 cm 
\left \{ \hskip -.2 cm 
\vcenter{\hbox{\scbox{p_1}{p_2}{p_3}{p_4}}}
,
\vcenter{\hbox{\scbox{p_1}{p_2}{p_4}{p_3}}} \hskip -.15 cm 
\right\} .
\end{align}
As discussed in footnote \ref{footnote:bub_tad}, in our construction, we exclude tadpole and massless-bubble-on-external-leg collapses such as the ones depicted in \fig{fig:1-loop-exclude-topos}. 
%
%
%
Furthermore, all non-1PI diagrams are embedded into representative sectors that share the same loop-momentum-dependent propagators, e.g. 
\begin{align}
\label{eq:eg_dangling_tree}
  \vcenter{\hbox{\sctridangling{p_4}{p_3}{p_1}{p_2}}} 
  \quad \longrightarrow \quad 
  \frac{1}{s_{12}}\vcenter{\hbox{\sctri{p_3}{p_1}{p_2}{p_4}}}\,,
\end{align}
where the relevant graph on the right hand side of \Eq{eq:eg_dangling_tree} appears as edge collapse of boxes in Eq.~\eqref{eq:triangle_from_box}. Thus, we can handle non-1PI contributions without the need to introduce additional representative graphs. The non-1PI origin of a particular contribution is signaled by the nonlocality (in the space of external kinematics) of the coefficient of the basis element---in the example above, the presence of the $1/s_{12}$ pole. 

Having determined the complete list of \emph{representative} graphs for our sample one-loop four-point problem, step \ref{algo:3_ISP} in our algorithm instructs us to choose a basis of Lorentz invariant scalar products for each representative. In principle, we have complete freedom to choose such a basis for each representative, but as commented above, it is advantageous for bookkeeping purposes to embed lower-level representatives into the labels of top-level representative graphs. Concretely, for the one-loop example at hand, we start with a basis choice for the box graphs,
\begin{align}
\label{eq:1L_box_1}
    I^{(1)}_{a_1,a_2,a_3,a_4} = \vcenter{\hbox{\scboxlabeled{p_1}{p_2}{p_3}{p_4}}} = \int \frac{\mathrm{d}^{D}\ell}{(2\pi)^D} \, \frac{1}{\big[\rho^{(1)}_1\big]^{a_1}\big[\rho^{(1)}_2\big]^{a_2}\big[\rho^{(1)}_3\big]^{a_3}\big[\rho^{(1)}_4\big]^{a_4}}\,,
\end{align}
where the momentum conventions are as follows:
\begin{align}
\label{eq:box_rho_explicit}
\begin{split}
    \rho^{(1)}_1 ={}& \ell^2\,, \qquad 
    \rho^{(1)}_2 = (\ell-p_1)^2\,, \qquad
    \rho^{(1)}_3 = (\ell- p_{12})^2\,, \qquad
    \rho^{(1)}_4 =  (\ell- p_{123})^2\,.
\end{split}    
\end{align}
We introduce the shorthand notation $p_{i\ldots j} = p_i + \cdots + p_j$. For a one-loop four-point process, there are no ISP's for the boxes, i.e. every Lorentz dot product between a loop momentum $\ell$ and one of the three independent (under momentum conservation) external momenta $\{p_1,p_2,p_3\}$ can be expressed in the inverse propagator basis of the $\rho^{(1)}_i$:\footnote{Here, we assume $D\geq 3$ so that the three external momenta are independent. We briefly address situations with restricted kinematics in section~\ref{subsec:restricted_kinematics}.}
\begin{align}
\label{eq:1loop_eg_dots_to_rhos}
    \ell^2 & = {} \rho^{(1)}_1\,,
 && \ell\cdot p_1 = \frac{1}{2}\left(\rho^{(1)}_1{-}\rho^{(1)}_2\right),  
 \nn \\[4pt]
    \ell\cdot p_2 & = \frac{1}{2}\left(\rho^{(1)}_2 {-} \rho^{(1)}_3{+}s_{12}\right) , 
 && \ell\cdot p_3 = \frac{1}{2}\left(\rho^{(1)}_3 {-} \rho^{(1)}_4 {-} s_{12}\right).
\end{align}
The remaining boxes are labeled according to our general naming conventions, where $n_G=1$ is the number of top-level representative graphs modulo graph isomorphisms without fixed external lines, and there are $n_M=4!=24$ possible maps of the external momenta $\{p_1,p_2,p_3,p_4\}$, indexed by $k$ in Eq.~\eqref{eq:general_index}, according to the symbols in the set
\begin{align}
\label{eq:map_ordering}
\begin{split}
\Big\{
&(1234),(1243),(1324),(1342),(1423),(1432),(2134),(2143),\\
&(2314),(2341),(2413),(2431),(3124),(3142),(3214),(3241),\\
&(3412),(3421),(4123),(4132),(4213),(4231),(4312),(4321)
\Big\}\, ,
\end{split}
\end{align}
where $k=0$ corresponds to the first entry.
The remaining two representative boxes from Eq.~(\ref{eq:1L_massless_representative_graphs}) are given by 
\begin{align}
\label{eq:1L_box_2}
    I^{(2)}_{a_1,a_2,a_3,a_4} & = \vcenter{\hbox{\scboxlabeled{p_1}{p_2}{p_4}{p_3}}} = \int \frac{\mathrm{d}^{D}\ell}{(2\pi)^D} \, \frac{1}{\big[\rho^{(2)}_1\bigr]^{a_1}\big[\rho^{(2)}_2\bigr]^{a_2}\bigl[\rho^{(2)}_3\bigr]^{a_3}\bigl[\rho^{(2)}_4\bigr]^{a_4}} \,,
    \\[5pt]
\label{eq:1L_box_3}    
    I^{(3)}_{a_1,a_2,a_3,a_4} & = \vcenter{\hbox{\scboxlabeled{p_1}{p_3}{p_2}{p_4}}} = \int \frac{\mathrm{d}^{D}\ell}{(2\pi)^D} \, \frac{1}{\bigl[\rho^{(3)}_1\bigr]^{a_1}\bigl[\rho^{(3)}_2\bigr]^{a_2}\bigl[\rho^{(3)}_3\bigr]^{a_3}\bigl[\rho^{(3)}_4\bigr]^{a_4}} \,,
\end{align}
and their inverse propagators $\rho^{(r)}_k$ are obtained from the ones in Eq.~(\ref{eq:box_rho_explicit}) by appropriate relabeling of the external momenta,
\begin{align}
    \rho^{(2)}_1 ={}& \ell^2\,, \quad 
    \rho^{(2)}_2 = (\ell-p_1)^2\,, \quad
    \rho^{(2)}_3 = (\ell- p_{12})^2\,, \quad
    \rho^{(2)}_4 =  (\ell- p_{124})^2\,, 
    \\[3pt]
    \rho^{(3)}_1 ={}& \ell^2\,, \quad 
    \rho^{(3)}_2 = (\ell-p_1)^2\,, \quad
    \rho^{(3)}_3 = (\ell- p_{13})^2\,, \quad
    \rho^{(3)}_4 =  (\ell- p_{132})^2\,.
\end{align}
In our labeling scheme, all lower-level representative graphs are embedded into (not necessarily representative) top-level graphs and inherit those labels, e.g., 
\begin{align}
 \label{eq:TriangleEg}
  I^{(13)}_{a_1,a_2,a_3,a_4} & = 
  \hskip -.3cm
  \vcenter{\hbox{\sctrilabeled{p_3}{p_1}{p_2}{p_4}}}
  \hskip -.3cm\,,\, \text{with }a_{2,3,4}>0,a_1\leq 0  
  && \hskip -.3cm \prec   
 \vcenter{\hbox{\scboxlabeled{p_3}{p_1}{p_2}{p_4}}} \,,
 \hskip -.4cm
 \\[5pt]
 I^{(1)}_{a_1,a_2,a_3,a_4} &= 
 \hskip -.3cm
 \vcenter{\hbox{\scbublabeled{p_1}{p_2}{p_3}{p_4}}}
 \hskip -.3cm \,, \text{with }a_{2,4}>0,a_{1,3}\leq 0  
  && \hskip -.3cm \prec 
 \vcenter{\hbox{\scboxlabeled{p_1}{p_2}{p_3}{p_4}}}\,.
 \hskip -.4cm
 \label{eq:StandardBubble} 
\end{align}
In a similar fashion, we label all other triangle and bubble standard representatives in terms of the labels of the box graphs. Our choice of representative sectors for the massless one-loop four-particle example is given in the ancillary file \texttt{RepresentativeSectors\textunderscore1L.m}. Note that the triangle in Eq.~(\ref{eq:TriangleEg}) has index $(13)$ which does not correspond to one of the representative boxes of Eqs.~(\ref{eq:1L_box_1}), (\ref{eq:1L_box_2}), or (\ref{eq:1L_box_3}). As explained above, our labeling scheme tracks permutations over external legs by simply shifting the index of diagrams. 

Another source of redundancy in the set of integrands is the freedom to assign \emph{internal} labels. It leads to additional relations between integrals with different assigned names whenever the underlying graph has a non-trivial automorphism. 
To be concrete, consider the one-loop bubble graph in Eq.~\eqref{eq:StandardBubble}. A generic integral of this type is written in the inverse propagator basis (\ref{eq:box_rho_explicit}) as $I^{(1)}_{a_1,a_2,a_3,a_4}$ with $a_{2,4}>0$, $a_{1,3}\leq 0$. The relabeling $(\ell-p_1)\to -(\ell-p_{123})$, leads to the following map of the propagators,
\begin{align}
\label{eq:bubble_map}
\begin{split}
\hskip -.5cm
\rho^{(1)}_1 {\to} \rho^{(1)}_2{-}\rho^{(1)}_1{+}\rho^{(1)}_4 {-}s_{23}\,,
 \ \ \ \ 
 \rho^{(1)}_2 {\to}  \rho^{(1)}_4\,,
 \ \ \ \ 
\rho^{(1)}_3 {\to} \rho^{(1)}_2{-}\rho^{(1)}_3{+}\rho^{(1)}_4 {-}s_{23}\,,
\ \ \ \ 
\rho^{(1)}_4 {\to}  \rho^{(1)}_2\,,
\hskip -.5cm
\end{split}
\end{align}
where $s_{ij}=p_{ij}^2$. One way to remove this redundancy is to assign an ordering to the different monomials of the graph, for example, lexicographic ordering of the indices $a_1,\ldots,a_4$; whenever an integral does not respect the chosen ordering, the transformation \eqref{eq:bubble_map} is used recursively to restore the ordering. 
Alternatively, in any amplitude with a fixed number of loops and external legs, the space of integrals is finite. Thus, based on power counting and the structure of Feynman rules, we can simply list all the possible combinations of exponents $a_1,\dots a_4$ that can appear and the relations between integrals and solve the resulting equations.
In general, the process of eliminating identical integrals within a sector introduces integrals in subsectors, which in turn require additional mapping, as described previously, including resolving automorphisms in lower sectors. 

The final type of graph that we may encounter is one where the propagators are not linearly independent. This leads to the appearance of higher-order poles when two or more of the propagators are identical upon the use of momentum conservation. The canonical example is shown in \Eq{eq:subleadingPole}. Since our definition accommodates propagators raised to higher powers, we can represent this topology within the collapsed topology depicted on the right-hand side of \Eq{eq:subleadingPole}. 

Summarizing this detailed discussion, a possible choice of representative sectors for one-loop four-point massless problems (that is aligned with the graphs in Eq.~(\ref{eq:1L_massless_representative_graphs})), that we  adopt in the following is
\begin{align}
\label{eq:1loopsectors1}
    &I_{1,1,1,1}^{{(k)}},\quad k=1,2,3\,,\\[3pt]
\label{eq:1loopsectors2}
    &I_{0,1,1,1}^{{(k)}},\quad  k=1,2,4,7,8,13\,,\\[3pt]
\label{eq:1loopsectors3}
    &I_{0,1,0,1}^{{(k)}},\quad  k=1,2,4\, .
\end{align}
The binary notation is to be understood as `$1$' representing a positive index and `$0$' representing a negative index, or zero. 

\medskip

We conclude this section by listing several explicit examples of the mapping of integrals to our chosen basis. 
Consider first a sample relation between isomorphic (with fixed external lines) boxes with misaligned internal labels:  
\begin{align}
\begin{split}
\hskip -1cm
 I^{(17)}_{1,1,1,2} = & \vcenter{\hbox{\boxegA{3}{4}{1}{2}}}   =  \int \frac{\mathrm{d}^D\ell}{(2\pi)^D} \frac{1}{\ell^2\, (\ell-p_3)^2 \,(\ell-p_{34})^2 \,\left[(\ell-p_{341})^2\right]^2}  
\hskip -1cm 
\\[-4pt]
\hskip -1cm 
= 
I^{(1)}_{1,2,1,1} = & \vcenter{\hbox{\boxegB{1}{2}{3}{4}}}  = \int \frac{\mathrm{d}^D\ell}{(2\pi)^D} \frac{1}{\ell^2\, \left[(\ell-p_{1})^2\right]^2 \, (\ell-p_{12})^2 \, (\ell-p_{123})^2} \,,
\hskip -1cm
\end{split}
\end{align}
which can be seen by the explicit shift of loop momentum $I^{(17)}_{1,1,1,2}\big|_{\ell \mapsto \ell -p_{12}} = I^{(1)}_{1,2,1,1}$. The `dot' ($\bullet$) represents the propagator raised to power two. 
Similarly, transformations of ISPs due to alignment of graph labels lead to relations of triangles and bubbles, e.g.:
\begin{align}
I^{(1)}_{1,1,1,-2}={}&I^{(13)}_{-2,1,1,1}+2 s_{12}I^{(13)}_{-1,1,1,1}+s_{12}^2 I^{(13)}_{0,1,1,1}\,,\\[3pt]
I^{(11)}_{1,-1,1,-2}={}&-\frac{1}{2}s_{13} I^{(2)}_{-2,1,0,1}+2 s_{13}I^{(2)}_{-1,1,-1,1}-\frac{1}{4}s_{13}^3 I^{(2)}_{0,1,0,1}\,.
\end{align}
While such relabelings generically lead to a spray of terms, they typically combine so that aligning graph labels does not lead to a proliferation of terms.

Any massless four-point integrand can be mapped to the global basis constructed here, with label maps constructed via graph isomorphisms. Moreover, since we are in a basis, different integrands of the same amplitude (e.g. in different gauges) will map to the same expression. A concrete example of a one-loop amplitude is discussed in~\sect{subsec:spinning_external_states}.

\subsubsection{A Two-Loop Four-Point Example
\label{subsubsec:twoLoopExample}}

\begin{figure}[ht!]
\centering
\begin{subfigure}[b]{0.3\textwidth}
\centering
    \includegraphicslog{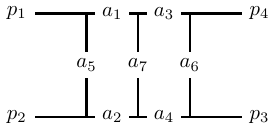}
    \caption{~~$I^{(1)}_{a_1,\dots,a_9}$}
    \label{subfig:IntFamily1}
\end{subfigure}
\quad
\begin{subfigure}[b]{0.3\textwidth}
\centering
    \includegraphicslog{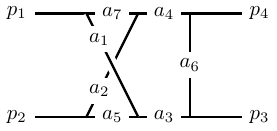}
    \caption{~~$I^{(2)}_{a_1,\dots,a_9}$}
    \label{subfig:IntFamily2}
\end{subfigure}
\quad
\begin{subfigure}[b]{0.3\textwidth}
\centering
    \includegraphicslog{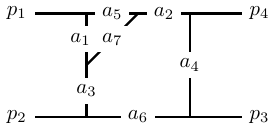}
    \caption{$~~I^{(3)}_{a_1,\dots,a_9}$}
    \label{subfig:IntFamily3}
\end{subfigure}
    \caption{Representative integral topologies for massless two-loop four-particle scattering modulo relabeling of external lines.}
    \label{fig:topologies_2L}
\end{figure}

Consider massless two-loop four-particle scattering. The three representative top-level graphs (modulo external leg relabeling, i.e. $n_G=3$) for the problem at hand are shown in \fig{fig:topologies_2L} and defined as 
\begin{align}
    I^{(1)}_{a_1,\dots,a_9}={}&\int\frac{\mathrm{d}^D\ell_1}{(2\pi)^D}\frac{\mathrm{d}^D\ell_3}{(2\pi)^D}\frac{1}{\bigl[\ell_1^2\bigr]^{a_1} \,\bigl[\left(\ell_1+p_{12}\right)^2\bigr]^{a_2} \bigl[\ell_3^2\bigr]^{a_3}\bigl[\left(\ell_3+p_{12}\right)^2\bigr]^{a_4}\bigl[\left(\ell_1+p_1\right)^2\bigr]^{a_5}}
    \nonumber\\
    &\hskip 2.5 cm \times\frac{\left(\ell_1\cdot p_3\right)^{-a_8}\left(\ell_3\cdot  p_1\right)^{-a_9}}{\bigl[\left(\ell_3{+}p_{123}\right)^2\bigr]^{a_6}\bigl[\left(\ell_1{-}\ell_3\right)^2\bigr]^{a_7}}\,,
    \label{eq:DefLadder}
    \\[5pt]
    I^{(2)}_{a_1,\dots,a_9}={}&\int\frac{\mathrm{d}^D\ell_1}{(2\pi)^D}\frac{\mathrm{d}^D\ell_2}{(2\pi)^D}\frac{1}{\bigl[\ell_1^2\bigr]^{a_1}\bigl[\ell_2^2\bigr]^{a_2}\bigl[\left(\ell_1-\ell_2-p_2\right)^2\bigr]^{a_3}\bigl[\left(\ell_1-\ell_2+p_1\right)^2\bigr]^{a_4}\bigl[\left(\ell_2+p_2\right)^2\bigr]^{a_5}}
    \nonumber\\
    &\hskip 2.5 cm \times\frac{(\ell_1\cdot \ell_2)^{-a_8}(\ell_1\cdot p_3)^{-a_9}}{\bigl[\left(\ell_1-\ell_2+p_{13}\right)^2\bigr]^{a_6}\bigl[\left(\ell_1+p_1\right)^2\bigr]^{a_7}}\,,
    \label{eq:DefxLadder}
    \\[5pt]
    I^{(3)}_{a_1,\dots,a_9}={}&\int\frac{\mathrm{d}^D\ell_1}{(2\pi)^D}\frac{\mathrm{d}^D\ell_2}{(2\pi)^D}\frac{1}{\bigl[\ell_1^2\bigr]^{a_1}\bigl[\ell_2^2\bigr]^{a_2}\bigl[\left(\ell_2+p_1\right)^2\bigr]^{a_3}\bigl[\left(\ell_2-p_2\right)^2\bigr]^{a_4}\bigl[\left(\ell_1+p_1\right)^2\bigr]^{a_5}}\nonumber\\
    &\hskip 2.5 cm \times\frac{(\ell_1\cdot p_2)^{-a_8}(\ell_1\cdot p_3)^{-a_9}}{\bigl[\left(\ell_2-p_{23}\right)^2\bigr]^{a_6}\bigl[\left(\ell_1+\ell_2+p_1\right)^2\bigr]^{a_7}}\,.
    \label{eq:DefYI}
\end{align}
In total, there are 181 representative sectors that are listed in the ancillary file\\ \texttt{RepresentativeSectors\textunderscore2L.m}. Our diagram conventions for massless scalar two-loop four-particle scattering are summarized in the ancillary file \texttt{StandardTopos\textunderscore SSSS\textunderscore2L.m}. Due to the symmetry of the cross ladder in Fig.~\ref{subfig:IntFamily2} under e.g. the exchange $p_3\leftrightarrow p_4$, integrals in sector $I^{(5)}_{1,1,1,1,1,1,1,0,0}$ can be mapped to the sector  $I^{(2)}_{1,1,1,1,1,1,1,0,0}$, e.g.
\begin{align}
    I^{(5)}_{1,1,1,1,1,1,1,-1,0} \hskip -.9 cm &\nonumber\\[4pt]
    ={}&
    \frac{1}{2} s_{12} {I}_{1,1,1,1,1,1,1,0,0}^{{(2)}}
    +\frac{1}{2} {I}_{0,1,1,1,1,1,1,0,0}^{{(2)}}
    +\frac{1}{2} {I}_{1,0,1,1,1,1,1,0,0}^{{(2)}}
    -\frac{1}{2} {I}_{1,1,0,1,1,1,1,0,0}^{{(2)}} \hskip 1.3 cm 
    \nonumber\\[3pt]
    {}&-\frac{1}{2} {I}_{1,1,1,0,1,1,1,0,0}^{{(2)}}
    -\frac{1}{2} {I}_{1,1,1,1,0,1,1,0,0}^{{(2)}}
    +\frac{1}{2} {I}_{1,1,1,1,1,1,0,0,0}^{{(2)}}
    -{I}_{1,1,1,1,1,1,1,-1,0}^{{(2)}}\,.
\end{align}
Recall that the ordering of external leg maps is given in Eq.~\eqref{eq:map_ordering} and the location of the map in that list gives the index $k$ in Eq.~\eqref{eq:general_index}, with the first entry corresponding to $k=0$. 
Because of the presence of the ISP in $I^{(5)}_{1,1,1,1,1,1,1,-1,0}$, the above mapping also produces contact terms, belonging to subsectors, in addition to integrals belonging to the original sector. These contact terms are not yet in our chosen basis of representative graphs, so we have to map these terms, as well. For the example at hand, we obtain
\begin{align}
    I^{(5)}_{1,1,1,1,1,1,1,-1,0}\hskip -.9 cm &\nonumber\\[4pt] 
    ={}&
    \frac{1}{2} {I}_{0,1,1,1,1,1,1,0,0}^{{(1)}}
    +\frac{1}{2} {I}_{0,1,1,1,1,1,1,0,0}^{{(19)}}
    +\frac{1}{2} s_{12} {I}_{1,1,1,1,1,1,1,0,0}^{{(2)}}
    -\frac{1}{2} {I}_{1,1,0,1,1,1,1,0,0}^{{(2)}}
    \nonumber\\[3pt]
    {}&-{I}_{1,1,1,1,1,1,1,-1,0}^{{(2)}}
    +\frac{1}{2} {I}_{0,1,1,1,1,1,1,0,0}^{{(22)}}
    +\frac{1}{2} {I}_{0,1,1,1,1,1,1,0,0}^{{(4)}}
    -\frac{1}{2} {I}_{1,1,0,1,1,1,1,0,0}^{{(5)}}\,,
\end{align}
in the labeling scheme introduced in \sect{subsec:map_algo_scalar}. All integrals are now written in terms of representative sectors and the algorithm terminates. In particular amplitude applications, putting an integrand in some representation in our basis entails performing similar mapping steps for all integrals that appear.

\subsection{Restricted Kinematics}
\label{subsec:restricted_kinematics}

\begin{figure}[thb!]
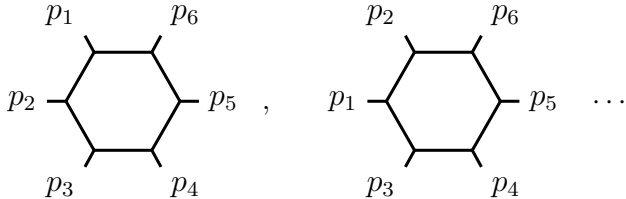

\centering
    $\vcenter{\hbox{\hexagon{p_1}{p_2}{p_3}{p_4}{p_5}{p_6}}}\,,$
    \quad 
    $\vcenter{\hbox{\hexagon{p_2}{p_1}{p_3}{p_4}{p_5}{p_6}}}$
    $\, \cdots $
\caption{\label{fig:eg_hexagons} Sample hexagons for massless one-loop six-particle scattering.}    
\end{figure}

For simplicity, in the previous section, we illustrated our construction using four-point processes as examples. We now briefly discuss the case of an arbitrary number $n$ of external legs. For such processes, without restricting the spacetime dimension, the top-level graphs contain $n$-gon maximal cut topologies at one-loop, e.g. hexagons for $n=6$, depicted in \fig{fig:eg_hexagons}.
However, if we are interested in situations with strictly four-dimensional internal and external kinematics, the maximal cut topologies are restricted to boxes. In four-dimensional (complexified) loop-momentum space, at one loop, we can set at most four propagators to zero simultaneously by adjusting each of the four components of loop momentum. From a practical perspective, this means that hexagons become \emph{reducible}, see, e.g.,~Refs.~\cite{Melrose:1965kb, vanNeerven:1983vr,Arkani-Hamed:2010pyv}, and can be eliminated for topologies with fewer propagators. 
Said differently, in strictly $D=4$, the six propagators of a hexagon are no longer independent but satisfy partial fraction identities that can be used to reduce all hexagons to possibly pentagons, boxes, triangles, and bubbles at the level of the integrand without using any properties of integration.

All the explicit examples we discuss here involve four-particle processes in a generic space-time dimension, so we do not need to take such reductions into account. More generally, if the number of external states in the scattering process is not too large, one can always work in a $D$-dimensional setup, without accounting for additional relations that can occur in specific integer dimensions, perform our basis construction algorithm, project the integrand to that basis, and subsequently restrict external and or internal kinematics to specific dimensions. 

This discussion highlights the general fact that a basis in a vector space is defined up to a priori specified equivalence relations, or identities. We constructed a basis in the space of integrands, i.e. the space of rational functions attached to graphs, and the equivalence relations are given by momentum conservation and graph isomorphism.
A similar philosophy can be used to construct a basis for amplitudes or, equivalently, for integrals of rational functions attached to graphs. In this case, in addition to momentum conservation and graph isomorphism, integrands that differ by total derivatives are also considered to be equivalent. These additional relations, due to Passarino-Veltman reduction and integration-by-parts identities, emphasize the differential form interpretation of loop integrands.

\subsection{Application: Gauge-Choice Independence of Feynman Integrands}
\label{subsec:xiDrop}
%
The advantages of our mapping procedure are best illustrated by complete physical examples. The four-scalar amplitudes in scalar electrodynamics contain all the relevant features and the expressions are sufficiently simple to be written in detail. The Lagrangian for this theory in $R_\xi$-gauge is 
\begin{align}
\label{eq:SQED_Lagrangian}
 \mathcal{L} = -\frac{1}{4} F_{\mu\nu}F^{\mu\nu} -\frac{1}{2\xi}(\partial_\mu A^\mu)^2+ \sum_{i=1}^2
                (D_\mu \phi_i)^\dagger (D^\mu \phi_i)\,,
\end{align}
where the covariant derivative acts on scalars of charge $Q_i$ via $D_\mu \phi_i = \left(\partial_\mu + \imath  Q_i A_\mu \right) \phi_i$, and $F_{\mu\nu} = \partial_\mu A_\nu - \partial_\nu A_\mu$ is the field strength tensor of the photon. While our procedure can demonstrate the equivalence of quantum field theories under local field redefinitions, we will omit this discussion here for the sake of brevity.\footnote{We mention, however, that, within generalized unitarity, it is sufficient to demonstrate the invariance of $D$-dimensional tree amplitudes under such redefinitions.}

\begin{figure}[t!]
\centering
\begin{subfigure}[b]{0.2\textwidth}
\centering
    \includegraphicslog{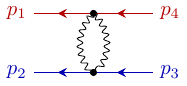}
    \caption{}
    \label{subfig:FeynmanSQED_Bubble}
\end{subfigure}
\begin{subfigure}[b]{0.2\textwidth}
\centering
    \includegraphicslog{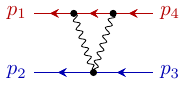}
    \caption{}
    \label{subfig:FeynmanSQED_Triangle}
\end{subfigure}
\begin{subfigure}[b]{0.2\textwidth}
\centering
    \includegraphicslog{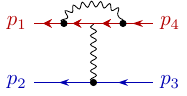}
    \caption{}
    \label{subfig:FeynmanSQED_Mushroom_Triangle}
\end{subfigure}
\begin{subfigure}[b]{0.2\textwidth}
\centering
    \includegraphicslog{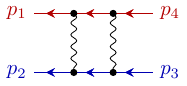}
    \caption{}
    \label{subfig:FeynmanSQED_Box_1L}
\end{subfigure}
    \caption{
    \label{fig:FeynmanSQED_1L}
    The four distinct Feynman diagram topologies contribute to the scattering amplitude of distinct-flavor massless charged scalars in QED, excluding closed scalar loops. The full set of seven diagrams is obtained through relabeling.  All external momenta are outgoing and the arrows represent charge flow. 
    }
\end{figure}

As a first example, we study the gauge-choice independence of a simple one-loop amplitude before integration in our basis of loop integrands. 
Without loss of generality and to keep the example relatively compact, we neglect terms containing a closed scalar loop (and thus proportional to the number of scalars), as they form a gauge-invariant subsector tagged by the number $N_s$ of scalar flavors. Retaining such contributions poses no conceptual challenge. 
The four representative diagrams are shown in Figure~\ref{fig:FeynmanSQED_1L}; the amplitude is determined by them together with three additional Feynman diagrams obtained by relabeling. 
%

The box diagram in \Fig{subfig:FeynmanSQED_Box_1L}, in $R_\xi$-gauge, evaluates to 
\begin{align}
\vcenter{\hbox{\includegraphicslog{figs/Box_1L.pdf}}}\hskip -.3cm ={}Q_1^2 Q_2^2 &\left[
\left(\xi^2{+}1\right) {I}_{{0,1,0,1}}^{{(1)}}
{+}{I}_{{0,1,0,1}}^{{(2)}}
{+}2 (1-\xi) (2 u{+}\xi  t) {I}_{{0,1,0,2}}^{{(1)}}
{+}(4 s{+}3 t) {I}_{{0,1,1,1}}^{{(1)}}\right. \nonumber \\[-9pt]
&
+\left.2 u {I}_{{0,1,1,1}}^{{(2)}}
+2 u {I}_{{0,1,1,1}}^{{(7)}}
+(4 s{+}3 t) {I}_{{0,1,1,1}}^{{(8)}}
+\frac{1}{2} (\xi {-}1)^2 t^2 {I}_{{0,2,0,2}}^{{(1)}}\right. \hskip 1 cm  \nonumber \\[3pt]
&
-\left. {I}_{{-1,1,1,1}}^{{(1)}}
-(\xi{-}1)^2 {I}_{{-1,2,-1,2}}^{{(1)}}
-{I}_{{-1,1,1,1}}^{{(8)}}
+4 u^2 {I}_{{1,1,1,1}}^{{(3)}}
\right],
\end{align}
where all external momenta are outgoing and the arrows on the scalar line denote charge flow. The Mandelstam invariants are
\begin{equation}
    s=(p_1+p_2)^2\,,\quad t=(p_2+p_3)^2\,,\quad  u=(p_1+p_3)^2\,.
\end{equation}
Our conventions for the various integrals $I^{(r)}_{a_1,a_2,a_3,a_4}$ were introduced in \sect{subsec:mapping_explicit} and are also collected in the ancillary file \texttt{SQED\textunderscore SSSS\textunderscore 1L.m}. In the process of mapping the Feynman diagram expressions to our global basis of loop integrands, according to our setup, we explicitly drop bubble-on-external leg and possible tadpole topologies. The crossed-box diagram is obtained by relabeling. 

The vertex-correction diagram in \Fig{subfig:FeynmanSQED_Mushroom_Triangle} evaluates to
\begin{align}
\vcenter{\hbox{\includegraphicslog{figs/Mushroom_Triangle_1L.pdf}}}=-4Q_1 Q_2^3 
\left[
    {I}_{-1,1,1,1}^{(1)}
    +\frac{t}{2} {I}_{0,1,1,1}^{(1)}
\right] ;
\end{align}
note that this diagram is by itself independent of the gauge choice, as one cannot pinch it to another diagram without generating tadpole or bubble-on-an-external line diagrams. We also need to include its image under $(p_1, p_4, Q_1)\leftrightarrow (p_2, p_3, Q_2)$. 

The triangle diagram in \Fig{subfig:FeynmanSQED_Triangle} evaluates to
\begin{align}
\hskip -.4cm
\vcenter{\hbox{\includegraphicslog{figs/Triangle_1L.pdf}}}
\hskip -.4cm
=Q_1^2 Q_2^2 \left[
    t\, I_{0,1,1,1}^{(8)}
    {-}(\xi {+}1)^2 I^{(1)}_{0,1,0,1}
    {+}2 (\xi {-}1)^2 t I^{(1)}_{0,1,0,2}
    {-}\frac{(\xi {-}1)^2 }{2} t^2 I^{(1)}_{0,2,0,2}
    \right],
\hskip -.2cm    
\end{align}
and we include its image under $(p_1, p_4, Q_1)\leftrightarrow (p_2, p_3, Q_2)$ in the final result. 

The bubble diagram in \Fig{subfig:FeynmanSQED_Bubble} is given by
\begin{align}
\hskip -.4cm
\vcenter{\hbox{\includegraphicslog{figs/Bubble_1L.pdf}}}
\hskip -.4cm 
=Q_1^2 Q_2^2 \left[
    \left(2 D {+}\xi(\xi {+}2) {-}3\right)  I^{(1)}_{0,1,0,1}
    +(\xi{-}1)^2 
    \left(
    \frac{1}{2} t^2 I^{(1)}_{0,2,0,2}
    {-}2 t I^{(1)}_{0,1,0,2}
    \right)
    \right].
\hskip -.2cm    
\end{align}
We note that in all cases the $\xi$ parameter multiplies integrals that have the topology of a bubble. This is due to the relative simplicity of the $R_\xi$ gauge and of the scalar QED Feynman rules. In more complicated theories, such as QCD and GR, the pollution due to gauge terms is much more significant and the cleanup described here becomes more dramatic. 

Adding up all seven Feynman diagrams, we find
\begin{align}
\label{eq:sqed_ssss_1L}
\begin{split}
\mathcal{A}^{(1)}_{\phi\phi\bar{\phi}\bar{\phi}}={}&
Q_2^2 Q_1^2 \left[2 (D-2) {I}_{{0,1,0,1}}^{{(1)}}+{I}_{{0,1,0,1}}^{{(2)}}+{I}_{{0,1,0,1}}^{{(4)}}+4 t {I}_{{0,1,1,1}}^{{(1)}}+2 s {I}_{{0,1,1,1}}^{{(13)}}\right.\\ 
&\left.+2 u {I}_{{0,1,1,1}}^{{(2)}}+2 s {I}_{{0,1,1,1}}^{{(4)}}+2 u {I}_{{0,1,1,1}}^{{(7)}}+4 t {I}_{{0,1,1,1}}^{{(8)}}+4 s^2 {I}_{{1,1,1,1}}^{{(1)}}+4 u^2 {I}_{{1,1,1,1}}^{{(3)}}\right]\\[2pt]
&
-4Q_2^3 Q_1 \left({I}_{{-1,1,1,1}}^{{(1)}}+\frac{t}{2} {I}_{{0,1,1,1}}^{{(1)}}\right)
-4Q_2 Q_1^3 \left({I}_{{-1,1,1,1}}^{{(8)}}+\frac{t}{2} {I}_{{0,1,1,1}}^{{(8)}}\right).
\end{split}
\end{align}
Crucially, the gauge parameter $\xi$ has canceled when adding up all the diagrams, signifying that the amplitude's integrand in this basis is indeed independent of the gauge choice. We emphasize that, when assembling Eq.~\eqref{eq:sqed_ssss_1L}, we did not use any identity that requires integration.
We also notice that the amplitude has fewer integrals to be computed and that the power-counting is reduced. In fact, other than the triangles in the odd-charge sectors, no integral reduction is necessary. Furthermore, all crossing symmetries are manifest due to our choice of symmetric basis.

We have verified the cancellation of the gauge parameter $\xi$ in our basis through two loops for the full quantum integrand and through four loops in classical electrodynamics in Ref.~\cite{Bern:2023ccb}. The result for the former is included in the ancillary file \texttt{SQED\textunderscore SSSS\textunderscore 2L.m} while the latter may be found in the ancillary file of \cite{Bern:2023ccb}. 
This cancellation prior to integration indicates a clean separation between the diagrams that integrate to zero in dimensional regularization (see footnote \ref{footnote:bub_tad} for conceptual issues related to them) and the diagrams captured in a unitarity-based approach.

\subsection{Tensor Integrals and On-Shell Ward Identities}
\label{subsec:spinning_external_states}

\begin{figure}[ht!]
\centering
\begin{subfigure}[b]{0.2\textwidth}
\centering
    \includegraphicslog{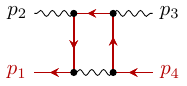}
    \caption{}
    \label{subfig:Compton_1_1L}
\end{subfigure}
\begin{subfigure}[b]{0.2\textwidth}
\centering
    \includegraphicslog{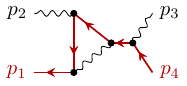}
    \caption{}
    \label{subfig:Compton_2_1L}
\end{subfigure}
\begin{subfigure}[b]{0.2\textwidth}
\centering
    \includegraphicslog{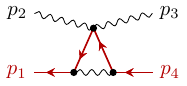}
    \caption{}
    \label{subfig:Compton_3_1L}
\end{subfigure}
\begin{subfigure}[b]{0.2\textwidth}
\centering
    \includegraphicslog{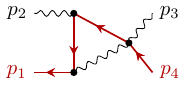}
    \caption{}
    \label{subfig:Compton_4_1L}
\end{subfigure}
\begin{subfigure}[b]{0.2\textwidth}
\centering
    \includegraphicslog{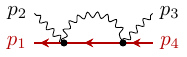}
    \caption{}
    \label{subfig:Compton_5_1L}
\end{subfigure}
\begin{subfigure}[b]{0.2\textwidth}
\centering
    \includegraphicslog{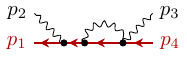}
    \caption{}
    \label{subfig:Compton_6_1L}
\end{subfigure}
\begin{subfigure}[b]{0.2\textwidth}
\centering
    \includegraphicslog{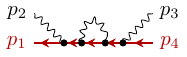}
    \caption{}
    \label{subfig:Compton_7_1L}
\end{subfigure}
    \caption{The 7 distinct Feynman diagram topologies contributing to the one-loop Compton scattering in scalar QED, excluding closed scalar loops. The full set of 15 diagrams is obtained by relabeling.
    }
    \label{fig:FeynmanSQED_1L_Compton}
\end{figure}

We have thus far covered integrands of amplitudes with external scalars and only briefly alluded to amplitudes with external spinning fields. In this 
section we discuss this topic in more detail. 
The new feature of such amplitudes is that loop momenta can be contracted with polarization vectors or tensors, or with vector and tensor currents 
constructed from spinor asymptotic states.  These new Lorentz scalars appear in the numerators of integrands, which now are tensor integrals (contracted with external-line data).\footnote{The amplitudes' dependence on external data restricts the dependence on these Lorentz scalars.}
The general algorithm described for scalar scattering in \sect{subsec:map_algo_scalar} can be readily adapted to accommodate this larger class of integrand expressions. In practice, tensor integrals can be treated in several different ways: (a) via Passarino-Veltman reduction, (b) via tensor projection, or (c) by including the additional scalar products of the form $v_i\cdot\ell_j$ as part of the list of ISPs, where $v_i$ are vectors constructed from external polarizations, momenta and spinors. Either approach has advantages and disadvantages, so the preferred methodology depends on the specifics of the problem at hand.

As a concrete example, we consider the one-loop Compton scattering amplitude in scalar QED. The seven distinct Feynman diagrams, excluding closed scalar loops, are shown in \fig{fig:FeynmanSQED_1L_Compton}. Here, we choose to work in setup (c) where we include the additional $(\epsilon_2\cdot\ell),\, (\epsilon_3\cdot\ell)$ contractions as part of the list of ISPs. 
A sample Feynman diagram expression in terms of the global basis:
\begin{align}
\label{eq:compton_1L_eg_diag}
\begin{split}
\hskip -.5cm
\vcenter{\hbox{\includegraphicslog{figs/Compton_2_1L.pdf}}} 
\hskip -.3cm =
Q^4_1\Bigg[& 
(p_1\cdot \epsilon_2) (p_1\cdot \epsilon_3) I_{0,1,1,1}^{{(4)}}
+\frac{(p_1\cdot \epsilon_2)(p_{12}\cdot \epsilon _3)}{2 s}I_{0,1,0,1}^{{(4)}}
\hskip -.3cm
\\[-5 pt]
&
+(\xi -1)(p_1\cdot \epsilon_2) \left[ I_{0,1,0,2,0,-1,0,0}^{{(4)}}
-(p_1\cdot \epsilon_3) I_{0,1,0,2}^{{(4)}}\right]
\\&
-(p_1\cdot \epsilon_2) I_{0,1,1,1,0,-1,0,0}^{{(4)}}\Bigg]\,.
\end{split}
\end{align}
The above expression is to be understood as follows: we append to the box propagators, defined in Eq.~\eqref{eq:box_rho_explicit}, four ISPs,
\begin{eqnarray}
    \rho^{(1)}_{5}=\epsilon_1\cdot \ell\,,\quad    \rho^{(1)}_{6}=\epsilon_2\cdot \ell\,,\quad 
    \rho^{(1)}_{7}=\epsilon_3\cdot \ell\,,\quad 
    \rho^{(1)}_{8}=\epsilon_4\cdot \ell\,,
\end{eqnarray}
which are understood to be mapped as $\{1{\to}1,\, 2{\to} 3,\, 3{\to} 4,\, 4{\to} 2\}$, 
\begin{eqnarray}
    \rho^{(4)}_{5}=\epsilon_1\cdot \ell\,,\quad    
    \rho^{(4)}_{6}=\epsilon_3\cdot \ell\,,\quad 
    \rho^{(4)}_{7}=\epsilon_4\cdot \ell\,,\quad 
    \rho^{(4)}_{8}=\epsilon_2\cdot \ell\,. 
\end{eqnarray}
For the integrals in Eq.~(\ref{eq:compton_1L_eg_diag}), we use a hybrid notation that suppresses the indices for the polarization-dependent ISPs whenever they are all zero.  
The fact that we added four additional ISPs instead of two is owed to our standard labeling procedure where integrals are indexed in terms of external kinematic maps. We find this setup convenient in order to impose the transversality constraint $\epsilon_i\cdot p_i=0$ throughout, as well as for reusing the integral basis for integrands with four external vectors. 

Because of the different external states, only a subset of the maps in Eq.~\eqref{eq:map_ordering} is relevant here.
We refrain from giving the values of the remaining diagrams but include them in the ancillary file \texttt{SQED\textunderscore SSAA\textunderscore1L.m}. The final assembled amplitude after mapping to the global basis is
\begin{align}
\label{eq:compton_1L}
\begin{split}
    \mathcal{A}^{(1)}_{\phi A A \bar \phi}={} 4Q_1^4
    \Bigg[
    &
    \epsilon _2\cdot \epsilon _3\left(\frac{1}{2} I_{0,1,0,1}^{{(1)}}+I_{0,1,0,1}^{{(2)}}+I_{0,1,0,1}^{{(4)}}+t\, I_{0,1,1,1}^{{(8)}}\right)\\
    &-2 (p_1\cdot \epsilon _2) (p_1\cdot \epsilon _3) 
        \left(t\, I_{1,1,1,1}^{{(1)}}+I_{0,1,1,1}^{{(1)}}+t \,I_{1,1,1,1}^{{(3)}}\right)\\
    &+ 2 (p_1\cdot \epsilon _2) \left(
    t\, I_{1,1,1,1,0,0,-1,0}^{{(1)}}
    +I_{0,1,1,1,0,0,-1,0}^{{(1)}}
    +t\, I_{1,1,1,1,0,-1,0,0}^{{(3)}}\right)\\
    &+ 2 (p_1\cdot \epsilon _3) \left(
     t\, I_{1,1,1,1,0,-1,0,0}^{{(1)}}
    +I_{0,1,1,1,0,-1,0,0}^{{(1)}}
    + t\, I_{1,1,1,1,0,0,-1,0}^{{(3)}}\right)\\
    &-2\left(
    t\, I_{1,1,1,1,0,-1,-1,0}^{{(1)}}
    +I_{0,1,1,1,0,-1,-1,0}^{{(1)}}
    +t\, I_{1,1,1,1,0,-1,-1,0}^{{(3)}}\right)
    \Bigg]\,,
\end{split}    
\end{align}
where, for compactness, we found it convenient to choose polarization vectors that obey $\epsilon_2\cdot p_3= \epsilon_3\cdot p_2=0$.
Notably, the gauge-choice parameter $\xi$ that is present in individual Feynman diagrams (e.g. in Eq.~(\ref{eq:compton_1L_eg_diag})), 
cancels again in the integrand. Moreover, there have been extensive cancellations between the various Feynman diagrams, e.g., most of the individual terms in Eq.~\eqref{eq:compton_1L_eg_diag} are absent in the final result.

When testing the on-shell Ward identity, $\left(\text{e.g.~}\mathcal{A}^{(1)}_{\phi A A \bar \phi}\,\big|_{\epsilon_2\to p_2}=0\right)$, for the one-loop Compton amplitude in Eq.~(\ref{eq:compton_1L}) we have to track the fate of the polarization-vector-dependent ISPs. For example,
\begin{equation}
    \rho_5^{(1)} =\epsilon_2\cdot\ell\to p_2\cdot\ell=\frac{1}{2}\left(s+\rho_2^{(1)}-\rho_3^{(1)}\right) ,
\end{equation}
and the resulting expression must be re-expressed in the global basis, i.e., the daughter topologies generated from the $\rho_{2,3}^{(1)}$ terms have to be mapped to their global representatives. We have checked that indeed the amplitude satisfies the expected on-shell Ward identities, $\mathcal{A}^{(1)}_{\phi A A \bar \phi}\,\big|_{\epsilon_i\to p_i}=0$ with $i=2,3$, at the level of the integrand. 

The discussion of the Compton amplitude shows how we can straightforwardly include external vector bosons (and their corresponding polarizations) in our setup. We have also checked that the gauge-choice parameter $\xi$ cancels and the Ward identities are satisfied at the integrand level for all the four-point amplitudes with up to four external photons through two-loop order. Similar to the discussion of external vectors, we can also treat more general tensor integrals and external fermions.
%

\subsection{Application: Integrand Cleanup}
\label{subsec:integrand_cleanup}

\begin{figure}[tbh!]
\centering 
\begin{subfigure}[b]{0.22\textwidth}
\centering
    \includegraphicslog{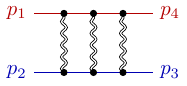}
    \caption{}
    \label{subfig:GR_2L_1}
\end{subfigure}
\begin{subfigure}[b]{0.22\textwidth}
\centering
    \includegraphicslog{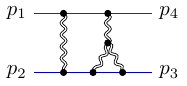}
    \caption{}
    \label{subfig:GR_2L_2}
\end{subfigure}
\begin{subfigure}[b]{0.22\textwidth}
\centering
    \includegraphicslog{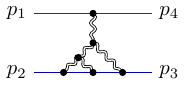}
    \caption{}
    \label{subfig:GR_2L_3}
\end{subfigure}
\begin{subfigure}[b]{0.22\textwidth}
\centering
    \includegraphicslog{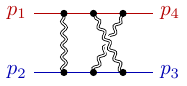}
    \caption{}
    \label{subfig:GR_2L_4}
\end{subfigure}\\[8pt]
\begin{subfigure}[b]{0.22\textwidth}
\centering
    \includegraphicslog{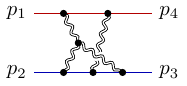}
    \caption{}
    \label{subfig:GR_2L_5}
\end{subfigure}
\begin{subfigure}[b]{0.22\textwidth}
\centering
    \includegraphicslog{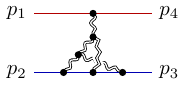}
    \caption{}
    \label{subfig:GR_2L_6}
\end{subfigure}
\begin{subfigure}[b]{0.22\textwidth}
\centering
    \includegraphicslog{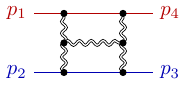}
    \caption{}
    \label{subfig:GR_2L_7}
\end{subfigure}
\begin{subfigure}[b]{0.22\textwidth}
\centering
    \includegraphicslog{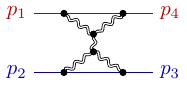}
    \caption{}
    \label{subfig:GR_2L_8}
\end{subfigure}
\caption{
The eight cubic diagram topologies relevant for extracting classical observables for conservative black hole scattering in general relativity in the conventions of Ref.~\cite{Bern:2019crd}. The non-1PI diagrams (3), (6), and (8) can be embedded into, e.g., diagrams (2), (5), and (8), respectively, by appropriate relabeling and keeping track of the pole in external kinematics resulting from the $1/(p_1+p_4)^2$ propagator. These diagrams are not included in the integrand of Ref.~\cite{Cheung:2020gyp}. 
\label{fig:2L_classical_GR_diags}}
\end{figure}

As a third application of our mapping procedure, we show that different integrand representations become identical when expressed in a basis of planar and nonplanar integrands. In other words, we demonstrate that the integrand is unique in the global basis.

To this end, we analyze two-loop four-point integrands for the scattering of massive scalar particles interacting via gravitons. These integrands are the starting point for extracting classical gravitational physics from quantum scattering amplitudes~\cite{Kosower:2018adc, Cheung:2018wkq, Bjerrum-Bohr:2018xdl} at order $G^3$ in a perturbative expansion in Newton's constant. Several variants are available. Ref.~\cite{Bern:2019nnu,Bern:2019crd} constructed an integrand using generalized unitarity, the double copy, and judiciously controlling the double-copy spectrum. Subsequently, Ref.~\cite{Cheung:2020gyp} derived three separate integrands by direct Feynman-diagram calculations in different gauges. 
Only a certain subset of diagrams contain classical long-range interactions, so for our purpose, all others can be ignored. Further focusing on conservative physics and ignoring radiative effects, a systematic approach to isolating the relevant diagrams~\cite{Bern:2019nnu, Bern:2019crd} is to impose the unitarity cuts shown in Fig.~\ref{fig:CutsGR_2L}. Any diagram not supported on one of these cuts can be dropped from the outset. Explicit calculations have shown that all available integrands lead to the same physical observables \emph{after} loop integration, and are therefore equivalent.
Here, we will show that, when written in our global basis, the classical truncations of all these integrands are identical even \emph{before} loop integration.

Both sets of integrands can be represented diagrammatically in terms of the cubic diagrams depicted in Fig.~\ref{fig:2L_classical_GR_diags} and can be obtained from the ancillary files of Refs.~\cite{Bern:2019crd}, and \cite{Cheung:2020gyp}, respectively. Note that Ref.~\cite{Cheung:2020gyp} only computes a particular subsector (first order in gravitational self-force, i.e.~the next-to-leading order term in the expansion in which the mass of one particle is much bigger than the other) of the full scattering amplitude. This has to be taken into account in the final comparison. In the language of cuts, Ref.~\cite{Cheung:2020gyp} only considers contributions with support on the cut depicted in Fig.~\ref{subfig:Cut_2L_N}, and drops the diagrams in Figs.~\ref{subfig:GR_2L_3} and \ref{subfig:GR_2L_6}. We could apply our integrand mapping procedure to the complete quantum amplitudes and show their equivalence. Here, however, we first carry out the classical expansion of both sets of integrands and perform the mapping afterward. 

The correspondence principle identifies the classical parts of the diagrams supported on the cuts in Fig.~\ref{fig:CutsGR_2L} as the term with leading logarithmic dependence in the expansion at small momentum transfer relative to the center-of-mass energy, see, e.g.,~Refs.~\cite{Cheung:2018wkq, Bern:2019crd}. Moreover, since loop momenta are the amount of transferred momentum at each interaction vertex, they must also be of the same order as the total momentum transfer.
The expansion of the integrand in this kinematic configuration is carried out in dimensional regularization, using the method of regions~\cite{Beneke:1997zp}.
For our purpose of demonstrating the power of our integrand mapping technique, many of the details of the classical limit are not essential. Two important details, however, are that (1) the classical expansion leads to linearized (eikonal) matter propagators, while the graviton propagators remain quadratic in loop momenta and (2) diagram \ref{subfig:GR_2L_8} can be neglected because it cannot yield the classical logarithmic dependence on the momentum transfer. We refer the reader to, e.g., Refs.~\cite{Cheung:2018wkq, Kosower:2018adc, Bern:2019crd, Parra-Martinez:2020dzs} for further details on the classical limit.

Carrying out the classical expansion on the various integrands constructed in Refs.~\cite{Bern:2019crd} and \cite{Cheung:2020gyp}, summing over the permutations of external legs, and consistently truncating to the classical conservative contributions, we end up with a set of apparently distinct integrands {\em before} mapping to the integrand basis. 
However, mapping each of the integrands to our global basis, we find that they all agree in the classical limit, irrespective of the gauge or the construction procedure. 
We include the mapped classical integrand and our diagram conventions in the ancillary file \texttt{GR\textunderscore SSSS\textunderscore 2L\textunderscore classical\textunderscore pot.m}.
The agreement before loop integration highlights a key usefulness of an integrand basis: no matter the starting point, upon mapping, physically-equivalent loop integrands become unique in a chosen global basis.

In addition to removing redundancy between different integrand representations, our mapping procedure has further important practical advantages. 
The large freedom in the construction of an integrand is sometimes used as a means to enforce properties relevant to the problem at hand. For example, in the construction of integrands with the goal of studying UV properties of gravitational theories, it is desirable that term by term, the UV power counting is as good as possible. This good power counting is a consequence of nontrivial zeroes, which are a consequence of momentum conservation and graph equivalence relations. 
Such an integrand, however, has {\em term by term} poor infrared or soft properties, and is difficult to use to study these properties of the amplitude. Indeed, we encountered such an example in the computation of the four-loop classical gravitational scattering in maximal supergravity \cite{Bern:2024adl}: the classical expansion, which probes the integrand when loop momenta are soft, exhibits a variety of spurious poles that cannot appear in a Feynman diagram computation which, if tackled directly, vastly increase the complexity of the analysis. 
In general, such features may place out of reach calculations that push the boundaries of integration capabilities. It is therefore beneficial to remove these spurious features before attempting the IBP reduction to master integrals. 
Provided that the chosen global basis has good soft properties, e.g. similar to those of Feynman rules, our mapping procedure removes these spurious poles and yields an integrand which, term by term, has the same properties as the basis elements.
More generally, by choosing a global basis with desired properties, we can probe whether an integrand manifestly exhibits them by simply attempting to project it onto that basis.

%
\section{Basis-Centric Unitarity Method}
\label{sec:unitarity}
%

Exploiting unitarity and analytic properties of scattering amplitudes has a long history dating back to the inception of quantum mechanics. During the 1960s, the concept of the analytic S-matrix theory~\cite{Cutkosky:1960sp, Eden:1966dnq} became a fundamental tool for constraining particle interactions.  In the traditional approach, this is done in terms of integrated amplitudes, by requiring that branch-cut discontinuities of the S-matrix elements are given by integrals over the phase-space of bilinears in S-matrix elements,  according to the optical theorem  
\begin{align}
\label{eq:basic_unitarity_relation}
-\imath (T-T^\dagger) = T T^\dagger\,.
\end{align}
Perturbatively expanding this unitarity relation implies that the discontinuity of a loop amplitude across a branch cut is given in terms of lower-loop and higher-point amplitudes.\footnote{Additional information, related to the treatment of the falloff of the integrand at large momenta (ultimately tied to renormalization), is required for these phase-space integrals to be well-defined. } 

The generalized unitarity method~\cite{Bern:1994zx, Bern:1994cg, Bern:1995db, Bern:1997sc, Britto:2004nc} further assumes that order by order in perturbation theory, S-matrix elements in quantum field theory {\em can} be constructed in terms of Feynman graphs, and thus they can be written as integrals of rational functions---or integrands---with established poles, residues, and factorization properties. With this and tree-level amplitudes as a starting point, loop integrands can be constructed so that the perturbative expansion of Eq.~\eqref{eq:basic_unitarity_relation} is satisfied at every step. The resulting expressions are equivalent to, but often simpler than, the ones obtained from Feynman diagrams because it is possible to directly leverage the simplicity of on-shell tree amplitudes into that of loop integrands.

\subsection{Summary of the Generalized Unitarity Method}
\label{subsec:unitarity_review}

\begin{figure}[tb]
\begin{center}
\includegraphicslog{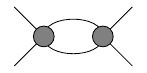}
\caption[a]{\small Representative spanning generalized unitarity cut for a massless one-loop four-point amplitude. Each blob represents a tree amplitude, and the exposed intermediate lines are on shell. The complete set is given by the independent relabelings of the four external legs.}
\label{OneLoopFourPtBubCutFigure}
\end{center}
\end{figure}

\begin{figure}[tb]
\begin{center}
\begin{subfigure}[b]{0.2\textwidth}
\centering
    \includegraphicslog[scale=1.2]{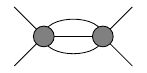}
    \caption{}
    \label{subfig:Cut_Gen_2L_1}
\end{subfigure}
\begin{subfigure}[b]{0.2\textwidth}
\centering
    \includegraphicslog[scale=1.2]{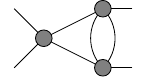}
    \caption{}
    \label{subfig:Cut_Gen_2L_2}
\end{subfigure}
\begin{subfigure}[b]{0.2\textwidth}
\centering
    \includegraphicslog[scale=1.2]{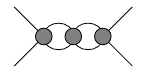}
    \caption{}
    \label{subfig:Cut_Gen_2L_3}
\end{subfigure}
\caption[a]{\small A spanning set of generalized cuts for a massless two-loop four-point amplitude.  The independent relabelings of the four external legs give the complete set.  
  }
\label{TwoLoopFourPtSpanCutFigure}
\end{center}
\end{figure}

The generalized unitarity method~\cite{Bern:1994zx, Bern:1994cg, Bern:1995db, Bern:1997sc, Britto:2004nc} is well-established (see, e.g., the review articles~\cite{Bern:1996je, Bern:2011qt}), with an extensive array of applications (see, e.g., Refs.~\cite{Bern:1998ug, Bern:1998sv, Bern:2007hh, Bern:2009kd, Drummond:2008bq, Berger:2008sj, Berger:2010zx, Abreu:2020xvt, Abreu:2021asb}) and refinements~\cite{Ossola:2006us, Bern:2007ct, Forde:2007mi, Bourjaily:2017wjl, Badger:2008cm, Bourjaily:2019iqr, Bourjaily:2019gqu}. Here we briefly review its salient points to set the stage for later sections. As we are going to discuss, augmenting traditional unitarity setups with our notion of global integrand bases leads to a new basis-centric unitarity framework that is characterized by its simplicity and power, especially at high loop orders.

The integrand-level generalized unitarity cuts iteratively employ the basic unitarity relation (\ref{eq:basic_unitarity_relation}) to compute residues of loop-level integrands as products of lower-loop (and ultimately tree) scattering amplitudes summed over the on-shell states that can be exchanged across the cut. Usually, one uses factorization into tree amplitudes where generalized unitarity cuts take the form, 
\begin{equation}
C \equiv \sum_{\rm states} 
\mathcal{A}^\tree_{n_1} 
\mathcal{A}^\tree_{n_2} 
\mathcal{A}^\tree_{n_3} \cdots 
\mathcal{A}^\tree_{n_m} \,.
\label{GeneralizedCut}
\end{equation}
The tree amplitudes entering the unitarity cuts can be computed in arbitrary perturbative quantum field theories. Most often, one is interested in either gauge or gravity amplitudes; in gauge theories, amplitudes can be color ordered or color dressed, implying that the state sum includes a summation over colors. Simple four-point examples of generic unitarity cuts at one and two loops are illustrated in \figs{OneLoopFourPtBubCutFigure}{TwoLoopFourPtSpanCutFigure}. To retain the full information for the loop integrands, one ought to compute the unitarity cuts in general spacetime dimensions~\cite{Bern:1995db, Badger:2008cm}, and only in specific circumstances (e.g. with large amounts of supersymmetry) it is legitimate to sew four-dimensional trees using simplifications of the spinor-helicity formalism~\cite{Xu:1986xb, Mangano:1990by, Dixon:1996wi}. 

To evaluate the state sums in \eqn{GeneralizedCut} we use completeness relations for the states that cross the cut. For example, for massless gauge fields, one has
\begin{align}
 \Pi^{\mu\nu}(\ell,q) =\!\! \sum_{{\rm states}} \!\!
  \epsilon^\mu(\ell)\epsilon^\nu(-\ell) = \eta^{\mu\nu} - \frac{q^\mu \ell^\nu + \ell^\mu q^\nu}{q\cdot\ell}\,,
  \label{StateSumGauge}
\end{align}
and for gravitons,  
\begin{align}
  \Pi^{\mu\nu \alpha \beta}(\ell,q)  =\!\! \sum_{{\rm states}} \!\!
  \epsilon^{\mu\nu}(\ell)\epsilon^{\alpha \beta}(-\ell) 
   = \frac{1}{2} \left[
      \Pi^{\mu \alpha}\Pi^{\nu \beta}
    + \Pi^{\nu \alpha}\Pi^{\mu \beta}
    -\frac{2}{D-2}\Pi^{\mu \nu}\Pi^{\alpha \beta}
    \right],
    \label{StateSumGravity}
\end{align}
both of which involve a null reference vector $q^\mu$ used to define the physical polarizations. Similar state sums exist for fermions. The null reference vector in these sums can be manifestly eliminated by judiciously choosing the form of the tree amplitudes that enter the cut. Indeed, $q$ always appears together with a linearized gauge transform, $\epsilon^\mu(\ell) \to \ell^\mu$, of a tree amplitude. While tree amplitudes are gauge invariant, this typically requires the use of transversality of the other polarizations. It is, however, possible to organize them such that they obey the \emph{generalized Ward identity} \cite{Kosmopoulos:2020pcd}, i.e., that they vanish under the single replacement $\epsilon^\mu(\ell) \to \ell^\mu$ without any further requirements on the other polarizations. Using such tree amplitudes guarantees that all $q$-dependent terms vanish from the outset and effectively simplifies the projectors
\begin{align}
\Pi^{\mu\nu}(\ell,q)\to \eta^{\mu\nu}\,, 
\qquad
\Pi^{\mu\nu \alpha \beta}(\ell,q)  \to
    \frac{1}{2} \left[
        \eta^{\mu \alpha}\eta^{\nu \beta}
    + \eta^{\nu \alpha}\eta^{\mu \beta}
    -\frac{2}{D-2}\eta^{\mu \nu}\eta^{\alpha \beta}
    \right]  ;
\end{align}
see Ref.~\cite{Kosmopoulos:2020pcd} for further details. 
For generic representations of the tree amplitudes, the $q$-dependent denominators generated by the state sum cancel once the unitarity cut is mapped to the integrand basis.  With the mapping, since the light-cone denominators cancel there is no need to apply an additional prescription, such as the principal part or the Mandelstam-Leibbrandt  prescription~\cite{Mandelstam:1982cb, Leibbrandt:1983pj, Leibbrandt:1987qv}, used for light-cone gauge.   

Being a rational function, the integrand can be reconstructed from the knowledge of its poles and the corresponding residues, i.e., the generalized unitarity cuts. While, in principle, all poles and residues must be considered, it suffices to focus on a subset of cuts such that any diagram that can appear in the Feynman diagram calculation of the amplitude appears in at least one cut. The resulting list of cuts is usually referred to as a \emph{spanning set of cuts}. 

The simplest example of a spanning set is that for massless one-loop four-point amplitudes: it is simply given by the two-particle cut shown in \fig{OneLoopFourPtBubCutFigure} together with the independent permutations of external legs. The figures here are agnostic on the QFT under consideration. When discussing particular amplitudes in specific theories, we often dress the cuts with additional labels and information about the particle species, see, e.g., Eq.\ (\ref{eq:sCut1L}) for four-scalar scattering in QED at one-loop. Any contribution not part of this cut (or its relabeling) has either only a single propagator (tadpole) or is a massless bubble on a massless external leg and thus vanishes in dimensional regularization, see also footnote \ref{footnote:bub_tad}. In non-supersymmetric theories, such terms require additional regularization prescriptions. In a first pass, it is convenient to ignore them and restore the missing information at the end by leveraging independent knowledge on the UV or IR properties of the amplitude, see, e.g., Ref.~\cite{Bern:1995db, Badger:2017gta, Bern:2021ppb}, or by attempting to make sense of single cuts of the amplitude \cite{Catani:2008xa, NigelGlover:2008ur, Bierenbaum:2010cy, Caron-Huot:2010fvq, Britto:2010um, Baadsgaard:2015twa}. As already mentioned, we drop tadpole and massless-bubble-on-external-leg terms in our integrand construction procedure, as well as from the cut discussion to follow. A less trivial example is the spanning set of cuts, illustrated in \fig{TwoLoopFourPtSpanCutFigure}, for a massless two-loop four-point amplitude. Again, the complete set is obtained by including all independent relabelings of external legs.

The spanning cuts typically do not yield completely disjoint data. Rather, their expressions contain overlapping information, corresponding to integrand terms that have poles that appear in two or more cuts. Even a single cut may double count integrand terms if those terms have two or more pole configurations that appear in that cut (see, e.g., Ref.~\cite{Bern:1998ug}, or \Eq{eq:2loop_sqed_eg_cut}). Thus, the construction of the integrand from the spanning set of cuts requires a careful \emph{merging} process of the partial information contained in individual cuts. This seemingly innocuous step can lead to considerable computational challenges, especially at high loop order or for gravitational theories where the cut expressions are complicated. Several methods have been developed for this purpose, both before and after reduction to master integrals.  

The method of maximal cuts~\cite{Bern:2007ct, Bern:2010tq, Bern:2008pv} organizes the information in the spanning set of cuts according to the number of on-shell propagators. First, the on-shell conditions are imposed in the maximal cuts, i.e. cuts with the maximal number of on-shell propagators. Since they correspond to different graphs, they are disjoint, and thus, they can be simply added together to obtain a partial integrand. This partial integrand has the correct maximal cuts by construction; it also has next-to-maximal cuts, i.e., cuts in which the next-to-maximal number of propagators are put on shell. The next-to-maximal cuts of the partial integrand differ from the correct answer (determined by, e.g., sewing tree amplitudes) by {\em local terms}. (Determining this difference analytically can become rather involved for complicated amplitudes.) These differences can be directly added to the partial integrand because the different local terms correspond to different graphs. The process continues iteratively, with increasingly fewer propagators put on shell, until the least singular terms in the spanning set are accounted for. 

Another approach to cut merging before IBP reduction calls for constructing an ansatz for the integrand, constructing its cuts and matching them against the expressions of the spanning cuts obtained by sewing together tree amplitudes. While this entails solving a linear system that can grow substantially at higher loop orders, it has the advantage that various desired properties can be imposed on the ansatz and, consequently, on the resulting amplitude. This includes, e.g., that a certain power counting is manifest in each diagram or that the diagrams' numerators satisfy the duality between color and kinematics. 

Prescriptive unitarity~\cite{Bourjaily:2017wjl} attempts to avoid the potentially difficult linear algebra problem by pre-computing specific integrand bases that have support only on a single spanning cut. The suggested procedure depends on the precise knowledge of the power counting of the theory and has mostly been worked out in the context of four-dimensional integrands, where it has been used to, e.g., systematically build maximally supersymmetric two-loop amplitudes with an arbitrary number of external legs~\cite{Bourjaily:2019gqu}.  

Perhaps the earliest method that addressed systematically the highly nontrivial cut overlaps was described in Ref.~\cite{Bern:2004cz} for the two-loop gluon splitting function in QCD. It starts by constructing a partial integrand as one of the cuts of the spanning set in which the cut conditions are released and then iteratively adding to it the other cuts in the set while subtracting the overlap and correcting for the diagram overcount. This, as well as the maximal-cut method, can also be used after the spanning cuts have been reduced to master integrals. Since the master integrals form a basis, the identification of the cut overlap or the construction of the next$^k$-maximal cuts can be accomplished by inspection. Consistency of the spanning cuts requires that the coefficient of a master integral is the same no matter what cut is used to determine it.

\subsection{Mapping of Generalized Cuts into Global Basis}
\label{subsec:unitarity_mapping}

As we will explain in this section, the realization that the {\em pre-IBP cuts} can be put in a global integrand basis greatly simplifies the cut-merging procedure in general quantum field theories. The process is straightforward: (1) put the trees in diagrammatic form; (2) use these trees to express the spanning cuts in diagrammatic form, (3) map the spanning cuts to the global integrand basis while keeping track of the cut legs (4) merge the cuts by reading off the coefficients of the global basis elements while checking mutual cut consistency. Our new cut-merging procedure can be summarized schematically as
\begin{align}
\label{eq:cutsTOintegrand}
  \text{Map}\Big[ \text{spanning cuts}\Big] \Rightarrow \text{loop integrand} \ .
\end{align}
Starting with step (1), the Feynman diagram representation of tree amplitudes guarantees that, regardless of their construction method, they have a diagrammatic form, e.g.,
\begin{align}
\label{eq:4pt_diags}
    \vcenter{\hbox{\includegraphicslog[scale=0.75]{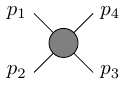}}}={}&\vcenter{\hbox{\includegraphicslog[scale=0.75]{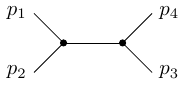}}}+\vcenter{\hbox{\includegraphicslog[scale=0.75]{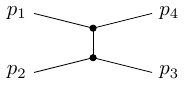}}}+\vcenter{\hbox{\includegraphicslog[scale=0.75]{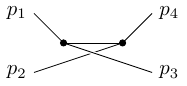}}}+\vcenter{\hbox{\includegraphicslog[scale=0.75]{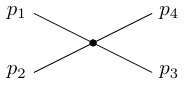}}}\,,\\[8pt]
    \vcenter{\hbox{\includegraphicslog[scale=0.75]{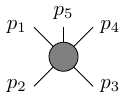}}}={}&
    \vcenter{\hbox{\includegraphicslog[scale=0.75]{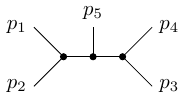}}}
    +\dots+\vcenter{\hbox{\includegraphicslog[scale=0.75]{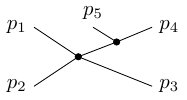}}}
    +\dots+\vcenter{\hbox{\includegraphicslog[scale=0.75]{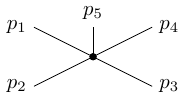}}}\, .
    \label{eq:5pt_diags}
\end{align}
Each of the tree diagrams on the right-hand side~of \Eqs{eq:4pt_diags}{eq:5pt_diags} is dressed with a numerator that depends on the external kinematics (momenta, polarizations, masses) and other data dictated by the QFT under consideration (charges, color factors, coupling constants, etc.). The denominator structure is encoded by the tree graph itself. Each internal line $i$ of the graph corresponds to a propagator assumed to be of the form $\frac{1}{\rho_i} = \frac{1}{s_{a\cdots b}-m^2_i}$, where the Feynman $\imath \varepsilon$ prescription is left implicit and $s_{a\cdots b} = (p_a+\cdots+p_b)^2$. 
Starting from such a representation, contact diagrams such as the last in Eqs.~\eqref{eq:4pt_diags} and \eqref{eq:5pt_diags} can be eliminated to obtain a purely cubic representation by multiplying and dividing the contact terms by products of $1=\frac{\rho_i}{\rho_i}$.  Schematically, the $n$-point tree amplitudes are then written as
\begin{align}
\label{eq:Atree_cubic_rep_generic}
 \mathcal{A}^{\tree}_n = \sum_{\Gamma} \frac{n_\Gamma}{\prod_i \rho_i}\,,
\end{align}
where $\{\Gamma\}$ is the set of cubic $n$-point tree diagrams contributing to the amplitude, $n_\Gamma$ is the theory-specific diagram numerator, and $\prod_i \rho_i$ are the inverse propagators of the tree graph $\Gamma$. 

Since the generalized cuts are sums of products of tree amplitudes, see~\Eq{GeneralizedCut}, they inherit a diagrammatic representation, e.g.
\begin{align}
    \vcenter{\hbox{\includegraphicslog[scale=.9]{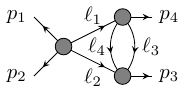}}} & = 
        \sum_{{\rm states}} 
            \mathcal{A}^{\tree}_4 (p_1,\ell_1,\ell_2,p_2)
            \mathcal{A}^{\tree}_4(-\ell_1,p_4,\ell_3,\ell_4) 
            \mathcal{A}^{\tree}_4(-\ell_3,p_3,-\ell_2,-\ell_4) \nonumber \\
            & 
          = 
         \sum_{{\rm states}} \left[  \vcenter{\hbox{\includegraphicslog{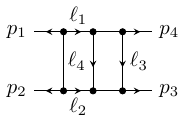}}}{+}  \vcenter{\hbox{\includegraphicslog{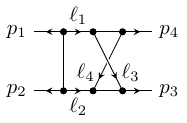}}}{+}  
         \text{7 more} \right],
         \hskip -.5cm 
\label{eq:cut_sum_diags}         
\end{align}
where the labeled internal legs on the second line of Eq.~(\ref{eq:cut_sum_diags}) are cut (on-shell). Evaluating the state sum leads to a diagrammatic representation of the unitarity cut in terms of Feynman-like diagrams with numerators $n_{\Gamma_i}$
\begin{align}
    \vcenter{\hbox{\includegraphicslog[scale=.9]{figs/Cut_Gen_2L_2_Labeled.pdf}}} & 
    = \,
        \left[  
            \frac{n_{\includegraphicslog[scale=0.2]{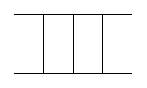}} (\ell_i,p_j,\ldots)}
            {(\ell_2 {+} p_2)^2(\ell_3{-}p_3)^2(\ell_3{+}p_4)^2} \right.  \nonumber\\
            & \hskip 2 cm 
      \left. \null  + \frac{n_{\includegraphicslog[scale=0.2]{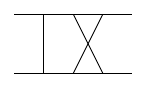}}(\ell_i,p_j,\ldots)}
             {(\ell_2 {+} p_2)^2(\ell_3{-}p_3)^2(\ell_4{+}p_4)^2}
        + \text{7 more} 
        \right] . 
\label{eq:cut_sum_num_den}         
\end{align}
Here we omitted the on-shell propagators shared between all diagrams. Notably, in the \emph{cut labels}, the cut legs are labeled uniformly in all diagrams contributing to a given cut. Below, we will explicitly mark cut propagators by a `$\,\widehat{\textcolor{white}{\rho}}\,$' on the corresponding index in the inverse propagator basis.  
The numerators $n_{\Gamma_i}$ depend on the loop and external momenta, as well as on possible external polarizations, charges, or color factors. In particular examples, the numerators can also be proportional to a subset or all (as is the case for the nonlinear sigma model discussed in \sect{subsec:NLSM}) of the uncut inverse propagators, limiting the number of graphs that need to be considered. 

Having expressed the unitarity cuts in a diagrammatic form, we can now map them to the global integrand basis. This can be done diagram by diagram, relabeling it so that it corresponds to the global integrand basis element with the topology of the diagram, writing subsequently the numerator in terms of inverse propagators, and iteratively repeating the relabeling for the resulting daughter graphs obtained by collapsing uncut propagators. Alternatively, we may start by writing all terms in the notation introduced in Eq.~\eqref{integralbasis_firenotation}, and then map the resulting integrals to the basis of integrands, as discussed in the previous section for off-shell integrands. In both strategies, it is important to explicitly keep track of the cut propagators.

With the spanning cuts (and, in fact, all unitarity cuts, if desired) mapped to the global integrand basis, cut merging is reduced to 
collecting (non-redundantly) all distinct integrand basis elements and reading off their coefficients. 
Consistency requires that whenever one integrand basis element appears in two different cuts, or in the same cut but with a different configuration of cut propagators, its coefficients have to agree. 
Once this is done for the spanning cuts, the resulting list of basis integrands and their coefficients sum to the off-shell loop integrand without performing extensive linear algebra manipulations. 
The strength of the global integrand basis is the simplicity in which terms that populate the overlap of multiple cuts are identified, in contrast to previous cut-merging procedures~\cite{Bern:2004cz}.

\subsection{Example: Cut Construction of Four-Scalar Amplitudes in QED
\label{subsec:SQED_cut_construction}
}

%

Having explained the general strategy for using the global integrand basis to streamline the construction of loop integrands from spanning cuts, we now proceed to discuss several illustrative examples: the one- and two-loop four-scalar amplitudes in massless scalar QED.  Below in \sect{subsec:NLSM} we present Adler-zero examples in the nonlinear sigma model (NLSM).

\noindent
\paragraph{One-Loop\\}
%
At one-loop order, all spanning cuts have the same topology and are obtained from the one cut shown in Fig.~\ref{OneLoopFourPtBubCutFigure} by relabeling the external legs. The evaluation calls for a sum over all the states that run on the exposed (cut) lines; we will list separately each contribution after mapping to the global integrand basis. The relevant tree amplitudes can be easily derived by, e.g., Feynman rules from the Lagrangian \eqref{eq:SQED_Lagrangian}.

We start by giving a concrete example. Adding loop-momentum labels to the cut lines, and sewing the tree-level amplitudes, we have
\begin{align}
\label{eq:sample_cut_explanation}
\begin{split}
\vcenter{\hbox{\includegraphicslog{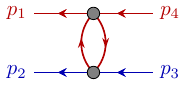}}}\!={}&\frac{1}{2}\left[  
\vcenter{\hbox{\includegraphicslog{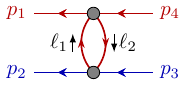}}}+
\vcenter{\hbox{\includegraphicslog{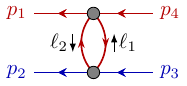}}}\right]
\\[5pt]
={}& \frac{1}{2}\left[
\vcenter{\hbox{\includegraphicslog{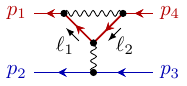}}}+
\vcenter{\hbox{\includegraphicslog{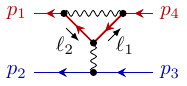}}}\right]
\\[4pt]
={} & \ Q^3_1 Q_2 \, 2\pi\delta(\ell^2_1) \, 2\pi\delta(\ell^2_2)\,
\big[\ell_1\cdot (p_2{-}p_3)\big] 
\left[\frac{1}{\ell_1\cdot p_1} - \frac{1}{\ell_1\cdot p_4}\right],
\end{split}
\end{align}
where we have used momentum conservation $\ell_2 = \ell_1 + p_{23}$ and the support of the on-shell delta functions of the cut. The factor $1/2$ is due to the symmetry of the two-particle-cut phase space and the external momenta are taken to be outgoing. Clearly, \Eq{eq:sample_cut_explanation} highlights the presence of two separate factorization channels of the cut encoded in the $1/(\ell_1 \cdot p_4)$ and $1/(\ell_1 \cdot p_1)$ singularity, respectively. We then proceed to write the expression in the inverse propagator basis
\begin{equation}
\label{eq:Bubble_Cut4_UnMapped}
\vcenter{\hbox{\includegraphicslog{figs/Bubble_Cut4_1L.pdf}}}=-2Q^3_1 Q_2 
\left[\left( {I}_{{-1,\cutindex{1},1,\cutindex{1}}}^{{(5)}}+\frac{t}{2} {I}_{{0,\cutindex{1},1,\cutindex{1}}}^{{(5)}}\right)
+\left( {I}_{{-1,\cutindex{1},1,\cutindex{1}}}^{{(19)}}+\frac{t}{2} {I}_{{0,\cutindex{1},1,\cutindex{1}}}^{(19)}\right)
\right],
\end{equation}
where a hat denotes a cut propagator, i.e. a replacement of a propagator by a delta function $\frac{1}{x}\to 2\pi\delta(x)$. The two expressions in the parentheses in \Eq{eq:Bubble_Cut4_UnMapped} are identical up to internal loop-momentum labels which are specified in the ancillary file \texttt{StandardTopos\textunderscore{}SSSS\textunderscore{}1L.m}. For the purpose of cut merging, we identify them, only keeping track of the position of the cut lines,
\begin{equation}
\label{eq:Bubble_Cut4_Mapped}
\vcenter{\hbox{\includegraphicslog{figs/Bubble_Cut4_1L.pdf}}}=-4Q^3_1 Q_2 \left( {I}_{{-1,\cutindex{1},1,\cutindex{1}}}^{{(8)}}+\frac{t}{2} {I}_{{0,\cutindex{1},1,\cutindex{1}}}^{{(8)}}\right),
\end{equation}
where the integrals $I^{(5)}$ and $I^{(19)}$ are mapped to the basis element $I^{(8)}$ as listed in the ancillary file \texttt{RepresentativeSectors\textunderscore{}SSSS\textunderscore{}1L.m} and in Eqs.~\eqref{eq:1loopsectors1}-\eqref{eq:1loopsectors3}.
 While there exists a uniform labeling of loop momenta across diagrams for individual cuts, in general, the same is not true for the off-shell integrand due to the label mismatch of the same diagram between different cuts. Therefore, in order to promote the cut to the off-shell integrand, in accordance with our global integrand construction, we have to identify equivalent terms in the cut. Upon the identification of the global basis elements, the cut in \Eq{eq:Bubble_Cut4_Mapped} uniquely fixes the coefficient off the integrals ${I}_{0,1,1,1}^{(8)}$ and ${I}_{-1,1,1,1}^{(8)}$, in agreement with the respective coefficients in the integrand obtained from Feynman diagrams in the last line of \Eq{eq:sqed_ssss_1L}.

Even though the different factorization channels have been identified, it is possible to go in the reverse and obtain the cut from the integrand. To do so, we take the sum of all integrals contributing to the given cut written in terms of Lorentz dot products and then reimpose the symmetry of the given cut.

For completeness we also list the remaining cuts mapped to the integrand basis that are required to construct the one-loop QED amplitude. The $s$-channel cut is given by
\begin{equation}
\vcenter{\hbox{\includegraphicslog{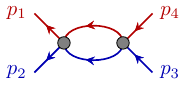}}}= Q_1^2Q_2^2\left(4s^2 I^{(1)}_{\cutindex{1},1,\cutindex{1},1}+2 s I^{(4)}_{0,\cutindex{1},1,\cutindex{1}}+2 s I^{(13)}_{0,\cutindex{1},1,\cutindex{1}}+I^{(4)}_{0,\cutindex{1},0,\cutindex{1}}\right),
\label{eq:sCut1L}
\end{equation}
where the propagators are defined in \Eq{eq:box_rho_explicit}. The $u$-channel cut is obtained from the $s$-channel cut via the relabeling $p_2\leftrightarrow p_3$:
\begin{equation}
\vcenter{\hbox{\includegraphicslog{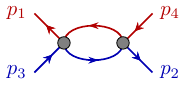}}}= Q_1^2Q_2^2\left(4u^2 I^{(3)}_{\cutindex{1},1,\cutindex{1},1}+2 u I^{(7)}_{0,\cutindex{1},1,\cutindex{1}}+2 u I^{(2)}_{0,\cutindex{1},1,\cutindex{1}}+I^{(2)}_{0,\cutindex{1},0,\cutindex{1}}\right).
\label{eq:uCut1L}
\end{equation}
This highlights the convenience of our labeling convention which greatly streamlines the map of the individual integrals.
The $t$-channel cut is a sum of three contributions, one of which we have explicitly computed in \Eq{eq:Bubble_Cut4_Mapped},  that should be added up to obtain the sum over states as in Eq.~\eqref{GeneralizedCut}:
\begin{align}
\label{eq:tCut1La}
\hskip -.7cm
\vcenter{\hbox{\includegraphicslog{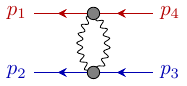}}}\!={}&
4Q_1^2Q_2^2\left[
       s^2 I^{(1)}_{1,\cutindex{1},1,\cutindex{1}}
    {+}u^2 I^{(3)}_{1,\cutindex{1},1,\cutindex{1}}
    {+}t I^{(1)}_{0,\cutindex{1},1,\cutindex{1}}{+} t I^{(8)}_{0,\cutindex{1},1,\cutindex{1}}
    {+}\frac{1}{2}(D{-}2)I^{(1)}_{0,\cutindex{1},0,\cutindex{1}}
\right],
\hskip -.6cm
\\[5pt]
\label{eq:tCut1Lb}
\hskip -.7cm
\vcenter{\hbox{\includegraphicslog{figs/Bubble_Cut4_1L.pdf}}}\!={}&
{-}4Q_1^3Q_2\left( {I}_{{-1,\cutindex{1},1,\cutindex{1}}}^{{(8)}}+\frac{t}{2} {I}_{{0,\cutindex{1},1,\cutindex{1}}}^{{(8)}}\right),
\hskip -.6cm
\\[5pt]
\label{eq:tCut1Lc}
\hskip -.7cm
\vcenter{\hbox{\includegraphicslog{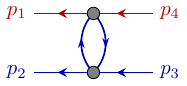}}}\!={}&
{-}4Q_1Q_2^3\left( {I}_{{-1,\cutindex{1},1,\cutindex{1}}}^{{(1)}}+\frac{t}{2} {I}_{{0,\cutindex{1},1,\cutindex{1}}}^{{(1)}}\right).
\hskip -.6cm
\end{align}
As in Sec.~\ref{subsec:xiDrop}, we dropped terms with a closed scalar loop that would contribute to the cuts in Eqs.~\eqref{eq:tCut1Lb} and \eqref{eq:tCut1Lc}. The cut in Eq.~(\ref{eq:tCut1La}), which splits the amplitude into two Compton amplitudes, has been worked out in detail before mapping in section III.A of Ref.~\cite{Kosmopoulos:2020pcd} as an example of the simplified sewing procedure with tree amplitudes obeying generalized Ward identities. Although we distinguished the three different $t$-channel cuts based on the particles crossing the cut, we could instead introduce a cut skeleton where the state sum is implicitly understood. We refrain from doing so here. Note that all the cuts have the correct symmetry under the flip $p_1\leftrightarrow p_4, p_2\leftrightarrow p_3$. 

As emphasized in our general discussion, an important advantage of mapping the unitarity cuts to a basis of (nonplanar) integrands is that it exposes the agreement of cuts in the overlap. That is, whenever an integral appears in multiple cuts, its coefficient is identical, up to combinatorial factors. For example, the overlap between the $s$-channel and the $t$-channel cuts is given by 
\begin{equation}
\vcenter{\hbox{\includegraphicslog{figs/Bubble_Cut2_1L.pdf}}}
    \cap
    \vcenter{\hbox{\includegraphicslog{figs/Bubble_Cut1_1L.pdf}}} 
    =\vcenter{\hbox{\includegraphicslog{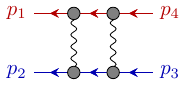}}} 
    =4 Q_1^2Q_2^2 s^2I^{(1)}_{\cutindex{1},\cutindex{1},\cutindex{1},\cutindex{1}}\, .
    \label{eq:1loop_overlap}
\end{equation}
Reading off the coefficients of the individual integrals from the cuts in Eqs.~\eqref{eq:sCut1L}--\eqref{eq:tCut1Lc}, we obtain the full amplitude for charged-scalar scattering in QED,
\begin{align}\begin{split}
\mathcal{A}^{(1)}_{\phi\phi\bar{\phi}\bar{\phi}}={}&
Q_2^2 Q_1^2 \left[2 (D-2) {I}_{{0,1,0,1}}^{{(1)}}+{I}_{{0,1,0,1}}^{{(2)}}+{I}_{{0,1,0,1}}^{{(4)}}+4 t {I}_{{0,1,1,1}}^{{(1)}}+2 s {I}_{{0,1,1,1}}^{{(13)}}\right.\\ 
&\left.+2 u {I}_{{0,1,1,1}}^{{(2)}}+2 s {I}_{{0,1,1,1}}^{{(4)}}+2 u {I}_{{0,1,1,1}}^{{(7)}}+4 t {I}_{{0,1,1,1}}^{{(8)}}+4 s^2 {I}_{{1,1,1,1}}^{{(1)}}+4 u^2 {I}_{{1,1,1,1}}^{{(3)}}\right]\\
&
-4Q_2^3 Q_1 \left({I}_{{-1,1,1,1}}^{{(1)}}+\frac{t}{2} {I}_{{0,1,1,1}}^{{(1)}}\right)
-4Q_2 Q_1^3 \left({I}_{{-1,1,1,1}}^{{(8)}}+\frac{t}{2} {I}_{{0,1,1,1}}^{{(8)}}\right).
\label{eq:1loopSQEDmerging}
\end{split}
\end{align}
The same result can also be obtained in the approach of Ref.~\cite{Bern:2004cz} in our basis, by adding together Eqs.~\eqref{eq:sCut1L}--\eqref{eq:tCut1Lc} and subtracting the overlap, e.g.~the box contribution in Eq.~\eqref{eq:1loop_overlap}.

\Eq{eq:1loopSQEDmerging} is in agreement with the amplitude in \Eq{eq:sqed_ssss_1L} obtained from a Feynman-diagram calculation. It is worth emphasizing again that the use of a basis of integrands greatly simplifies the cut-merging procedure: we simply count each integrand basis element once and add it with its coefficient (that was determined by one of the spanning cuts) to form the off-shell loop integrand. This procedure allows us to turn cuts into off-shell loop integrands in an algorithmic manner, without having to solve systems of linear equations.  While at one loop, the basis is, of course, a bit too simple to illustrate all the features, the same procedure continues to hold at higher loops, as we now show at two loops. 

\begin{figure}
\centering
\begin{subfigure}[b]{0.2\textwidth}
\centering
    \includegraphicslog{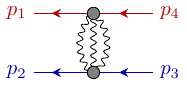}
    \caption{}
    \label{subfig:Cut_2L_A}
\end{subfigure} \hskip .8 cm 
\begin{subfigure}[b]{0.2\textwidth}
\centering
    \includegraphicslog{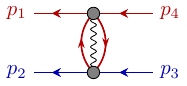}
    \caption{}
    \label{subfig:Cut_2L_A2}
\end{subfigure} \hskip .8 cm 
\begin{subfigure}[b]{0.2\textwidth}
\centering
    \includegraphicslog{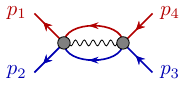}
    \caption{}
    \label{subfig:Cut_2L_B}
\end{subfigure}
\\[8pt]
\begin{subfigure}[b]{0.2\textwidth}
\centering
    \includegraphicslog{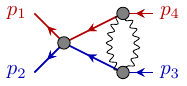}
    \caption{}
    \label{subfig:Cut_2L_C}
\end{subfigure}  \hskip .8 cm 
\begin{subfigure}[b]{0.2\textwidth}
\centering
    \includegraphicslog{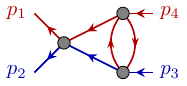}
    \caption{}
    \label{subfig:Cut_2L_C2}
\end{subfigure} \hskip .8 cm 
\begin{subfigure}[b]{0.2\textwidth}
\centering
    \includegraphicslog{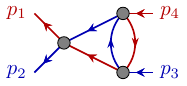}
    \caption{}
    \label{subfig:Cut_2L_C3}
\end{subfigure}
\\[8pt]
\begin{subfigure}[b]{0.2\textwidth}
\centering
    \includegraphicslog{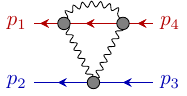}
    \caption{}
    \label{subfig:Cut_2L_D}
\end{subfigure} \hskip .6 cm 
\begin{subfigure}[b]{0.2\textwidth}
\centering
    \includegraphicslog{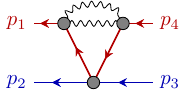}
    \caption{}
    \label{subfig:Cut_2L_D2}
\end{subfigure} \hskip .6 cm 
\begin{subfigure}[b]{0.2\textwidth}
\centering
    \includegraphicslog{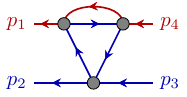}
    \caption{}
    \label{subfig:Cut_2L_D3}
\end{subfigure} \hskip .6 cm 
\begin{subfigure}[b]{0.2\textwidth}
\centering
    \includegraphicslog{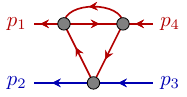}
    \caption{}
    \label{subfig:Cut_2L_D4}
\end{subfigure}\\[8pt]
\begin{subfigure}[b]{0.2\textwidth}
\centering
    \includegraphicslog{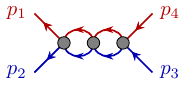}
    \caption{}
    \label{subfig:Cut_2L_E}
\end{subfigure} \hskip .6 cm 
\begin{subfigure}[b]{0.2\textwidth}
\centering
    \includegraphicslog{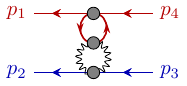}
    \caption{}
    \label{subfig:Cut_2L_F}
\end{subfigure} \hskip .6 cm 
\begin{subfigure}[b]{0.2\textwidth}
\centering
    \includegraphicslog{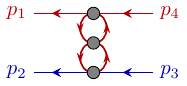}
    \caption{}
    \label{subfig:Cut_2L_F2}
\end{subfigure} \hskip .6 cm 
\begin{subfigure}[b]{0.2\textwidth}
\centering
    \includegraphicslog{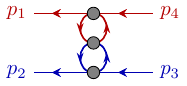}
    \caption{}
    \label{subfig:Cut_2L_F3}
\end{subfigure}
    \caption{The 14 representative unitarity cuts for scalar scattering in QED at two loops (in the $N_s=0$ sector where closed internal scalar loops are consistently dropped). The full set of 36 cuts is obtained by relabeling the external legs. The exposed lines in the diagrams are all cut.}
    \label{fig:CutsSQED_2L}
\end{figure}
%

\noindent
\paragraph{Two-Loops\\}
%
To go beyond the one-loop example considered above and illustrate features not present in that simple situation, we discuss here the construction of the four-scalar integrand in QED. The topology of the spanning cuts is shown in Fig.~\ref{TwoLoopFourPtSpanCutFigure}. As at one loop, we must include all distinct labelings of external lines and all possible assignment of states to the internal exposed lines.

Excluding contributions from closed scalar loops, there are 36 (flavor-dressed) cuts that are needed in order to fix the integrand at two-loop order. These cuts can be obtained by relabeling of a smaller set of 14 representative cuts displayed in Figure~\ref{fig:CutsSQED_2L}. Each of the cuts is straightforward to compute from tree amplitudes. The resulting expressions for the cuts are detailed in the ancillary file \texttt{SQED\textunderscore SSSS\textunderscore 2L\textunderscore Cuts.m} in the global two-loop four-point integrand basis. 

To avoid misidentifying the coefficients of basis elements at two loops (and above), it is critical to track the position of the cut legs. While they can sometimes be identified a posteriori by considering the number of ways a basis element can be cut, or by adding appropriate symmetry factors, it is perhaps more systematic to carry this out from the beginning, by starting from the construction of the diagrammatic form of the spanning cuts discussed in Sec.~\ref{subsec:unitarity_mapping}. For example, considering the field configuration corresponding to Fig.~\ref{subfig:Cut_2L_B}, we have
\begin{equation}
\label{eq:2loop_sqed_eg_cut}
\vcenter{\hbox{\includegraphicslog{figs/Cut_2L_B.pdf}}}=-8s^3Q_1^3Q_2^3 \left(I^{(1)}_{1,\cutindex{1},\cutindex{1},1,1,1,\cutindex{1},0,0}+I^{(1)}_{\cutindex{1},1,1,\cutindex{1},1,1,\cutindex{1},0,0}\right)+\dots
\end{equation}
In this expression, the ellipsis denotes terms that are irrelevant to the present discussion. The scalar ladder integral is
\begin{equation}
\hskip -.5cm
I^{(1)}_{1,1,1,1,1,1,1,0,0}=\!\int\!\!\frac{\mathrm{d}^D\ell_1}{(2\pi)^D}\frac{\mathrm{d}^D\ell_3}{(2\pi)^D}\frac{1}{\ell_1^2(\ell_1{+}p_{12})^2\ell_3^2(\ell_3{+}p_{12})^2(\ell_1{+}p_1)^2(\ell_3{+}p_{123})^2(\ell_1{-}\ell_3)^2}\,,
\hskip -.4cm
\end{equation}
and, as before, a hat denotes the replacement $\frac{1}{x}\to 2\pi\delta(x)$. 
Graphically the two cuts of the ladder in \eqref{eq:2loop_sqed_eg_cut} are 
\begin{align}
    I^{(1)}_{a_1,\cutindex{a_2},\cutindex{a_3},a_4,a_5,a_6,\cutindex{a_7},a_8,a_9}={}&\vcenter{\hbox{\includegraphicslog[scale=0.75]{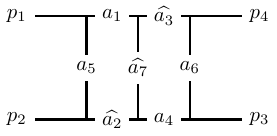}}}\,,\\
     I^{(1)}_{\cutindex{a_1},a_2,a_3,\cutindex{a_4},a_5,a_6,\cutindex{a_7},a_8,a_9}={}&\vcenter{\hbox{\includegraphicslog[scale=0.75]{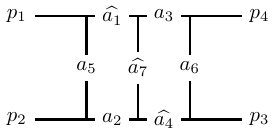}}}\,.
\end{align}
The planar double-box contribution to the full amplitude is 
\begin{equation}
\mathcal{A}_{\phi\phi\bar{\phi}\bar{\phi}}^{(2)}=-8s^3Q_1^3Q_2^3\, {I}^{(1)}_{1,1,1,1,1,1,1,0,0}+\dots \,.
\end{equation}
We see that, to determine the numeric coefficient correctly, we can either distinguish the two different cuts of the ladder diagram (as was done in Eq.~\eqref{eq:2loop_sqed_eg_cut}), or we could introduce a symmetry factor of 1/2. We note that the coefficient can also be uniquely fixed by the maximal cut
\begin{equation}
\vcenter{\hbox{\includegraphicslog{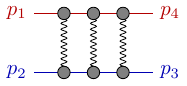}}}=-8s^3Q_1^3Q_2^3 \,{I}^{(1)}_{\cutindex{1},\cutindex{1},\cutindex{1},\cutindex{1},\cutindex{1},\cutindex{1},\cutindex{1},0,0}\,,
\end{equation}
which lifts the degeneracy. 

We obtain the full integrand by computing the 14 cuts displayed in Figure~\ref{fig:CutsSQED_2L}. We then carry out the sum over internal states by adding cuts that correspond 
to the same topology, explicitly the sets $\{1,2\},\{3\},\{4,5,6\},\{7,8,9,10\},\{11\}$ and $\{12,13,14\}$. We then compute the remaining cuts that are obtained by relabeling 
the external kinematics. 
As in the direct construction of the initial 14 cut contributions, when carrying out this relabeling and the subsequent map to the integrand basis it is important 
to track the cut legs and ensure and discard contributions in which a cut leg is mapped to a collapsed leg, as such terms vanish by definition. A test of these transformations 
is cut consistency as well as realization of symmetries of cuts.
Finally, we add up the expressions for the individual cuts, avoiding adding more than once the contributions from overlapping cuts. 
The resulting integrand agrees with the Feynman-diagram integrand expressed in the global basis computed in \sect{subsec:xiDrop}, also given in a computer-readable format in the ancillary file \texttt{SQED\textunderscore SSSS\textunderscore 2L.m}

While the two-loop integrand in scalar QED is simple, at sufficiently high loop orders, or in more nonlinear theories already at two loops, cut expressions can be quite involved. We note, however, that when combined with the non-planar integrand basis, the cut construction can avoid large expressions at intermediate steps by, e.g., focusing on a single basis element at a time.
Furthermore, the number of basic building blocks represented by the 14 cuts in Figure~\ref{fig:CutsSQED_2L} is smaller than the number of 31 representative Feynman diagrams---a scaling that is amplified at higher loop orders. Similar to traditional unitarity methods, since the building blocks are on-shell, they are substantially simpler, especially when compared to Feynman diagrams in a generic gauge.

%
\subsection{Example: Adler Zero of the NLSM at Loop Level}
\label{subsec:NLSM}
%

Recently, there has been a renewed interest in the study of the amplitudes of the NLSM, which, for a suitable symmetry group, describe the low-energy behavior of QCD. 
The NLSM is the typical example of a theory with a shift symmetry that leads to the famous Adler zero, i.e.,~the vanishing of amplitudes when one of the external momenta is taken to zero.  Furthermore, the NLSM is an integral part of a web of double-copy relations \cite{Bern:2019prr}, is amenable to CHY \cite{Cachazo:2014xea} and ambitwistor string methods \cite{Casali:2015vta}, and has appeared in the surfacehedron picture \cite{Arkani-Hamed:2023lbd, Arkani-Hamed:2023mvg,Arkani-Hamed:2023jry,Arkani-Hamed:2024nhp, Arkani-Hamed:2024yvu}. In Ref.~\cite{Bartsch:2022pyi, Bartsch:2024ofb}, the authors describe a prescription for extending the Adler zero to one-loop \emph{integrands} when phrased in the surfacehedron formalism. However, starting at two-loop level, this method had failed to produce an integrand with Adler zero property. In contrast to Ref.~\cite{Bartsch:2024ofb}, we find an integrand-level version of the Adler zero. The essential difference between our construction and the analysis performed in Ref.~\cite{Bartsch:2024ofb} is that we self-consistently drop all tadpole and massless-bubble-on-external-leg contributions throughout. 

We have explicitly constructed\footnote{We thank Clifford Cheung and Jaroslav Trnka for raising the question about the fate of the integrand level Adler zero in our framework.} the one and two-loop four-point integrands of the non-linear sigma model from unitarity cuts (applying the same restrictions as Ref.~\cite{Carrasco:2023qgz}). Importantly, as in our general philosophy, we consistently drop tadpole and massless bubble on external leg topologies because they integrate to zero in dimensional regularization. 
The topologies of the diagrams in the one- and two-loop spanning cuts in general theories are shown in Figs.~\ref{OneLoopFourPtBubCutFigure} and \ref{TwoLoopFourPtSpanCutFigure}. The absence of five-point (and more generally odd-point) amplitudes in NLSM implies that the sunrise cut vanishes identically so only two of the three two-loop spanning cuts are relevant. 
Thus, we only need four-point tree amplitudes to construct the one- and two-loop four-point integrands. The NLSM Lagrangian is,
\begin{align}
 \mathcal{L} = \frac{f_\pi}{4} \mathrm{tr}\Big[ (\partial_\mu U)^\dagger(\partial^\mu U)\Big]\,,
\end{align}
where $U(x) = \exp({\imath \, {\pi(x)}/{f_\pi}})$ is the nonlinear sigma model field, $f_\pi$ is the pion decay constant, $\pi(x) = t^a \pi^a(x)$ depends on the symmetry-group generators $t^a$ in the fundamental representation that satisfy $[t^a,t^b]= 2 \imath f^{abc}\,t^c$ and $\mathrm{tr}[t^a t^b] = 2 \delta^{ab}$. The four-point tree is (see, e.g.,~Eq.~(4.3) of Ref.~\cite{Carrasco:2023qgz}.)
\begin{align}
\label{eq:NLSM_4pt_tree}
\begin{split}
\mathcal{A}^{{\rm tree}}_{\pi \pi\pi \pi} = 
\vcenter{\hbox{\includegraphicslog[scale=1.17]{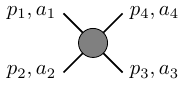}}}
=
\frac{1}{6 f_\pi} 
\Big[
 \phantom{+} 
  &(p_1-p_2)\cdot(p_3-p_4) f^{a_1a_2 b}f^{a_3a_4 b}\\[-15pt]
 +&(p_1-p_3)\cdot(p_2-p_4) f^{a_1a_3 b}f^{a_2a_4 b}\\
 +&(p_1-p_4)\cdot(p_3-p_2) f^{a_1a_4 b}f^{a_3a_2 b}
\Big]\,,
\end{split}
\end{align}
where the color\footnote{We refer to the group theory factors as `color' despite them not being gauged, to maintain the analogy with gauge theory.} factors are not all independent but satisfy the Jacobi relation
\begin{align}
\label{eq:NLSM_jacobi_eg}
f^{a_1a_2 b}f^{a_3a_4 b}=  f^{a_1a_3 b}f^{a_2a_4 b} + f^{a_1a_4 b}f^{a_3a_2 b}\,.
\end{align}
Since the tree amplitude in Eq.~(\ref{eq:NLSM_4pt_tree}) is local and contains no additional propagators, the cut-merging procedure is even more 
straightforward than in the scalar QED examples discussed in Sec.~\ref{subsec:SQED_cut_construction}. We work with color-dressed amplitudes and cuts, and eliminate redundant color structure by repeated use of the Jacobi relation \Eq{eq:NLSM_jacobi_eg} throughout. At one-loop, we start from the three inequivalent permutations of the bubble cut shown in \fig{OneLoopFourPtBubCutFigure}. At two loops, the set of spanning cuts is given by the six inequivalent external leg permutations of \fig{subfig:Cut_Gen_2L_2} and the three inequivalent permutations of \fig{subfig:Cut_Gen_2L_3}; see also Eq.~(4.28) of Ref.~\cite{Carrasco:2023qgz}. There are no overlapping diagrams between different elements of the spanning set, so we directly add the cuts to obtain the one and two-loop four-point integrands. 
In the cases considered, the Adler zero of the cuts directly extends to the Adler zero of the integrand
\begin{align}
 \mathcal{A}^{(1)}_{\pi \pi\pi \pi}\,\Big|_{p_i\to0} = \mathcal{A}^{(2)}_{\pi \pi\pi \pi}\,\Big|_{p_i\to0} = 0\,.
\end{align}
More generally, independent of the loop order or the number of external states, all spanning unitarity cuts are sums of products of tree-level amplitudes. Based on this observation, in the soft limit where one of the momenta of an external state vanishes, $p_i\to0$, one tree-level amplitude in each spanning cut always vanishes, since the Adler zero condition holds at tree-level. Subsequently, generalized unitarity enforces that the entire loop integrand has to match `$0$' in the soft limit. Therefore, cut consistency in the global integrand basis requires that the loop integrand vanishes in the soft limit. 

To conclude that the integrand-level zero readily transfers to a zero of the integrated result, one has to additionally show that other loop-momentum regions (such as collinear regions where loop momenta become collinear to massless external lines) \cite{Beneke:1997zp} do not contribute.

Finally, we remark that our approach is insensitive to the choice of field parametrization. In fact, since the tree-level amplitudes are uniquely fixed through the soft bootstrap~\cite{Cheung:2015ota}, we can derive (unique) loop-level NLSM integrands without reference to a field parametrization. Similar results hold for the Dirac-Born-Infeld theory and the Galileon model.

%
\section{The Double Copy to All Loop Orders}
\label{sec:double_copy}
%

As reviewed in the introduction, color-kinematics duality and the associated double-copy construction~\cite{Bern:2008qj, Bern:2010yg, Bern:2019prr} have been the main tools for constructing higher-loop integrands in gravitational theories. The direct application of the double copy requires the availability of color-kinematics duality satisfying gauge-theory amplitudes at the desired loop order~\cite{Bern:2010yg, Bern:2012uf}.  As the loop order, or the number of external legs increases, it becomes more difficult to find explicit forms of gauge-theory integrands that have this property~\cite{Bern:2015ooa, Mogull:2015adi, Bern:2012uc}. Moreover, since the spectra of double-copy theories are the tensor product of the two theories entering the construction, obtaining a desired spectrum may require that some states be removed either via a projection or by some other methods~\cite{Johansson:2014zca}. For example, to obtain general relativity via double copy it is necessary to project out the dilaton and antisymmetric tensor generated in the product of two vectors.

In contrast, tree-level amplitudes that manifest color-kinematics duality are readily available in diagrammatic form, and therefore, double-copy tree-level amplitudes with essentially any number of external legs can straightforwardly be constructed. 
In this section, we explain how to leverage these tree-level amplitudes and the streamlined cut-merging procedure based on a global integrand basis, which we discussed and illustrated in previous sections, to reorganize the double-copy construction such that it can be applied to all loop orders while seamlessly incorporating a projection onto the desired double-copy states using generalized Ward identities~\cite{Kosmopoulos:2020pcd}. Indeed, this is how integrands in Einstein's GR have been built for use in gravitational-wave physics~\cite{Bern:2019nnu, Bern:2019crd, Bern:2021dqo, Bern:2021yeh}. The novel ingredient here is the improved cut merging based on a global integrand basis.

\subsection{Basics of Color-Kinematics Duality and Double Copy}
\label{cktrees}

We first review salient aspects of color-kinematics duality, and double copy~\cite{Kawai:1985xq,  Bern:2008qj, Bern:2010yg, Bern:2019prr} that connect to earlier sections and will be relevant to establish how double copy can be applied to all loop orders.
For this purpose it suffices to discuss gauge theories with massless matter in the adjoint representation of the gauge group. For theories with matter in other representations, we refer the reader to reviews~\cite{Bern:2019prr, Adamo:2022dcm} and to the original literature~\cite{Chiodaroli:2013upa, Johansson:2014zca, Johansson:2015oia}. 

Consider the color-dressed tree-level amplitudes of such a gauge theory. As discussed more generally in \sect{subsec:unitarity_mapping}, by suitably multiplying and dividing by propagators, the amplitudes can be generically reorganized as a sum over diagrams with only three-point vertices,
\begin{equation}
\imath {\cal A}^\tree_m(1,2,3,\ldots,m)\,= g^{m-2} \sum_{i}
                \frac{n_i c_i }{\prod_{\alpha_i} p^2_{\alpha_i}}\,,
\label{Anrep}
\end{equation}
where the summation label $i$ runs over $(2m-5)!!$ diagrams with only cubic vertices. The $c_i$ are color factors obtained by attaching a structure constant $\f^{abc} = \imath\sqrt{2} f^{abc}$ (the $f^{abc}$ conventions are defined in \sect{subsec:NLSM}) to each vertex of the diagram and $\delta^{ab}$ to each internal line, $n_i$ are kinematic numerators depending on external and loop momenta as well as polarizations and spinors, and the $p^2_{\alpha_i}$ in the denominator are the inverse propagators for the $(m-3)$ internal lines $\alpha_i$ of diagram $i$. The gauge-theory coupling constant is $g$.

\begin{figure}[tb]
\centering
\begin{subfigure}[b]{0.3\textwidth}
\centering
\renewcommand\thesubfigure{(i)}
    $\vcenter{\hbox{\includegraphicslog[]{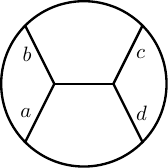}}}$
    \caption{}
    \label{subfig:4ptGluon_s_blob}
\end{subfigure}
\begin{subfigure}[b]{0.3\textwidth}
\centering
    \renewcommand\thesubfigure{(j)}
    $\vcenter{\hbox{\includegraphicslog[]{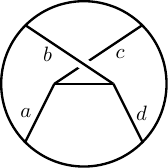}}}$
    \caption{}
    \label{subfig:4ptGluon_u_blob}
\end{subfigure}
\begin{subfigure}[b]{0.3\textwidth}
\centering
    \renewcommand\thesubfigure{(k)}
    $\vcenter{\hbox{\includegraphicslog[]{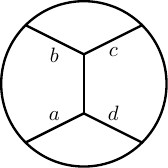}}}$
    \caption{}
    \label{subfig:4ptGluon_t_blob}
\end{subfigure}

\caption{
\label{GeneralJacobiFigure}
The color factors of the three indicated diagrams form a color Jacobi identity, see \Eq{jacobic}. The circle surrounding the exposed lines represents an embedding in a larger diagram. In a color-dual representation, the kinematic numerators of the diagrams satisfy the same relation as the color factors, see \Eq{jacobin}. 
}
\renewcommand\thesubfigure{(\arabic{subfigure})}
\end{figure}

The two fundamental properties of the $\f^{abc}$ structure constants are their complete antisymmetry and the Jacobi relation, 
\begin{align}
    \f^{abe} \f^{ecd} - \f^{ace} \f^{ebd} - \f^{aed} \f^{ebc}=0 \,,
\label{ColorJacobi}
\end{align}
corresponding to the color factors of the three diagrams in Fig.~\ref{GeneralJacobiFigure}. For every internal line of every diagram, there is one Jacobi relation connecting it with two others.  If these three diagrams are embedded in a larger diagram with the remaining parts identical, the Jacobi identity continues to hold, as it is independent of kinematics.  Labeling the three diagrams appearing in such 
a Jacobi triplet by $i,j,k$,  Eq.~\eqref{ColorJacobi} translates to 
\begin{equation}
    c_i - c_j - c_k = 0\, .
\label{jacobic}
\end{equation}
where the $c_{i}$, $c_{j}$ and  $c_{k}$ are the color factors of the larger diagrams.

The color Jacobi relations allow us to modify the numerator factors $n_i$ without modifying the amplitude. The amplitude is said to obey the duality between color and kinematics if the kinematic numerators $n_i$ are arranged to obey algebraic relations in one-to-one correspondence with those obeyed by the color factors $c_i$.
Thus, we require that they are antisymmetric and that they obey the same Jacobi identities as the color factors~\footnote{In more general situations the duality between color and kinematics only requires that the kinematic numerators obey the same algebraic relations as the relations between color factors that are required by the on-shell Ward identities~\cite{Chiodaroli:2015rdg, Bern:2019prr}.}:
\begin{align}
c_i \rightarrow -c_i \quad \Rightarrow \quad  n_i \rightarrow -n_i\,,  
\hskip 1.5 cm 
c_i =  c_j + c_k  \quad \Rightarrow \quad n_i =  n_j + n_k \, .
\label{jacobin}
\end{align}
As in \fig{GeneralJacobiFigure}, the kinematic Jacobi identity requires that lines common to the three participating graphs be labeled identically. This implies that as one lists all the kinematic Jacobi identities, the graphs need to be constantly relabeled. It is also important that no relations be imposed that are valid only for particular gauge groups or representations. 

The duality between color and kinematics has been proven to hold at tree level in gauge theories~\cite{Cachazo:2012uq, Chen:2011jxa, Feng:2010my, Stieberger:2009hq, Bjerrum-Bohr:2009ulz, Bern:2010yg}, and amplitudes manifesting it are available for all multiplicities~\cite{Edison:2020ehu}.

Perhaps the most remarkable aspect of amplitudes manifesting color-kinematics duality is that an amplitude of a gravitational theory can be constructed by replacing the color factors of one gauge-theory amplitude with the color-dual kinematic numerators of another:
\begin{equation}
\label{cTon}
c_i\rightarrow n_i \, .
\end{equation}
To be specific and focusing on tree-level amplitudes, given a tree-level $m$-point gauge-theory amplitude in Eq.~\eqref{Anrep} together with a second $m$-point amplitude in another gauge theory, possibly with a different field content or gauge group, then the corresponding gravity amplitude is
\begin{eqnarray}
&& \imath {\cal M}^\tree_m(1,2,\ldots,m)\,=
 \Big(\frac{\kappa}{2}\Big)^{m-2}  \sum_{i} \frac{n_i \n_i}{ \prod_{\alpha_i} p^2_{\alpha_i}} \, ,
\label{squaring}
\end{eqnarray}
where the sum runs over the same set of $(2m-5)!!$ diagrams as in Eq.~\eqref{Anrep}, and the $\n_i$ are the kinematic numerators of the second gauge theory. The gravitational coupling is related to Newton's constant via $\kappa^2 = 32 \pi G$.

Requiring that a loop-level gauge-theory amplitude obeys the duality is more involved.  The obvious generalization of Eq.~\eqref{Anrep} to loop level is
\begin{equation}
\imath {\cal A}^{(L)}_{m} = \imath^{L} g^{m-2+2L} \sum_{i}\,
 \int \Bigl(\prod_{j=1}^L \frac{\mathrm{d}^{D}\ell_j}{(2\pi)^{D}} \Bigr) \frac{1}{S_i} \frac{n_i\, c_{i}}{\prod_{\alpha_i} p^2_{\alpha_i}} \, ,
 \label{gaugeAmpLoops}
 \end{equation}
where the sum runs over distinct $L$-loop $m$-point diagrams with only cubic vertices and the $S_i$ are internal symmetry factors associated with the diagram. As at tree level, writing a gauge-theory amplitude in this form is straightforward and amounts to multiplying and dividing by suitable inverse propagators. 
Imposing the Jacobi identities in Eq.~\eqref{jacobic} requires that for each triplet of diagrams, we relabel the loop momenta so that all but one propagator in each diagram is identical to the other two diagrams.  Such a choice is implicit in the three diagrams in \fig{GeneralJacobiFigure}.  Whenever a color-dual form of a gauge-theory amplitude can be found, then the double copy follows immediately, giving a gravity amplitude~\cite{Bern:2010ue},
\begin{equation}
\imath {\cal M}^{(L)}_{m} 
  =  \imath^{L}\;\!\Big(\frac{\kappa}{2}\Big)^{m-2+2L} \sum_{i}\, \int \Bigl(\prod_{j=1}^L \frac{\mathrm{d}^{D}\ell_j}{(2\pi)^{D}} \Bigr) 
  \frac{1}{S_i} \frac{n_i \tilde{n}_i}{\prod_{\alpha_i} p^2_{\alpha_i}} \,.
\label{DCformulaLoops}
\end{equation}
While a large number of examples are available, see, e.g.,~Refs.~~\cite{Bern:2012uf, Boels:2012ew, Carrasco:2012ca, Bjerrum-Bohr:2013iza, Bern:2013yya, Chiodaroli:2013upa, Chiodaroli:2014xia, Badger:2015lda, Mafra:2015mja,  He:2015wgf, Mogull:2015adi, Chiodaroli:2015rdg, Yang:2016ear, Chiodaroli:2017ngp, Chiodaroli:2017ehv, He:2017spx, Johansson:2017bfl, Chiodaroli:2018dbu, Carrasco:2020ywq, Lin:2021kht, Li:2022tir, Edison:2022smn, Edison:2022jln}), a proof that all loop amplitudes in gauge theories can be organized to manifest the duality is lacking.

In fact, it can sometimes be rather difficult to find explicit forms that realize the duality~\eqref{jacobin}, making it challenging to obtain the double-copy amplitudes.   Indeed, no color-dual form of the five-loop four-point amplitude for $\mathcal N = 4$ super-Yang-Mills theory has been found~\cite{Bern:2012uc, Bern:2017ucb}. Even at two loops for pure Yang-Mills theory or at one loop and eight external particles, the task is not straightforward~\cite{Bern:2015ooa, Mogull:2015adi, Edison:2022jln}.

There are two approaches to bypassing the difficulty of finding gauge-theory integrands where the duality holds. In the first~\cite{Bern:2017yxu, Bern:2017ucb}, a naive double copy, constructed from a gauge-theory amplitudes not obeying color-kinematics duality, is corrected order by order in the number of collapsed propagators such that the unitarity cuts of the corrected expression reproduce those from a spanning set of cuts. The correction formulae depend on the topology of the cut and the violation of the kinematic Jacobi relations by the naive double-copy expression.
This procedure has been successfully employed to obtain the five-loop four-point integrand of $\NeqEight$ supergravity and for its UV properties to be determined~\cite{Bern:2018jmv}. However, this procedure does not scale well with the number of loops and legs, making it much more difficult to apply in even more complicated cases.

Here we use the observation made in Ref.~\cite{Bern:2015ooa} that because the duality holds at the tree level, it also holds on (any spanning set of) generalized cuts built from tree amplitudes. This perspective was used, in particular, to construct the two- and three-loop amplitudes used for extracting the conservative contributions to black hole scattering~\cite{Bern:2019nnu, Bern:2019crd, Bern:2021dqo, Bern:2021yeh}. Combining it with the global integrand basis discussed in previous sections and with projectors selecting the desired propagating states, we will see that complete integrands in gravitational theories with specific spectra at any loop fixed order can be read off.

\subsection{Double Copy of Cuts and Mapping to the Global Integrand Basis}
\label{subsec:DoubleCopy}

\begin{figure}
\centering
\begin{subfigure}[b]{0.3\textwidth}
\centering
    \includegraphicslog{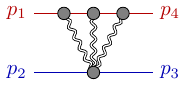}
    \caption{}
    \label{subfig:Cut_2L_W}
\end{subfigure}
\begin{subfigure}[b]{0.3\textwidth}
\centering
    \includegraphicslog{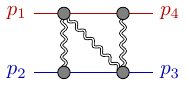}
    \caption{}
    \label{subfig:Cut_2L_N}
\end{subfigure}
\begin{subfigure}[b]{0.3\textwidth}
\centering
    \includegraphicslog{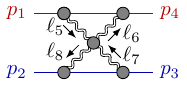}
    \caption{}
    \label{subfig:Cut_2L_X}
\end{subfigure}
    \caption{The three distinct unitarity cuts to obtain the potential region classical scattering amplitude at 2-loops for massive scalars interacting via gravitons. The complete set of unitarity cuts is given by including the additional independent relabelings of the displayed cuts.  
    \label{fig:CutsGR_2L}
    }
\end{figure}

Following the review of the tree-level double copy in the previous section, it is hardly necessary to belabor the double copy of generalized cuts built from such tree amplitudes. We will instead discuss an example and focus on the removal of double-copy states, such as the dilaton and the antisymmetric tensor, which are unwanted in,  e.g., Einstein's gravity. Removing these undesired states amounts to sewing the generalized cuts with graviton physical-state projectors~\eqref{StateSumGravity}, as described in Ref.~\cite{Bern:2019crd, Kosmopoulos:2020pcd}.
The result is a spanning set of cuts in diagrammatic form. Each cut can be written in the global integrand basis and subsequently merged into an integrand by simply reading off the coefficients of the basis elements according to our more general discussion in \sect{sec:unitarity}.

To this end, consider the unitarity cuts in \fig{fig:CutsGR_2L} for the two-loop four-scalar amplitude with two distinct scalar fields used in Refs.~\cite{Bern:2019nnu, Bern:2019crd} to extract the classical two-body potential of two compact astrophysical objects in general relativity. These cuts and their distinct relabeling of external legs are the spanning cuts that determine the classical interaction potential of two scalar particles, dropping terms that contribute only to the quantum amplitude.
The cuts in \fig{subfig:Cut_2L_X}, which have been analyzed in some detail in Ref.~\cite{Bern:2019crd}, are representative for our current purpose, so we will focus on it. Its evaluation requires that we sew together four two-scalar one-graviton tree amplitudes and one four-graviton tree amplitude. 

\begin{figure}
\centering
\begin{subfigure}[b]{0.3\textwidth}
\centering
    \renewcommand\thesubfigure{(s)}
    \includegraphicslog{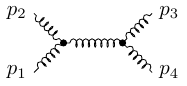}
    \caption{}
    \label{subfig:4ptGluon_s}
\end{subfigure}
\begin{subfigure}[b]{0.3\textwidth}
\centering
    \renewcommand\thesubfigure{(u)}
    \includegraphicslog{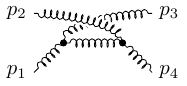}
    \caption{}
    \label{subfig:4ptGluon_u}
\end{subfigure}
\begin{subfigure}[b]{0.3\textwidth}
\centering
    \includegraphicslog{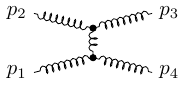}
    \renewcommand\thesubfigure{(t)}
    \caption{}
    \label{subfig:4ptGluon_t}
\end{subfigure}
 \caption{
 \label{fig:FourGluon}
 The $s$, $t$ and $u$-channel diagrams defining the four-gluon amplitude in \eqn{GaugeFourAmplitude}.
 }
\renewcommand\thesubfigure{(\arabic{subfigure})}
\end{figure}

Because the double copy holds for tree amplitudes, instead of starting with gravity amplitudes, we start with gauge-theory amplitudes.  
The three-point  amplitude with one gluon and two massive scalars is given by
\begin{equation}
\imath {\cal A}_3^\tree(1^s,2,3^s) = \frac{1}{\sqrt{2}}\, g\, \f^{a_1 a_2 a_3} (p_1 - p_3)\cdot\pol_2 \, ,
\label{YMThreeAmplitude}
\end{equation}
and a convenient organization of the four-gluon amplitude is 
\begin{equation}
\imath {\cal A}^\tree_4(1,2,3,4) = g \biggl[\frac{ n_s c_s}{s}+ \frac{n_t c_t}{t} + \frac{n_u c_u}{u} \biggr]\,,
\label{GaugeFourAmplitude}
\end{equation}
where the three terms correspond to the $s$, $t$ and $u$ channel diagrams as drawn in~\fig{fig:FourGluon}, respectively. 
As reviewed in the previous section, the color factors $c_x$ are directly read off from the diagrams, with a factor of $\f^{abc}$ 
at each vertex, and satisfy the color Jacobi identity $c_t = c_s - c_u$. With this normalization, the $s$-channel kinematic numerator, $n_s$, is
\begin{align}
 n_s &= n(1,2,3,4) \nn \\
 & = -\frac{1}{2}\Bigl\{ \Big[(\varepsilon_1 \cdot \varepsilon_2) p_1^\mu+2
(\varepsilon_1 \cdot p_2) \varepsilon_2^\mu-(1 \leftrightarrow 2)\Big]
\Big[ (\varepsilon_3 \cdot \varepsilon_4) p_{3\mu}+2(\varepsilon_3 \cdot p_4)
\varepsilon_{4\mu}-(3 \leftrightarrow 4)\Big] \nn  \\
& \null \hskip 2 cm
+ s \Bigl[ (\varepsilon_1 \cdot \varepsilon_3)(\varepsilon_2 \cdot \varepsilon_4)
- (\varepsilon_1 \cdot \varepsilon_4)(\varepsilon_2 \cdot \varepsilon_3)\Bigr] \Bigr\}\,,
\label{s_num}
\end{align}
where the momenta and polarization vectors satisfy on-shell conditions $p_i^2 = \varepsilon_i \cdot p_i=0$. The $u$- and $t$-channel numerators are given by relabeling $2\leftrightarrow 3$ and the kinematic Jacobi relations,
\begin{align}
n_u =  n(1,3,2,4)\,, \hskip 1cm 
n_t =  n(1,2,3,4) - n(1,3,2,4) \, ,
\end{align}
respectively. The corresponding gravity tree amplitudes are related to these via the double copy. The three-point amplitude, up to overall normalization, is simply the square of the color-stripped three-point amplitude in \eqn{YMThreeAmplitude}, 
\begin{equation}
\imath \mathcal M^\tree_3(1,2,3) = \frac{\kappa}{4} \bigl[(p_1 - p_3) \cdot \pol_2\bigr]^2 \ ,
\end{equation}
while the four-point amplitude is, 
\begin{equation}
\imath\mathcal M^\tree_4(1,2,3,4) = \Bigl( \frac{\kappa}{2} \Bigr)^2\biggl[\frac{n_s^2}{s}+ \frac{n_t^2}{t} + \frac{n_u^2}{u}\biggr] \,,
\label{GravitFourAmplitude}
\end{equation}
with the same numerator factors, $n_x$, as for gauge theory.

The generalized cut in Fig.~\ref{subfig:Cut_2L_X} is given by a product of tree amplitudes summed over the states of the exposed intermediate lines,
\begin{align}
 C_{\rm GR}^{\ref{subfig:Cut_2L_X}} = & \sum_{\text{graviton states}}
{\mathcal M}_3^\tree(p_1, \ell_5, -\ell_5 - p_1) \,
{\mathcal M}_3^\tree(p_2, -\ell_8, \ell_8 - p_2) \,
{\mathcal M}_3^\tree(p_3, \ell_7, -\ell_7 - p_3) \, \nn \\
& \hskip 1 cm \times
{\mathcal M}_3^\tree(p_4,  -\ell_6, \ell_6 - p_4) \,
{\mathcal M}_4^\tree(-\ell_5, \ell_8, -\ell_7, \ell_6) \, ,
\end{align}
and the sum over graviton states is realized in terms of sums over the single-copy gluon states.
For each color, gluons have $D-2$ states in $D$ dimensions, so the double copy naturally yields $(D-2)^2 = \frac{1}{2}D(D-3) + \frac{1}{2}(D-2)(D-3) + 1$ states, corresponding to a graviton, an antisymmetric tensor, and a dilaton. To obtain Einstein's gravity, we need to remove the antisymmetric tensor and dilaton from the cuts, which is accomplished by using the graviton physical-state projector~\eqref{StateSumGravity}. 
As discussed in \sect{subsec:unitarity_mapping}, if we arrange the tree amplitudes to satisfy generalized Ward identities~\cite{Kosmopoulos:2020pcd}, the much simpler de Donder gauge projector, with no reference momenta, can be used (see discussion below~\eqn{StateSumGravity}). In fact, the three cuts in  \fig{fig:CutsGR_2L} are even a bit simpler because the antisymmetric tensor field does not appear in loops.  This is because the antisymmetric tensor does not directly couple to scalars, and all massless lines terminate on a massive scalar line; there is no need to symmetrize the indices on the graviton physical state projector. 
A further subtlety stems from the fact that amplitudes can be organized to satisfy the generalized Ward identity on all but two external lines. To use this simplified sewing procedure for all four legs of the four-graviton amplitude, it is necessary to carry it out in two stages. We first make legs $\ell_7$ and $\ell_8$ obey the identity and sum over the states with these momenta. Subsequently, we make the cut one-loop amplitude with external legs $p_2, p_3, \ell_5$ and $\ell_6$ obey the generalized Ward identity for legs $\ell_5$ and $\ell_6$ and carry out the remaining sum over states.

With these clarifications of the sewing procedure described in \sect{sec:unitarity}, the cut in \fig{subfig:Cut_2L_X} evaluates to a surprisingly compact expression (see Ref.~\cite{Bern:2019crd} for details),
\begin{align}
C_{\rm GR}^{\ref{subfig:Cut_2L_X}}  = &
  -2\imath \biggl[  t^2 m_1^4 m_2^4 +
  \frac{1}{t^6} \Bigl(
  {\cal E}_1^4 + {\cal O}_1^4 + 6 {\cal O}_1^2 {\cal E}_1^2 +
  {\cal E}_2^4 + {\cal O}_2^4 + 6 {\cal O}_2^2 {\cal E}_2^2 \nonumber \\
& \hskip 3.3 cm \null 
   -2 t G_5  ( {\cal E}_1^2 + {\cal O}_1^2  +  {\cal E}_2^2 + {\cal O}_2^2)
   + t^2 G_5^2 \Bigr) \nonumber \\
& \qquad
   - \frac{1}{t}  ( 2 m_1^2 m_2^2 + (\t_{1 2} - \t_{1 8} + \t_{2 5} - \t_{5 8})^2 ) \, G_5
 \biggr]  \biggl(\frac{1}{(-\t_{58})} + \frac{1}{\t_{68}} \biggl) \,,
 \label{eq:CutH}
\end{align}
where we have taken $ \eta_\mu^{\ \mu} = 4$ for simplicity, and  
\begin{align}
{\cal E}_1^2 & =
\frac{1}{4}  t^2 \bigl(\t_{1 2} \t_{5 8} - \t_{1 8}\t_{2 5} \bigr)^2 \,,
\qquad
{\cal O}_1^2 = {\cal E}_1^2 - \t_{5 8}^2 m_1^2  m_2^2 \, t^2  \,,
\label{EvenOddOne}
\\[3pt]
{\cal E}_2^2 & =
\frac{1}{4} t^2 \bigl( \t_{24} \t_{5 7} + \t_{1 7} \t_{57} \bigr)^2 \,,
\qquad
{\cal O}_2^2 = {\cal E}_2^2 - \t_{5 7}^2 m_1^2 m_2^2 \, t^2 \,,
\label{EvenOddTwo}
\end{align}
where the momentum labels can be read off from \fig{subfig:Cut_2L_X}, $t = (p_2 + p_3)^2$, $\t_{i j} = 2 p_i \cdot \ell_j$, and $G_5$ is the $5 \times 5$ Gram determinant $16 \det(a \cdot b)$ (where $a, b \in \{p_1,p_2,p_3,\ell_5,\ell_8\}$), which vanishes in four dimensions but does not vanish when using dimensional regularization because of either infrared or ultraviolet singularities.

\begin{figure}
\centering
\begin{subfigure}[b]{0.3\textwidth}
\centering
    \includegraphicslog{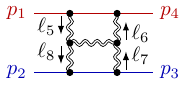}
    \caption{}
    \label{subfig:Cut_2L_X_Labelled_Blowup1}
\end{subfigure}
\begin{subfigure}[b]{0.3\textwidth}
\centering
    \includegraphicslog{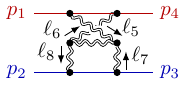}
    \caption{}
    \label{subfig:Cut_2L_X_Labelled_Blowup2}
\end{subfigure}
\begin{subfigure}[b]{0.3\textwidth}
\centering
    \includegraphicslog{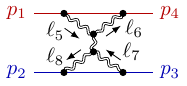}
    \caption{}
    \label{subfig:Cut_2L_X_Labelled_Blowup3}
\end{subfigure}
\caption{The three top-level graphs that appear in the cut in \fig{subfig:Cut_2L_X}.  
\label{fig:CutHParentDiags}
}
\end{figure}

With the gravity cuts in hand, the remaining steps are as described in the previous section.  For our example, the cut in  \fig{subfig:Cut_2L_X} given in \eqn{eq:CutH} can be written in terms of the top-level integrands corresponding to the graphs in \fig{fig:CutHParentDiags} and their associated daughter integrands obtained by collapsing the horizontal graviton propagator. 
The terms containing $1/(-\tau_{58})$ in the cut \eqref{eq:CutH} are assigned to the first diagram in \fig{fig:CutHParentDiags}, while the $1/\tau_{68}$ terms are assigned to the second diagram in that figure.
Terms with neither of these denominators are local with respect to loop momenta and thus are assigned to the common daughter representative graph of the two diagrams in \fig{fig:CutHParentDiags}, in which the horizontal graviton propagator is collapsed. 

The resulting expression for the cut is complete. We may either expand it in the soft-graviton limit, as discussed in \sect{subsec:integrand_cleanup}, if we are interested in extracting its classical content, or we may keep it complete if we are interested in quantum mechanical results. Either way, with expressions for all the cuts in the spanning set obtained through the double copy, the construction of complete integrands in the global basis follows the discussion in \sect{sec:unitarity}.

\begin{figure}
\centering
\begin{subfigure}[b]{0.22\textwidth}
\centering
    \includegraphicslog{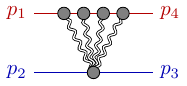}
    \caption{}
    \label{subfig:Cut_3L_A}
\end{subfigure}
\begin{subfigure}[b]{0.22\textwidth}
\centering
    \includegraphicslog{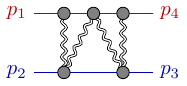}
    \caption{}
    \label{subfig:Cut_3L_B}
\end{subfigure}
\begin{subfigure}[b]{0.22\textwidth}
\centering
    \includegraphicslog{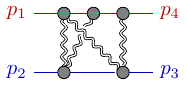}
    \caption{}
    \label{subfig:Cut_3L_C}
\end{subfigure}
\begin{subfigure}[b]{0.22\textwidth}
\centering
    \includegraphicslog{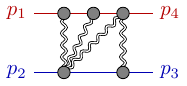}
    \caption{}
    \label{subfig:Cut_3L_D}
\end{subfigure}
\begin{subfigure}[b]{0.22\textwidth}
\centering
    \includegraphicslog{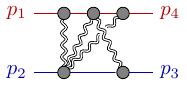}
    \caption{}
    \label{subfig:Cut_3L_E}
\end{subfigure}
\begin{subfigure}[b]{0.22\textwidth}
\centering
    \includegraphicslog{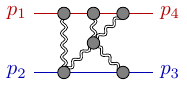}
    \caption{}
    \label{subfig:Cut_3L_F}
\end{subfigure}
\begin{subfigure}[b]{0.22\textwidth}
\centering
    \includegraphicslog{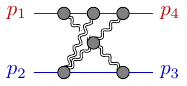}
    \caption{}
    \label{subfig:Cut_3L_G}
\end{subfigure}
\begin{subfigure}[b]{0.22\textwidth}
\centering
    \includegraphicslog{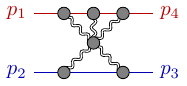}
    \caption{}
    \label{subfig:Cut_3L_H}
\end{subfigure}
   \caption{The eight distinct unitarity cuts needed to obtain the potential region classical scattering amplitude at three loops for massive scalars interacting via gravitons. The additional cuts are given by the independent relabelings of the external lines. }
    \label{fig:CutsGR_3L}
\end{figure}

While we chose the two-loop example above for simplicity, the same ideas extend without modifications to all loop orders. By carrying out the double-copy construction on cuts, we bypass the difficulties of finding color-kinematics-satisfying representations of gauge-theory amplitudes at higher loops and also have a detailed control on the spectrum of the double-copy theory.
We explicitly checked that the three-loop integrand for the potential-mode contributions to the four-scalar scattering obtained from the eight cuts in \fig{fig:CutsGR_3L} reproduces 
the integrand obtained in Ref.~\cite{Bern:2021dqo} constructed via the method of maximal cuts.  
The global integrand basis advocated here streamlines the cut-merging procedure and commutes with, e.g., the soft expansion, which thus can be carried out directly for the spanning cuts. We have verified this process in the classical limit through four-loop order in general relativity. 

%
\section{Conclusions and Outlook}
\label{sec:conclusions}
%

In this paper, we described a general construction of bases for planar and nonplanar scattering amplitude integrands to all loop orders and multiplicity. A key feature is that such bases are complete, not overcomplete, and unique once a choice---any suitable choice---is made for irreducible scalar products of external and loop momenta for top-level and daughter representative graphs. 
Writing loop integrands in a complete basis has a number of immediate important consequences. 
For example, all integrands become identical (assuming that no integration-by-parts relations are used in their construction). Thus, gauge parameters or spurious 
singularities that may be present in the integrand cancel manifestly.
For the same reason, the use of an integrand basis facilitates the verification of, for example, the on-shell Ward identities and the Adler zero at the integrand level for the NLSM.

One of the main consequences of using an integrand basis is that it greatly simplifies the construction of complete integrands via generalized unitarity, by merging generalized cuts. When written in a basis, terms in cuts equate directly with terms in off-shell integrands, up to combinatorial factors. 
Our organization of integrand numerators in terms of inverse propagators and ISPs trivializes the evaluation of cuts, which simply sets the appropriate inverse propagators to zero. 
It is thus natural to arrange the basis elements according to the 
number of collapsed propagators and determine their coefficients iteratively, using the method of maximal cuts~\cite{Bern:2007ct}: maximal cuts determine the coefficients of top-level integrals, next-to-maximal cuts determine the coefficients of integrals with one collapsed propagator, etc. This approach can be particularly helpful at high loop orders, where analyzing cuts in the minimal spanning set becomes increasingly complex.
We also outlined several approaches to integrands of amplitudes with external spinning particles, including treating external data as part of the ISPs and, thus, as part of the definition of the basis elements.

While the off-shell double copy constructs directly gravitational integrands from gauge-theory integrands, suitable representations of the latter are sometimes difficult to find at higher loops or at sufficiently high multiplicity.
The streamlined cut-merging procedure enabled by using an integrand basis allows for the efficient application of the double-copy construction at all loop orders while simultaneously controlling the spectrum of the resulting theory. Indeed, the readily-available color-dual tree-level amplitudes~\cite{Bjerrum-Bohr:2010pnr, Mafra:2011kj, Edison:2020ehu} lead straightforwardly to cuts of gravitational amplitudes which are then directly converted into off-shell integrands by reading off the coefficients of the basis elements.
This side-steps difficulties~\cite{Bern:2012uc, Mogull:2015adi, Bern:2017yxu, Bern:2017ucb} with finding gauge-theory integrands that satisfy the duality between color and kinematics, and at the same time allows us to project out unwanted states, e.g.~the dilaton and antisymmetric tensor, by simply not including them in cuts. 

We demonstrated these concepts through several examples, including a detailed illustration of how scalar QED amplitudes are constructed from generalized cuts. Another example involved the manifestation of Adler zeros in the integrand of NLSM amplitudes at one and two loops. Additionally, we provided a detailed example in general relativity to illustrate that having the double copy at tree level directly implies having the double copy to all loop orders with straightforward means to project out the dilaton and antisymmetric tensor from the spectrum. 
Apart from the examples described in this paper, we have used this procedure to construct the four-loop integrand needed for determining the $\mathcal{O}(G^5)$ classical two-body interactions in general relativity and to eliminate the spurious terms in the analogous calculation in ${\cal N}=8$ supergravity~\cite{Bern:2024adl}. 

In our discussion, we systematically discarded contributions containing bubbles on external lines and tadpoles. In massless amplitudes, while such terms are potentially ill-defined and require additional regularization in addition to dimensional regularization, the corresponding integrals vanish identically. Massless amplitudes are, therefore, complete. 
In massive theories, such contributions are not zero. Within generalized unitarity, they can be determined from single-particle cuts or/and in the presence of further off-shell regularization of external data. They can also be determined by means other than generalized unitarity, by demanding that {\em integrated} amplitudes have the correct ultraviolet and infrared singularities in the massless limit. 
Thus, in any chosen basis, generalized unitarity uniquely determines the well-defined terms, which in turn, together with consistency conditions, uniquely determine the potentially problematic ones.

Our definition and approach to the construction of an integrand basis assumes that each term has a graph interpretation: in each term, the propagators are specified in terms of a graph, and graph information is used to map each term to a unique combination of basis elements. 
In general, a graph interpretation may not be necessary. Given an integrand without such an interpretation, i.e. a rational function, a possible approach to constructing a basis is to first partial fraction it to terms with least-complicated denominators, akin to diagram topologies, Feynman-parameterize each term and then use isomorphism algorithms based on Symanzik polynomials to identify a basis in the space of rational functions with those (and less complex) denominators. The same data is also used to construct the mapping or, equivalently, the decomposition of the original rational function in this basis. 
Integrands without a diagrammatic interpretation, perhaps exhibiting spurious poles, arise, e.g., in the application of recursion relations. They might also be useful---or even necessary---to manifest properties of amplitudes which are visible only when multiple graphs are combined. Examples are enhanced ultraviolet cancellations in supergravity theories~\cite{Bern:2014sna}, which are not visible diagram by diagram.
While appealing, we leave it for future investigation.

Global planar and nonplanar integrand bases, as described in this paper, exist in all quantum field theories. We look forward to using them for novel high-order calculations. These ideas are particularly well-suited for problems in (super)gravity, as they align seamlessly with the double copy. Besides applications to the classical two-body problem, another promising use is to high-loop order calculations of ultraviolet properties of supergravity theories with various amounts of supersymmetry, which could shed light on the origin and generality of enhanced ultraviolet cancellations~\cite{Bern:2014sna}.

\section*{Acknowledgements}
\vskip -.4cm
We thank Clifford Cheung and Jaroslav Trnka for raising the interesting question about the fate of integrand-level Adler zeroes for the nonlinear sigma model. We thank Calvin Chen, Tushar Gopalka, and Callum Jones for collaboration on related work.
Z.B., E.H., and M.R.~are supported in part by the U.S. Department of Energy (DOE) under award number DE-SC0009937.
R.R.~is supported by the U.S.  Department of Energy (DOE) under award number~DE-SC00019066.
M.Z.'s work is supported in part by the U.K.\ Royal Society through Grant
URF\textbackslash R1\textbackslash 20109. For the purpose of open access,
the author has applied a Creative Commons Attribution (CC BY) license to any
Author Accepted Manuscript version arising from this submission.
The authors acknowledge the Texas Advanced Computing Center (TACC) at The University of Texas at Austin for providing high performance computing resources that have contributed to the research results reported within this paper (URL: http://www.tacc.utexas.edu).
In addition, this work used computational and storage services associated with the Hoffman2 Shared Cluster provided by UCLA Office of Advanced Research Computing’s Research Technology Group.
We are also grateful to the Mani L. Bhaumik Institute for Theoretical Physics for support.
\vskip -1.1cm
\textcolor{white}{.}
\bibliographystyle{JHEP}
\bibliography{refs.bib}
\end{document}